\documentclass{emulateapj}

\newcommand{\fun}{\ \mathrm{erg} \ \mathrm{cm}^{-2} \ \mathrm{s}^{-1}}
\newcommand{\CLhz}{RDCS J1252.9-2927}

\shorttitle{Spectroscopic survey of \CLhz}
\shortauthors{Demarco et al.}

\begin{document}

\title{VLT and ACS observations of \CLhz: dynamical structure and
galaxy populations in a massive cluster at z=1.237\footnote{Based on
observations carried out with the ESO VLT under programs 166.A-0701,
69.A-0683, 73.A-0832 and 76.A-0889}}

\author{R. Demarco\altaffilmark{1,2}, P. Rosati\altaffilmark{3},
C. Lidman\altaffilmark{4}, M. Girardi\altaffilmark{5},
M. Nonino\altaffilmark{6}, A. Rettura\altaffilmark{1},
V. Strazzullo\altaffilmark{3}, A. van der Wel\altaffilmark{1},
H. C. Ford\altaffilmark{1}, V. Mainieri\altaffilmark{3},
B. P. Holden\altaffilmark{7}, S.~A. Stanford\altaffilmark{8},
J.~P. Blakeslee\altaffilmark{9}, R. Gobat\altaffilmark{3},
M. Postman\altaffilmark{10}, P. Tozzi\altaffilmark{6,11},
R.~A. Overzier\altaffilmark{1}, A.~W. Zirm\altaffilmark{1},
N. Ben\'{\i}tez\altaffilmark{12}, N.~L. Homeier\altaffilmark{1},
G. D. Illingworth\altaffilmark{7}, L. Infante\altaffilmark{13},
M.~J. Jee\altaffilmark{1}, S. Mei\altaffilmark{14,15},
F. Menanteau\altaffilmark{16}, V. Motta\altaffilmark{17},
W.~Zheng\altaffilmark{1}, M.~Clampin\altaffilmark{18},
G. Hartig\altaffilmark{10}}

\altaffiltext{1}{Department of Physics and Astronomy, Johns Hopkins
University, Baltimore, MD 21218. USA}

\altaffiltext{2}{demarco@pha.jhu.edu}

\altaffiltext{3}{ESO - European Southern
Observatory. Karl-Schwarzschild-Str. 2. D-85748 Garching bei
M\"unchen. Germany}

\altaffiltext{4}{ESO - European Southern Observatory. Alonso de
Cordova 3107, Casilla 19001, Santiago, Chile}

\altaffiltext{5}{Dipartimento di Astronomia, Universit\`a degli Studi
di Trieste, via Tiepolo 11, 34131 Trieste, Italy; INAF - Osservatorio
Astronomico di Trieste, via Tiepolo 11, 34131 Trieste, Italy}

\altaffiltext{6}{INAF - Osservatorio Astronomico di Trieste, via
G.B. Tiepolo 11, 34131 Trieste, Italy}

\altaffiltext{7}{UCO/Lick Observatory, University of California, Santa
Cruz, 1156 High Street, Santa Cruz, CA 95065}

\altaffiltext{8}{University of California, Davis, CA 95616; Institute
of Geophysics and Planetary Physics, Lawrence Livermore National
Laboratory, Livermore, CA 94551}

\altaffiltext{9}{Department of Physics and Astronomy, Washington State
University, Pullman, WA 99164-2814}

\altaffiltext{10}{Space Telescope Science Institute, 3700 San Martin
Drive, Baltimore, MD 21218}

\altaffiltext{11}{NFN, National Institute for Nuclear Physics, Trieste,
Italy}

\altaffiltext{12}{Instituto de Astrof\'{\i}sica de Andaluc\'{\i}a
(CSIC), Camino Bajo de Hu\'etor 50, Granada 18008, Spain}

\altaffiltext{13}{Departamento de Astronom\'{\i}a y Astrof\'{\i}sica,
Pontificia Universidad Cat\'olica de Chile, Casilla 306, 22 Santiago,
Chile}

\altaffiltext{14}{University of Paris Diderot, 75205 Paris Cedex 13, France}

\altaffiltext{15}{GEPI, Observatoire de Paris, Section de Meudon,
92195~Meudon Cedex, France}

\altaffiltext{16}{Department of Physics and Astronomy, Rutgers, the
State University of New Jersey, 136 Frelinghuysen Road, Piscataway, NJ
08854-8019}

\altaffiltext{17}{Departamento de F\'{\i}sica y Meteorolog\'{\i}a,
Universidad de Valpara\'{\i}so, Avda. Gran Breta\~na 1111,
Valpara\'{\i}so, Chile}

\altaffiltext{18}{NASA Goddard Space Flight Center, Code 680,
Greenbelt, MD 20771}

\begin{abstract}
We present results from an extensive spectroscopic survey, carried out
with VLT FORS, and from an extensive multiwavelength imaging data set
from the HST Advanced Camera for Surveys and ground based facilities,
of the cluster of galaxies \CLhz. We have spectroscopically confirmed
38 cluster members in the redshift range $1.22 < z < 1.25$. A cluster
median redshift of $z=1.237$ and a rest-frame velocity dispersion of
$747^{+74}_{-84}$ {\rm km~s$^{-1}$} are obtained. Using the 38
confirmed redshifts, we were able to resolve, for the first time at $z
> 1$, kinematic structure. The velocity distribution, which is not
Gaussian at the 95\% confidence level, is consistent with two groups
that are also responsible for the projected east-west elongation of
the cluster. The groups are composed of 26 and 12 galaxies with
velocity dispersions of $486^{+47}_{-85}$ km s$^{-1}$ and
$426^{+57}_{-105}$ km s$^{-1}$, respectively. The elongation is also
seen in the intracluster gas and the dark matter distribution. This
leads us to conclude that \CLhz\ has not yet reached a final virial
state. We extend the analysis of the color-magnitude diagram of
spectroscopic members to more than $1$ Mpc from the cluster
center. The scatter and slope of non-[OII]-emitting cluster members in
the near-IR red sequence is similar to that seen in clusters at lower
redshift. Furthermore, most of the galaxies with luminosities greater
than $\sim K_s^*+1.5$ do not show any [OII], indicating that these
more luminous, redder galaxies have stopped forming stars earlier than
the fainter, bluer galaxies. Our observations provide detailed
dynamical and spectrophotometric information on galaxies in this
exceptional high-redshift cluster, delivering an in-depth view of
structure formation at this epoch only 5 Gyr after the Big Bang.
\end{abstract}


\section{Introduction}

Clusters of galaxies are tracers of the peaks of matter density in the
universe. Their study over a wide range in redshift provides an
insight into the process of mass assembly of structures through cosmic
history, from galactic to Mpc scales. They are suitable laboratories
in which to study galaxy populations, providing clues to better
understand the effects of the local environment on galaxy properties
and galaxy evolution. Cluster X-ray luminosity and temperature
indicate the existence of massive systems ($M > 10^{14} M_{\odot}$)
already when the universe was half its present age
\citep{jcb01,mje03,mje04,rte04,mrl05}. While some of these clusters
have a relaxed X-ray morphology by $z \sim 0.8$ \citep{mje04}, others
show clear signatures of being still in a formation stage, as
indicated by their filamentary morphology and the presence of
substructure \citep{ghm99,gbb04,drl05,gdrb05,tka06}.

The dense cluster environment has a profound effect on the properties
of the cluster galaxies. Most notably, early-type galaxies are more
prevalent in clusters than in low-density, field environments
\citep{d80}, and this morphology-density relation has existed since
$z\sim1$ \citep{doc97,ste05,pfc05}. The existence of the
morphology-density relation suggests that the properties of the
stellar populations of galaxies also depend on their environment,
which is corroborated by the relation between star-formation history,
color, and environment
\citep{bsm98,hp03,kwh04,wcn05,pfc05,hpm06,tk06}.

Measurements of the evolution of the mass-to-light ratio of massive
cluster early-type galaxies have demonstrated that their stellar
populations were largely formed at $z>2$
\citep*[e.g.,][]{vds03,hwf05}. The slow evolution of the
color-magnitude relation
\citep*[e.g.,][]{bfp03,hse04,bhf06,mbs06,mhb06,hmb06} shows that this
is the case for early-type galaxies well below the characteristic
luminosity although there seems to be a lack of faint, red galaxies at
redshifts higher than $z \sim 0.5$ \citep{tka05}. Remarkably, massive
field early-type galaxies have been shown to be not much younger, in
terms of their stellar populations, than cluster early-type galaxies
with the same mass \citep{vdwf05,tel05,vdvdm06}. This raises the
question whether galaxy mass or environment is the most important
driver of galaxy evolution.

Most of our knowledge of cluster evolution comes from observations at
redshift lower than unity and a few sparse data sets at $z = 1$-$1.3$.
Observing galaxy clusters at $z > 1$ is difficult and only a small
number of galaxy clusters at such a high redshift have been confirmed
to date
\citep{see97,rse99,shr02,rte04,hbb04,mrl05,seb05,srsd06,bba06,egm06}.
With the primordial activity of cluster and massive galaxy formation
happening at $z \gtrsim$ 1, the observations of systems at those
redshifts offer the possibility of learning more about their physical
properties at an epoch when baryons are first being assembled into
gravitationally bound systems within massive dark matter halos. One of
those high redshift clusters, \CLhz\ \citep{rte04}, is a massive X-ray
luminous system at $z=1.24$ selected from the ROSAT Deep Cluster
Survey \citep*[RDCS;][]{rdcs98}, which has been the center of an
intensive multi-wavelength campaign in the past 5 years.

\CLhz\ ($\alpha_{J2000}=12^h 52^m 48^s$, $\delta_{J2000}=-29^o 27^{'}
00^{''}$) was discovered in the $15.7$ ksec ROSAT PSPC field with ID
WP300093 at an off-axis angle of 13\farcm9 with 31 net counts,
corresponding to a flux of $(2.5\pm0.9) \times 10^{-14} \fun$ in the
0.5-2 keV band. I-band imaging (30 minutes exposure), obtained with
the Prime Focus camera at the CTIO 4-m telescope in February 1997,
revealed a faint ($I\simeq 21.7$) galaxy pair very close to the X-ray
centroid position. In November 1998, $J$- and $K_s$-band imaging
obtained with SofI \citep{mcl98} on the ESO New Technology Telescope
(NTT) showed a clear overdensity of red galaxies with $J-K_s\simeq
1.9$, typical of early-type galaxies at $z>1$ \citep{lrd04}.

X-ray imaging with Chandra and XMM-Newton has been obtained allowing a
complete modeling of the X-ray surface brightness of the cluster
together with an unprecedented accuracy in the estimation of its
temperature, metallicity and dynamical mass \citep{rte04}. \CLhz\ is a
massive structure with a luminosity of $L_x(bol)=(6.6\pm1.1) \times
10^{44} \ \mathrm{erg} \ \mathrm{s}^{-1}$ ($H_0$ = 70 {\rm km
s$^{-1}$} Mpc$^{-1}$, $\Omega_m$=0.3, and $\Omega_{\lambda}$=0.7)
measured within an aperture of 60\arcsec (500 kpc) radius and a total
mass of $M=(1.9\pm0.3) \times 10^{14} M_{\odot}$ within a $R_{500}$
radius\footnote{$R_{\Delta}$ is defined as the radius within which the
mean density is $\Delta$ times the critical density of the universe at
a given redshift.} of $536\pm40\ \mathrm{kpc}$ \citep{rte04}. Its
intracluster medium (ICM) is characterized by a temperature
$T_x=6.0^{+0.7}_{-0.5}\ \mathrm{keV}$ and a metallicity
$Z=0.36^{+0.12}_{-0.10} Z_{\odot}$ \citep{rte04}, and the surface
brightness profile shows a discontinuity typical of cold fronts
systems found in Chandra observations of low redshift clusters.

High angular resolution optical observations of the cluster with the
Advanced Camera for Surveys \citep*[ACS;][]{F98} on the Hubble Space
Telescope (HST) delivered unprecedented morphological information for
the cluster galaxy populations as well as accurate photometry
\citep{bfp03}. Additional ground based photometry in the near-IR
\citep{lrd04} has been obtained with ISAAC\footnote{ISAAC stands for
Infrared Spectrometer And Array Camera.} on the ESO Very Large
Telescope (VLT), while Spitzer/IRAC observations of \CLhz\ (Stanford
et al., in preparation) have made it possible to sample the rest-frame
near-IR light of the cluster.

During the past few years we have carried out an extensive
spectroscopic survey of \CLhz\ with the ESO VLT aimed at confirming a
large fraction of the cluster population. This effort has yielded the
most complete spectroscopic dataset so far on this cluster. In this
paper we present the results from this ESO VLT/FORS\footnote{FORS
stands for FOcal Reducer and low dispersion Spectrograph.}
spectroscopic program in combination with ACS and VLT/ISAAC imaging
data. This investigation is aimed at spectroscopically identifying the
cluster galaxy populations in order to study the cluster dynamics from
galaxy kinematics and the spectrophotometric properties of cluster
members. Unless otherwise indicated, we assume a $\Lambda$CDM
cosmology with $H_0$ = 70 {\rm km s$^{-1}$} Mpc$^{-1}$,
$\Omega_M$=0.3, and $\Omega_{\Lambda}$=0.7.

\section{Data set and data reduction}

\subsection{ACS and ground based imaging data}\label{imaging}

Imaging observations of \CLhz\ in the optical have been carried out
from the ground and from space. The ACS Wide Field Camera (WFC) on HST
was used to obtain imaging with the F775W and F850LP filters
(hereafter $i_{775}$ and $z_{850}$, respectively), as part of a
guaranteed time observation program (ID: 9290). Three orbits in
$i_{775}$ and five orbits in $z_{850}$ were completed during May and
June of 2002, distributed in a $2\times2$ mosaic pattern. The
pointings overlapped about 1\arcmin, producing exposures of 12 orbits
in $i_{775}$ and 20 orbits in $z_{850}$ of the cluster core. A more
detailed description of the ACS data on \CLhz\ is presented in
\citet{bfp03}.

The ground based optical data were collected with FORS2 \citep{ar92}
on the VLT, under ESO program 169.A-0458(A). The imaging data in the
$B$-, $V$- and $R$-band\footnote{The $B$-, $V$- and $R$-band
correspond to the B\_BESSEL, V\_BESSEL and R\_SPECIAL (FORS2) or
R\_BESSEL (FORS 1) filters, respectively.} were obtained between March
1st and March 3rd, 2003, in very good seeing conditions. The co-added
images have FWHM of 0\farcs66 in the $B$-band, 0\farcs60 in the
$V$-band, and 0\farcs56 in the $R$-band. The photometric calibration
was done using the many Landolt field stars that were observed during
the same nights. The images have been corrected for the instrumental
response, bias and flat, in a standard manner. All images were
photometrically aligned by using about one hundred bright but
unsaturated stars in the field, which showed that the nights were
indeed photometric. The images were astrometrically flattened and then
finally co-added. As a further check, the same point sources used for
the single images photo alignment were compared with FORS1 data
acquired in May 1999, in the same $B$-, $V$-, and $R$-band. The mean
differences in magnitude for the selected objects were
-0.006$\pm$0.030 in $B$ and 0.015$\pm$0.021 in $V$. In the $R$-band, a
$V-R$ color term correction was found giving a mean difference of
0.002$\pm$0.032.

Ground based near-IR imaging observations of \CLhz\ were taken with
ISAAC \citep{mcb99} on the ESO VLT and are described in
\citet{lrd04}. In the central regions of the cluster the data reach
limiting magnitudes ($5\sigma$ limit over a 0\farcs9 diameter
aperture) of 26.5 (AB) and 26.0 (AB) in the $J$- and $K_s$-bands,
respectively. Since then, these data have been reprocessed to optimise
image quality and to facilitate the computation of aperture
corrections over the entire region covered by near-IR
observations. The image quality in the reprocessed images varies from
0\farcs32 to 0\farcs43. The FWHM of point sources in a single tile is
relatively constant and this makes it simpler to compute these
aperture corrections, whereas the FWHM of point sources in the mosaic
varies from tile to tile. Hence, rather than combining the images into
a single mosaic, each tile was processed and analysed separately.
This means that some objects appear multiple times as there is
considerable overlap between different tiles. To handle this, we
create catalogs for each tile and if a source is in more than one
catalog, we average the result. We also compute an error from the
different measurements and compare it to the error computed in
SExtractor. If the source is in only one catalog, we use the error
computed in SExtractor, otherwise we use either the SExtractor error
or the error derived from the multiple measurements, whichever is the
largest. Magnitudes are measured in 10 pixel diameter apertures
($\sim$1\farcs5) and then corrected to larger apertures (4\arcsec
diameter) using stars in each tile to compute aperture corrections.

\subsection{The photometric catalog}\label{photcat}

A first multi-color catalog was built out of the above data sets. This
catalog contained the photometry computed from the ISAAC mosaic as
described in \citet{lrd04}, without including the more recently
reprocessed ISAAC photometry (see \S\ref{imaging}). We note that the
original ISAAC photometry in \citet{lrd04} and the reprocessed one
presented here (see below) are both in good agreement with each other,
showing median differences of 0.002 in $K_s$ and 0.007 in $J_s$ with
dispersion ($\sigma$) values of 0.064 and 0.033, respectively. The
ACS, FORS2 and ISAAC images have different PSF FWHM. In order to avoid
source blending due to a plain smoothing of the best seeing images to
match the worst one, we adopt a different strategy to correct for
these different PSFs.  Using the same bright point-like sources as
before (see \S\ref{imaging}), we construct growth curves. We then
computed the aperture magnitude for all ground based and ACS images
using an aperture of 1\farcs5 diameter. An aperture correction from
1\farcs5 to 4\arcsec\ diameter has been first computed using point
sources, and then applied to all sources in the optical (ACS and
FORS2) and near-IR (ISAAC) bands. Although this aperture correction is
not strictly equivalent to the smoothing technique, we have checked,
using the ACS data, that the two approaches give consistent
results. The main advantage of using this approach instead of the
smoothing one in this analysis is that we avoid significant source
blending, especially in the central region of the cluster. This
catalog was used to select candidates for spectroscopy while preparing
most of the FORS2 masks.

The final photometric catalog published in this work and showing
only spectroscopic cluster members is presented in Table
\ref{phot_tab}. It contains the $B$-, $V$- and $R$-band photometry
from FORS2, the $i_{775}$- and $z_{850}$-band photometry from ACS and
the newly processed $J_s$ and $K_s$ ISAAC photometry, all of them as
described above. The magnitudes are aperture corrected to a radius of
2\arcsec\ and are in the AB system \citep{o74}. To transform our
photometry from the Vega system to the AB system, we recomputed AB
corrections adopting the latest available Vega spectrum. The AB
corrections for the B, V, R, $J_s$, $J$ and $K_s$ filters are -0.088,
0.052, 0.244, 0.968, 0.964 and 1.899 respectively. For the $i_{775}$
and $z_{850}$ filters we use AB correction values of 0.401 and 0.569,
respectively \citep{btr03}. Values in Table \ref{phot_tab} are
corrected for galactic extinction. In order to take into account the
galactic reddening in our analyses, we used the extinction maps of
\citet{sfd98}. The corrections for the $B$-, $V$-, $R$-, $i_{775}$-,
$z_{850}$-, $J_s$- and $K_s$- bands are 0.323, 0.248, 0.200, 0.145,
0.127, 0.067 and 0.027, respectively. Near-IR magnitudes presented in
Fig. \ref{color_selection} and Fig. \ref{cmd} are placed on the 2MASS
$J$,$K_s$ system \citep{c01,cwm03}. The corresponding transformation
including the galactic reddening correction is given by:

\begin{equation}
{(J-K_s)}_{2MASS} = 1.038 {(J_s-K_s)}_{ISAAC} - 0.048
\label{2mass_col}
\end{equation}

\noindent
and

\begin{equation}
{K_s}_{2MASS} = {K_s}_{ISAAC} - 0.028 \\ .
\label{2mass_mag}
\end{equation}

\noindent
Unless otherwise stated, magnitudes presented throughout this paper
are in the AB system.

\subsection{Photometric selection}\label{selection}

The above photometric information was used to compute photometric
redshifts \citep{b00} for all the objects within the field of view
(FoV) covered by the ISAAC mosaic \citep{lrd04}. This region was
chosen in order to have optical ACS, optical FORS2 and near-IR ISAAC
photometry available (see Fig. \ref{galdist}) to compute photometric
redshifts. The set of galaxy templates for the photometric redshift
calculations is presented in \citet{bfb04}. Photometric redshifts
together with color-color diagrams, were used to select candidate
galaxies for spectroscopy. In our first spectroscopic campaign,
galaxies with $K_s< 21$, $J_s-K_s<2.1$ and $R-K_s>3$ were
targeted. This selection of the spectroscopic sample, indicated by the
dashed lines in Fig. \ref{color_selection}, minimizes the pollution by
field galaxies, although it can be slightly biased against very red
($J_s-K_s \gtrsim 2.2$) cluster galaxies. This primary selection was
complemented by another one on the $V-I$ vs $I-z_{850}$ plane in order
to improve the selection of Balmer-Break and [OII] ($\lambda 3727$)
galaxies. The color tracks in Fig. \ref{color_selection} correspond to
different evolutionary models, computed with the BC03 code
\citep{bc03}, reproducing galaxy colors observed at $z=1.237$ for
different ages (from 1 to 5 Gyr). These models were employed together
with the available multicolor information to guide the selection of
targets. The orange lines represent single burst, solar metallicity
models with burst durations of 0.5 Gyr (dot-dashed), 1.0 Gyr (dashed)
and 2.0 Gyr (solid). The green lines represent exponentially declining
star formation rate (SFR), solar metallicity models with
characteristic times of $\tau=1$ Gyr (dot-dashed), $\tau=2$ Gyr
(dashed) and $\tau=5$ Gyr (solid). Model points, indicated by small
open circles, are spaced by 0.5 Gyr in age and are in the ISAAC
system. The blue line corresponds to 6 starburst galaxy templates from
\citet{kcb96}, hereafter KC templates, spanning the mean color excess
range $0.0 < E(B-V) < 0.70$ and marked by open diamonds. In subsequent
campaigns, we took advantage of photometric redshifts in the range
$1.1 \lesssim z_{phot} \lesssim 1.3$ as they became available.

\subsection{VLT Spectroscopy}\label{spectroscopy}

A total of 15 masks were designed to be observed with FORS1 and
FORS2. The data were obtained as part of an ESO Large Program
(LP-166.A-0701) and two subsequent ESO proposals (69.A-0683 and
073.A-0832). The spectroscopic observations started in March 2000 and
ended in April 2004. The first mask was observed in MOS mode with
FORS1 and the 300I grism, whereas the other 14 masks were all observed
with FORS2. Spectroscopy with FORS2 was carried out with both the 300I
and 150I grisms. All the masks designed for MXU spectroscopy were
observed with the 300I grism while all but one MOS mode observations
were carried out with the 150I grism. The slit width for MXU
observations was set to 1\arcsec\ while the MOS masks were designed
with slit widths of 1\farcs4 and 1\farcs7. A summary of our
spectroscopic observations is presented in Table \ref{spec_tab} and
more detailed information about the MOS and MXU modes as well as the
grism characteristics are given in \citet{drl05}. From October 2002
on, the observations were carried out with a CCD mosaic composed of
two 2k $\times$ 4k MIT/LL detectors with increased sensitivity ($\sim$
30\%) in the red, allowing us to obtain a better success rate compared
to previous observations with the old FORS2 CCD. The data were reduced
following the same procedure described in \cite{drl05}.

The efficient selection criteria of targets based on high quality
imaging data has been a key piece in the success of our spectroscopic
campaign. From a total number of 418 galaxies targeted with the 15
masks on FORS (see Table \ref{spec_tab}) we were able to obtain 282
(67\%) redshifts, from which 227 (54\%) correspond to secure
measurements. We consider as cluster members all galaxies in the
redshift range $1.22<z<1.25$, which corresponds to about
$\pm3\sigma_v$ around the cluster median velocity, where $\sigma_v$
the cluster velocity dispersion (see \S\ref{redshifts}). With this
criterion 38 galaxies are classified as spectroscopic cluster members
(see Table \ref{spec_phot_tab}). The signal-to-noise ratio (SNR) per
resolution element of cluster member galaxies in the range $3995$ \AA\
$< \lambda < 4085$ \AA\ rest-frame is observed to vary between $\sim$1
to $\sim$10. The redshifts and the corresponding error bars were
obtained from the cross-correlation between the object and a template
spectra as described in \citet{drl05}. In the cases of very poor SNR
(SNR $\sim 1$), the redshift was determined from the identification of
[OII]($\lambda$3727) in the spectrum. In the case of passive galaxies,
i.e., galaxies with an absorption line spectrum and no visible
emission lines, the redshift was obtained from the cross-correlation
technique as implemented in the RVSAO/XCSAO task in IRAF. In 23
objects (all with SNR $> 2.5$), CaII (H+K) features were identified as
well. Yet, further spectroscopic classification, such as the one
proposed by \citet{dsp99}, was not possible with the current data due
to the lack of an accurate measurement of the $H_{\delta}(\lambda
4102)$ feature. In general, we avoided observing the same object more
than once, however, we obtained more than one redshift measurement for
a small number of objects. This allowed us to estimate total
(random+systematic) errors, yielding typical values of $\delta z \sim
12 \times 10^{-4}$. This value represents a more rigorous estimate of
the redshift errors since the formal errors given by the
cross-correlation \citep{td79} are known to be smaller than the true
errors \citep*[e.g.,][]{mkd92,bzv94,ey94,qcr00}, and will be used in
computations.

Those spectroscopic members for which FORS2 and ISAAC photometry is
available (see table \ref{phot_tab}) are shown in
Fig. \ref{color_selection}. The red circles indicate spectroscopically
confirmed non-emission line members, while blue triangles indicate
emission line [OII]($\lambda 3727$) confirmed members. The emission
line object in the upper right side of the plot, marked with a square,
corresponds to an X-ray source at the cluster redshift \citep*[ID=174;
see][]{rte04,mmt07}. Note the unusual morphology of this source (see
Fig. \ref{mosaic}) with a faint red nucleus and an irregular, diffuse
structure \citep*[also see][]{mmt07}. In Table \ref{xray_tab} we
present the X-ray information corresponding to this source and other
X-ray sources in the cluster FoV, for which a redshift measurement was
obtained. The X-ray fluxes and rest frame X-ray luminosities in the
soft ([0.5-2] keV) and hard ([2-10] keV) bands were extracted from the
190 ksec Chandra observations \citep{rte04} assuming a power law
spectrum with photon index 1.4.

We estimated the success rate of our spectroscopic survey as the ratio
of the number of objects with spectroscopic redshifts to the number of
objects that were targeted for spectroscopy as a function of $K_s$
(AB) magnitude. The data were binned in $\Delta K_s=0.5$ mag intervals
and include only objects for which an estimate of their $K_s$
magnitude is available, as shown in Fig. \ref{ssr}. Beside the overall
success rate of our survey (solid histogram), the success rate for
galaxies from the primary color-color selection (see section
\S\ref{imaging}) is shown as the dashed histogram. A secondary
color-color selection, defined by $I-z_{850} > 0.4$, $I-z_{850} <
0.85$, $V-I > 0.2$, $V-I < 1.2$, $V-I > [2.4\ (I-z_{850})-1.12]$ and
$V-I < [7.0\ (I-z_{850})-2.3]$, was used to identify possible
Balmer-Break and star-forming galaxies.  The success rate of this
additional cut in color-color space is shown in Fig. \ref{ssr} as
the dotted histogram. For galaxies within the primary color-color
cut, the success rate decreases dramatically from about 85\% to about
20\% for galaxies with magnitudes fainter than 21.5 in $K_s$. At
fainter magnitudes, the secondary color-color selection was more
effective, as intended, at detecting sources with emission lines,
increasing the number of galaxies per mask for which a redshift could
easily be obtained. This secondary selection allowed us to obtain the
redshift for $23$ faint objects with $K_s > 21.5$. This corresponds to
$\sim 12\%$ of all (no color restriction) targeted sources with $K_s
> 21.5$. We note that all the sources within the $V-I$ vs $I-z_{850}$
cut with $K_s > 21.5$ and spectroscopic confirmation show some
star-forming activity, while there are no targeted objects in this
secondary color-color region with $K_s < 20.2$. The $V-I$ vs
$I-z_{850}$ selection was intended to target star-forming galaxies at
the cluster redshift, however only 8 confirmed members with [OII] were
confirmed this way. In addition, no secure redshift was obtained for
objects with $K_s < 20.5$ in this secondary color-color selecting
area.

\section{Analysis}

\subsection{Member selection and global properties}\label{redshifts}

The distribution in redshift space of the 227 galaxies with a secure
estimate of their redshift is shown in Fig. \ref{z_survey}. The bin
size of the histogram is $\Delta z=0.01$. We perform the
adaptive-kernel method \citep*[hereafter
DEDICA;][]{p93,p96,fgg96,gfg96,gm01} to search for the significant
($>$99\% confidence level [c.l.])  peaks in the velocity
distribution. This procedure detects \CLhz\ as the strongest peak at
$z\sim1.24$ (see Fig. \ref{z_survey}) populated by 38 galaxies
(hereafter referred to as cluster members), the largest number of
spectroscopic members discovered so far in a cluster of galaxies at $z
> 1$, distributed over a region of $1.8$ Mpc in radius. Out of
non-member galaxies, 169 and 20 are foreground and background
galaxies, respectively (see Table \ref{spec_field_tab}). In
particular, 33 foreground galaxies belong to a dense peak at $z\sim
0.74$. This structure, in the redshift range $0.70 < z < 0.79$ (see
Table \ref{spec_field_tab}), has a 3$\sigma$-clipped mean redshift of
$\bar{z}=0.7429\pm0.0024$. The projected distribution of these sources
extends over a region of about 5\farcm6 a side in front of the
cluster, showing a small concentration of about 10 galaxies at about
1\arcmin\ north from the cluster center. There is no X-ray extended
emission associated to this structure indicating the lack of a hot gas
component, although we note that there are 3 X-ray point sources at
this redshift (see Table \ref{xray_tab}). This ``group'' may
contribute to overestimate the cluster mass obtained by weak lensing
analyses as discussed in \citet{lrb05}.

By applying the biweight estimator to the cluster members
\citep{bfg90}, we compute a mean cluster redshift of
$\left<z\right>=1.2370\pm$ 0.0004. We estimate the line-of-sight (LOS)
velocity dispersion, $\sigma_{v}$, by using the biweight estimator and
applying the cosmological correction and the standard correction for
velocity errors \citep{ddd80}. We obtain $\sigma_{v}=747_{-84}^{+74}$
{\rm km \ s$^{-1}$}, where errors are estimated through a bootstrap
technique. The spectroscopic information available on the cluster
member sample is given in Table \ref{spec_phot_tab}, while the ACS and
ground based photometry of cluster members is presented in Table
\ref{phot_tab}. Fig.  \ref{mosaic} show 5\arcsec$\times$5\arcsec\
thumbnail images of the 37 spectroscopic cluster galaxies in the FoV
of ACS together with the corresponding FORS spectrum. The left and
middle panel correspond to the $i_{775}$ (rest-frame U) and $z_{850}$
(rest-frame B) data. The most prominent spectral features identified
in each spectrum in the displayed wavelength range are indicated (see
figure's caption for details). One cluster member, ID=7001, is out of
the ACS FoV and its spectrum is shown in Fig. \ref{g7001}. For
completeness, in Fig. \ref{mosaic_agn} we show the optical spectrum of
the 10 non cluster member AGN listed in Table \ref{xray_tab}. We do
not present any analysis of the spectroscopic properties of these
sources for being out of the scope of this work. Fig. \ref{z_cluster}
shows the distribution of \CLhz\ member galaxies in velocity and
redshift space. The hatched area indicates the distribution of
star-forming members. The secondary selection in $V-I$ vs $I-z_{850}$
allowed the discovery of 8 of the 17 cluster members with
[OII]($\lambda$3727) emission lines. If we separate cluster members
into two categories, passive and star-forming galaxies, there is no
significant offset in median velocity between the two categories (see
Fig. \ref{z_cluster}). The mean redshift of passive and emission line
[OII] spectroscopic members is $1.2373\pm0.0057$ and
$1.2369\pm0.0054$, respectively. The velocity dispersion of these
populations are also consistent with each other and with the overall
cluster velocity dispersion, within the uncertainties.

In Fig. \ref{radial_profile} we show the rest-frame velocity
vs. projected distance from the cluster center (lower
panel). Hereafter we consider as cluster center the position of the
X-ray center \citep{rte04}:
R.A.=$12^{\mathrm{h}}52^{\mathrm{m}}54.4^{\mathrm{s}}$,
Dec.=$-29^{\mathrm{o}} 27\arcmin 17\farcs5$ (J2000.0). We also show
the rest-frame velocities of the three brightest cluster members (IDs
291, 247, and 289, hereafter BCM1, BCM2, and BCM3). BCM1 is about
$0.3$ magnitudes brighter in $K_s$ than BCM2 and BCM3 (upper
panel). BCM1 and BCM3 are closely located near the position of the
X-ray center and the peak of the galaxy distribution as recovered from
2D DEDICA method.

Assuming that \CLhz\ is in dynamical equilibrium (this will be
discussed in \S\ref{discussion}) we can compute global virial
quantities.  Following the prescriptions of \citet{gm01}, we assume
for the radius of the quasi-virialized region R$_{\rm
vir}=$R$_{178}=0.17\times \sigma_{v}/H(z) = 1.61$ Mpc (see their
equation~1 after introducing the scaling with $H(z)$; see also
equation~ 8 of \citet{cye97} for R$_{200}$). Therefore, our
spectroscopic catalog samples the whole cluster virialized
region. We can compute the mass using the virial theorem
\citep*[][]{lm60,ggm98} under the assumption that the galaxies trace
the total mass:

\begin{equation}
M=M_{\rm svir}-\rm{SPT},
\end{equation}

\noindent where 
\begin{equation}
M_{\rm svir}=\left(\frac{3\pi}{2}\right) \times \left(\frac{\sigma_{v}^2\ {\rm R}_{\rm PV}}{G}\right)
\end{equation}

\noindent 
is the standard virial mass, R$_{\rm PV}$ a projected radius (equal to
two times the harmonic radius), and SPT is the surface pressure term
correction \citep{tw86}.  The estimate of $\sigma_{v}$ is generally
robust when computed within a large cluster region as shown in Fig.
\ref{sigma_profile}. At radii larger than $1.4$ Mpc the profile
becomes flatter, consistent with observations of low redshift clusters
\citep{gfg96,fgg96}, suggesting that at large cluster radii any
velocity anisotropy of cluster galaxies does not affect the value of
$\sigma_v (<R)$. We thus consider the above global value,
$\sigma_{v}$, as the cluster velocity dispersion.  The value of
R$_{\rm PV}$ depends on the size of the considered region, so that the
computed mass increases (but not linearly) when increasing the
considered region. Using the 37 galaxies within R$_{\rm vir}$ we
obtain R$_{\rm PV}=1.09\pm 0.15$ Mpc.  As for the SPT correction, we
assume a correction factor of 20\%, which we obtained by combining
data on many clusters \citep*[e.g.,][]{cye97,ggm98}. This leads to a
virial mass $M(<{\rm R}_{\rm vir}=1.61\,Mpc)=5.3_{-1.4}^{+1.3} \times
10^{14} M_{\odot}$.

We also used an alternative estimate of the virial mass \citep*[see
eq.~13 of][]{ggm98}.  This alternative estimate is based on the
knowledge of the galaxy distribution and, in particular, a galaxy
King-like distribution with parameters typical of
nearby/medium-redshift clusters: a core radius R$_{\rm
core}=1/20\times {\rm R}_{\rm vir}$ and a slope-parameter $\beta_{\rm
fit}=0.8$ \citep*[which gives a volume galaxy density at large radii
as $r^{-3 \beta_{\rm fit}}=r^{-2.4}$;][]{gm01}.  For the whole
virialized region we obtain R$_{\rm PV}=1.20$ Mpc where a 25\% error
is expected due to the fact that typical, rather than individual,
galaxy distribution parameters are assumed.  This leads to a virial
mass $M(<{\rm R}_{\rm vir}=1.61$ Mpc)=$5.9_{-2.0}^{+1.9} \times
10^{14} M_{\odot}$, in good agreement with the above value.

To compare our result with the estimate obtained from the X-ray data
for a smaller cluster region \citep{rte04} we assume that the cluster
is described by a King-like mass distribution (see above) or,
alternatively, a NFW profile where the mass-dependent concentration
parameter is taken from \citet{nfw97} and rescaled by the factor $1+z$
\citep{bks01,dbp04}. We obtain $M_{\rm proj}(<\rm{R}=0.536$
Mpc)=$(1.6-2.3) \times 10^{14} M_{\odot}$ in good agreement with that
found by \citet{rte04}. This value is also in agreement with that
found by \citet{etb04} within a similar projected radius. Using the
same mass distributions we also compute the projected mass within 1
Mpc for comparison with the weak lensing analysis by
\citet{lrb05}. The cluster mass distribution is truncated at one
virial radius or, alternatively, at two virial radii. The range of our
results is $M_{\rm proj}(<\rm{R}=1$ Mpc)=$(4.5-6.3) \times 10^{14}
M_{\odot}$, in agreement with \citet{lrb05} when taking into account
the uncertainties on the mass values, although systematically
lower. 

By comparing the cluster X-ray luminosity \citep{rte04} and the
cluster velocity dispersion derived above with the $L_X-\sigma_v$
relation presented in Fig. 2 of \citet{xw00}, we see a good agreement
between the observed $L_X$ and $\sigma_v$ values in \CLhz\ and those
from more than a hundred galaxy clusters selected from the
literature. Using the best fit $\sigma_v-T$ relation from
\citet{xw00}, without taking into account the fit errors, we obtain a
temperature of $T=3.89^{+0.61}_{-0.65}$ keV for a velocity dispersion
of $\sigma_v=747^{+74}_{-84}$ km s$^{-1}$. This temperature is more
than $3\sigma$ away from the observed value of $T_x=6.0^{+0.7}_{-0.5}$
keV \citep{rte04}, and still $\sim 0.5 \sigma$ lower if the fit errors
are considered.

\subsection{Substructures and projected distribution}\label{vel_struct}

We analyze the velocity distribution to look for possible deviations
from Gaussianity that could provide important signatures of complex
dynamics. For the following tests the null hypothesis is that the
velocity distribution is a single Gaussian.

We compute three shape estimators, i.e. the kurtosis, the skewness,
and the scaled tail index \citep*[see, e.g.,][]{bgf91}. The value of
the normalized kurtosis (-0.754) shows evidence that the velocity
distribution differs from a Gaussian, being lighter-tailed, with a
c.l. of $\sim 95-99\%$ \citep*[see Table~2 of][]{bb93}.

Then we investigate the presence of gaps in the distribution.  A
weighted gap in the space of the ordered velocities is defined as the
difference between two contiguous velocities, weighted by the location
of these velocities with respect to the middle of the data. We obtain
values for these gaps relative to their average size, precisely the
midmean of the weighted-gap distribution. We look for normalized gaps
larger than 2.25 since in random draws of a Gaussian distribution they
arise at most in about $3\%$ of the cases, independent of the sample
size \citep{ws78,bgf91}. One significant gap in the ordered velocity
dataset is detected dividing the dataset in two subsets containing 26
and 12 galaxies from ``low'' to ``high'' velocities (hereafter WGAP1
and WGAP2 groups; see top panel in Fig. \ref{radial_profile}).  We
compare these two subsets applying the the 2D Kolmogorov-Smirnov tests
to the galaxy positions \citep*[2DKS-test; see][]{ff87}, as
implemented by \citet{ptv92}. In spite of the modest statistics we
find a marginal significance (89\% c.l.)  for the difference. The
``high'' velocity WGAP2 group is located towards East with respect to
the ``low'' velocity group WGAP1 (see Fig.~\ref{figkmm}).

To further investigate the substructure membership we use the Kaye's
mixture model (KMM) test as implemented by \citet{abz94}.  The KMM
algorithm fits a user-specified number of Gaussian distributions to a
dataset and assesses the improvement of that fit over a single
Gaussian.  In addition, it provides the maximum-likelihood estimate of
the unknown $n$-mode Gaussians and an assignment of objects into
groups.  However, one of the major uncertainties of this method is the
optimal choice of the number of groups for the partition.  Using the
results of the gap analysis we decide to fit two groups and determine
the first guess for the group partition.  We do not find any group
partition which provides a significantly better description of the
velocity distribution with respect to a single Gaussian in the 1D
analysis. However, since the 3D diagnostic is, in general, the most
sensitive indicator of the presence of substructure
\citep*[e.g.,][]{prb96}), we apply the 3D version of the KMM software
using simultaneously galaxy velocity and positions. In the 3D case we
find that a two-group partition of 26 and 12 galaxies (hereafter KMM1
and KMM2 groups) better describes the velocity distribution at the
$95.6\%$ c.l., according to the likelihood ratio test (see
Fig. \ref{radial_profile}). These groups are located at $-354\pm159$
km s$^{-1}$ and $845\pm182$ km s$^{-1}$ with respect to the cluster
mean velocity, and to the West and East of the cluster center,
respectively. The velocity dispersions of KMM1 and KMM2 are estimated
to be $486^{+47}_{-85}$ km s$^{-1}$ and $426^{+57}_{-105}$ km
s$^{-1}$, respectively. This partition corresponds to that indicated
by the weighted gap analysis with the difference of two galaxies (see
Fig.~\ref{figkmm}). These velocity dispersion values are less robust
than the overall cluster velocity dispersion due to uncertainties in
the membership of galaxies to KMM1 and KMM2, although the detection
and location of the substructures are reliable. By assuming dynamical
equilibrium and theoretically estimating $R_{PV}$ for each KMM group
as done in section \S\ref{redshifts} for the whole cluster, we obtain
masses of $M (<\rm{R_{200}}=1.05$ Mpc)=$1.6^{+0.5}_{-0.7} \times
10^{14} M_{\odot}$ for KMM1 and $M (<\rm{R_{200}}=0.92$
Mpc)=$1.1^{+0.4}_{-0.6} \times 10^{14} M_{\odot}$ for KMM2. We note
that the sum of these masses is about the half of the cluster mass; we
will come back to this in section \S\ref{discussion}.

To look for further evidence that \CLhz\ is still not completely
dynamically relaxed, we analyzed the velocity of the brightest galaxy
BCM1. In fact, since its location coincides with the X--ray center,
one expects that BCM1 is at the center of the cluster potential and
thus at the center of the velocity distribution. We find that BCM1
shows evidence of peculiarity at the $>95\%$ c.l. according to the
Indicator test by \citet{gb91}, while it lies very close to the peak
of the velocity distributions of KMM1 (see Fig. \ref{radial_profile}).
We finally note that the close couple of luminous galaxies BCM1 and
BCM3 are separated by $\sim 550$ km s$^{-1}$ in rest-frame velocity
and, as first noted by \citet{bfp03}, in interaction as supported by
\citet{rrs06} who show signs of interaction in the form of an S-shaped
residual after galaxy subtraction linking the two galaxy centers. In
addition, BCM2 is near the edge of the coma-like structure of the
X-ray surface brightness reported by \citet{rte04}.

In Fig. \ref{galdist} we show the projected distribution of \CLhz\
spectroscopic members on the plane of the sky\footnote{The background
image corresponds to the ACS combined $i_{775}$- and $z_{850}$-bands
data.}. Circles are passive galaxies and triangles are emission line
members. The overall shape of the distribution of confirmed galaxy
members is clearly elongated in the East-West direction, as previously
reported in \citet{tmr04} based on the $K_s$-band light distribution
of photometric cluster members, although more uniform than that of
other high redshift clusters \citep*[see,
e.g.,][]{ghm99,drl05,gdrb05}. The projected distribution of the KMM
groups is consistent with the overall cluster galaxy distribution,
suggesting that the observed elongation is caused by the merger of
both groups.

\subsection{Spectral properties of cluster galaxies}\label{spec_prop}

The spectral features of cluster members indicate the stellar content
of galaxies in \CLhz\ and for some members, the [OII] line was used to
estimate their SFR. The spectrum of each one of the 37 cluster members
in the ACS FoV is shown in the right panel of Fig.  \ref{mosaic} while
the spectrum of the spectroscopic member outside the FoV of ACS is
shown in Fig. \ref{g7001}. The flux calibration of the spectra is not
accurate in the very red. Following \citet{hdr05}, we estimated the
SFR for the 17 member galaxies with [OII]$\lambda$3727 in their
spectra. We measured the integrated flux of the [OII] line using the
bandpass defined in \citet{tfi03}. In two cases with a very low
signal-to-noise continuum (IDs 174 and 9000), the [OII] line flux was
underestimated due to sky over-subtraction. Previous to these
measurements, all the spectra were doppler corrected by using the task
DOPCOR in IRAF. These line fluxes were converted into $L_{[OII]}$
luminosities by assuming that all these galaxies are at the same
distance, corresponding to the median redshift of the cluster
($z=1.237$). SFRs were derived from the $L_{[OII]}$ values by
following the prescription of \citet{kgj04}:

\begin{equation}
SFR_{[OII]}(M_{\odot} \ yr^{-1})=(6.58\pm1.65)\times 10^{-42}L_{[OII]}(ergs \ s^{-1}),
\end{equation}

\noindent
which takes into account the mean reddening corrected [OII]/$H
\alpha=(1.2\pm0.3)$ ratio from the Nearby Field Galaxies Survey
\citep{jff00}. We do not correct for metallicity and dust extinction
effects on the [OII] flux measurements, therefore our SFRs estimates
should be considered as lower limits to the true SFR. The [OII] line
can be considerably affected by dust absorption, giving SFRs one or
two orders of magnitude lower than SFRs unaffected by dust at 15
$\mu$m \citep{dpf02,cmm05}. A more robust SFR indicator is the $H
\alpha$ line \citep*[see, e.g.,][]{cl01}, however, at the cluster
redshift this line is not observed in the optical. IR luminosities can
also be used as reasonable tracers of SFR in galaxies
\citep{kgj02,ckb05}. Integrated [OII] line flux measurements, [OII]
equivalent widths and SFR values derived from $L_{[OII]}$ of
star-forming cluster galaxies are given in Table
\ref{spec_phot_tab}. Error bars are estimated by taking into account
the RMS fluctuation in flux within the two sidebands at both sides of
the [OII] bandpass, as defined in \citet{tfi03}. The median value of
the derived SFRs is 0.7 $M_{\odot}$ yr$^{-1}$, with a few galaxies
reaching SFRs greater than 2.0 $M_{\odot}$ yr$^{-1}$ (see Table
\ref{spec_phot_tab}). Typical errors in the SFR are about 47\%,
reaching values greater than 60\% in a couple of cases. Our median SFR
value is consistent within the errors with the mean SFR (0.17$\pm$0.02
$h_{100}^{-2} M_{\odot}\ yr^{-1}$) derived by \citet{bsm98} from the
[OII]($\lambda$3727) emission line for cluster galaxies at large
(R$_{200} \sim$1.5-2 $h^{-1}_{100}$ Mpc) clustercentric radii and in
the redshift range $0.18 < z < 0.55$. These values are, however, much
lower than that of $\sim 3 h_{100}^{-2} M_{\odot}\ yr^{-1}$ derived
for cluster galaxies at $z\sim 0.75$ from $H\alpha$ \citep{fzm05}, and
even lower than the $H\alpha$ SFR of field galaxies at $z\sim 1$
\citep{gbe99,dbs04}. In spite of the little or no evolution in SFR
between the sample of \citet{bsm98} and our sample, this comparison
can be affected by dust (see section \S\ref{colourcolour}) and
metallicity effects on the [OII] line, therefore, a fair comparison
requires H$\alpha$ and IR measurements of the true SFR in order to
quantify the amount of evolution. The lack of SFR measurements in
cluster galaxies at $z >1$ and the limitations of our data prevent us
from studying the evolution of the SFR in clusters from $z\sim1.3$
down to the local universe.

The co-added spectrum of the 17 star-forming cluster members is shown
in Fig. \ref{sf_coadded}. The presence of young stellar populations
(A- and F-type stars) is inferred by the detection of Balmer
absorption lines. Most of these galaxies present irregular, extended
features, most likely the home for the young stars and where most of
the star-forming activity takes place. In particular, we note that
object ID=6306, with an $i_{775}-z_{850} = 0.85$ color, presents a
spectrum showing a young post-starburst component (prominent Balmer
absorption features) in addition to an on-going star-formation
activity ([OII] in emission; see Fig. \ref{mosaic}).  This galaxy can
therefore be classified as an E+A+[OII] galaxy
\citep*[e.g.,][]{drl05}. We also note that ID=7001 has similar
characteristics.

On the other hand, there is an indication of the remaining traces of
the latest episode of star formation in some of the massive, passive
early-type galaxies in the cluster. Due to the lower signal-to-noise
ratio of our spectra at wavelengths larger than about 9000 \AA, an
accurate measurement of the EW of the $H_{\delta}$ absorption feature
is not possible. Therefore, an identification of post-starburst
stellar populations in individual early-type cluster galaxies cannot
be properly done with the existing data. However, by co-adding the
spectra of the 10 brightest passive galaxies in \CLhz\ \citep{r04}, a
prominent $H_{\delta}$ feature emerges and some other higher-order
Balmer features become visible. In Fig. \ref{stack_membs} we show the
co-added spectra of the 10 (top panel) and 20 (bottom panel) brightest
(in $K_s$) passive cluster members. An increase of the $H_{\delta}$
(about 36\%) and other Balmer absorption features equivalent width can
be seen when including fainter early-type members. We note that the
$H_{\delta}$ absorption line can indeed clearly be seen in the
individual spectrum of the first, second, third and fifth brightest
cluster members (in $z_{850}$) after a total integration time of 24 hr
\citep{hwf05}. A more detailed quantitative analysis of this
observation is underway (Rosati et al., in preparation).

\subsection{Color-magnitude distribution of cluster galaxies}\label{colourmag}

\citet{lrd04} included all galaxies within 20\arcsec\ ($0.17$ Mpc) of
the cluster center in their $J-K_s$ versus $K_s$ color-magnitude
diagram of \CLhz. In this paper, we produce a second color-magnitude
diagram using the reprocessed ISAAC data, but here we only use
galaxies with measured redshifts. The stellar symbols in
Fig. \ref{cmd} represent spectroscopically confirmed stars. Crosses
are non-cluster members, i.e., objects with redshift $z \leq 1.22$ or
$z \geq 1.25$. Filled circles are cluster members without detectable
[OII]($\lambda$3727) emission. The filled triangles, on the other
hand, are cluster members with [OII] emission. The dotted red line is
the fit published in \citet{lrd04}. At that time, only a limited
number of redshifts were available, so, to limit the effect of
contamination from non-cluster members, only galaxies within
20\arcsec\ of the cluster center and within the blue rectangle in
Fig. \ref{cmd} were used in fitting the CM relation. Here, only
cluster members without [OII] are used in the fit, which is shown as
the solid red line. No other restrictions are used. The slope and
scatter about the two fits are listed in Table \ref{cmfit} and are
calculated as explained in \citet{lrd04}. Within the statistical error
both fits are the same. Therefore, as in \citet{lrd04}, the values
here obtained for the slope and scatter about the new fit imply that
passive galaxies in \CLhz\ are home to old stellar populations with a
mean age of $\sim3$ Gyr \citep*[according to solar-metallicity, single
stellar population models of][]{bc03}. This corresponds to a formation
redshift of $z_*\sim3$ for the bulk of the stars in these
galaxies. The values of the slope and scatter presented in this work
were obtained using all passive members up to more than $0.5$ Mpc from
the center, without restricting ourselves to the central 40\arcsec\
(diameter; $\sim$ 0.3 Mpc) as in \citet{lrd04}, and are consistent
with measurements in lower redshift clusters \citep{sed98}. 

Looking at morphology and colors, we do not see any indication of the
existence of a population of S0 galaxies with a bluer (with respect to
elliptical galaxies) color-magnitude sequence, in contrast to what has
been observed in the galaxy cluster RDCS J0910+5422 at $z=1.1$
\citep{mbs06}. As pointed out by \citet{mbs06}, the existence of a
bluer color-magnitude sequence of the S0 galaxies with respect to the
elliptical galaxies in RDCS J0910+5422 could be the result of a still
forming cluster, with these bluer S0 galaxies being part of a group
infalling from the field onto a more evolved red cluster
population. \CLhz\ presents a more evolved structure having a more
evolved early-type population, with elliptical and S0 galaxies
distributed over a common red sequence.

From Fig. \ref{cmd}, we observe that all spectroscopic members without
detectable [OII]($\lambda$3727) emission have magnitudes brighter than
$K_s=21.5$, corresponding to $\sim K^*_s+1.5$ (${K^{*}_s}_{,rest}=-24$)
at z=1.237 \citep{srs06}. This indicates the spectroscopic limit for
galaxies with no detectable [OII] emission, while star-forming objects
can be confirmed down to $K_s\sim24$ ($\sim K^*_s+4$). Star-formation
is active in faint ($ > K^*_s+1.5$), less massive objects with only a
few star-forming members with $K_s < 21.5$. One of these objects is
ID=174, the cluster member AGN, which is also the reddest ($J-K_s \sim
1.3$) member in Fig. \ref{cmd}. The other star-forming galaxies (IDs
309, 339, 445 and 345) are well located within the near-IR CM
relation.

If the fit of the cluster red sequence were restricted to cluster
members brighter than $\sim K^*_s$, the resulting slope would be
considerably flatter. This steepening of the slope when including
fainter passive members could be due in part to the selection effect
imposed by our inability of detecting passive members below the
spectroscopic limit of our survey. Star-forming objects, having bluer
colors, would contribute to produce a fit slope steeper than the one
obtained if only more luminous and passive (redder) objects were
considered. The selection effect caused by the spectroscopic limit of
passive galaxies prevent us from properly investigating the faint end
of the red cluster sequence, in particular its build-up at this high
redshift. As indicated by \citet{tka05}, the faint end of the red
cluster sequence seems to be still under construction at $z\sim0.8$.
The observed deficiency of galaxies at the faint end in $z\sim0.8$
clusters suggests that part of the blue cluster galaxy population may
well be the progenitors of the present day faint end population of the
red sequence \citep{dlpa04}, as supported by recent high-resolution
simulations \citep{dlsw06} which, however, find it difficult to
reproduce such tight color sequences at this redshift. More
observations are needed to firmly establish this.

In Fig. \ref{cmd_acs} we show the color-magnitude diagram using the
ACS filters. Here we include spectroscopically confirmed passive (red
circles) and emission line (blue triangles) cluster members. Squares
are objects with photometric redshifts \citep{b00} in the range $1.12
< z_{phot} < 1.35$ and within an aperture of 1 Mpc centered in the
cluster (we excluded known spectroscopic non-members). These
photometric redshifts have been corrected for a systematic deviation
$\left<z_{spec}-z_{phot}\right> = 0.17$. The red sequence can be
traced down to a magnitude of $z_{850}=24.5$ in these ACS bandpasses,
vanishing quickly at fainter magnitudes. However, this cutoff
magnitude can be affected by uncertainties in the photometric
redshifts.

\subsection{Color-Color distribution of cluster galaxies}\label{colourcolour}

The distribution of spectroscopic cluster members in color-color
space is shown in Fig. \ref{colcol}. Red filled circles correspond to
passive galaxies and blue filled triangles to [OII] emission
galaxies. The open square indicates the only known AGN at the cluster
redshift (ID=174; see section \S\ref{spectroscopy}). A clear
separation between passive and star-forming galaxies is seen, with the
exception of three emission line galaxies, IDs 309, 345 and 445, which
are located in the locus of passive members. These galaxies are also
located in the red sequence of both $i_{775}-z_{850}$ vs $z_{850}$ and
$J_s-K_s$ vs $K_s$ color-magnitude diagrams. The estimated SFR for
these objects varies from 0.5 to 2.9 $M_{\odot}$ yr$^{-1}$ (see table
\ref{spec_phot_tab}), although the error bars are large.

The shape of the continuum and the [OII]($\lambda$3727) line of IDs
345 and 445 are consistent with their red ($i_{775}-z_{850} \sim
0.95$, $J_s-K_s \sim 0.8$) colors and star-forming nature. ID=309, on
the other hand, shows a rather flat continuum up to 9000 \AA\
suggesting that we are looking at a dust-rich star-forming galaxy. The
increase in number of sky lines, a relatively low SNR (SNR$\sim$3) and
therefore a poorer extraction of the continuum beyond 8500 \AA\ may
have caused the flattening of the spectrum towards red wavelengths, in
contrast to its red $i_{775}-z_{850}$ color. These galaxies are
located within an annulus spanning from $\sim 0.4$ to $\sim 0.8$ Mpc
in radius from the two central bright elliptical galaxies, which
corresponds to $R/R_{200}\simeq0.43 - 0.85$ in \citet{pfc05}. The
morphology-radius relation obtained for a composite cluster sample of
7 galaxy clusters (including \CLhz) at $z \sim 1$ observed with the
ACS \citep{pfc05} shows that these objects are in a region where the
average fraction of early-type galaxies ($\sim$39\%) is about half of
that of late-type galaxies ($\sim$61\%). Red, possibly obscured,
star-forming galaxies in clusters at low redshift have already been
reported in the literature \citep{dpf02,cmm05,wgm05,pbr06}, while
similar objects have also been observed at redshift $z>0.8$
\citep{vds03,drl05} and in the field up to $z\sim2$ \citep{fsm06}.

Finally, we note that the two other [OII] galaxies with colors
$J_s-K_s > 0.9$ (IDs 248 and 339), excluding the X-ray source ID=174,
can be starburst galaxies with $E(B-V) > 0.39$ \citep{kcb96} as shown
in Fig. \ref{color_selection} and Fig. \ref{colcol}, while the colors
of the rest of the member star-forming galaxies can be reproduced by
galaxies with $Z=Z_{\odot}$ and exponential SFRs (see
Fig. \ref{color_selection} and Fig. \ref{colcol}). We also note that
ID=339, with an elongated and irregular morphology, has a blue color
in the optical ($i_{775}-z_{850}=0.64$) but a red color in the
near-IR ($J_s-K_s=0.92$), suggesting that this star-forming galaxy has
a significant amount of dust obscuring an important fraction of its
``blue'' light.

\subsection{Morphologies and spectrophotometric properties}\label{morphologies}

The morphological T-type class \citep{dvdvc76} as determined by
\citet{pfc05} for cluster members is given in Table
\ref{phot_tab}. T-type values range from -5 for elliptical galaxies to
8 for disk/irregular galaxies. A value of -2 corresponds to S0
galaxies, values between -1 and 1 are assigned to morphologies between
S0 and Sa and a value of 6 corresponds to an Sd morphology. In
general, we observe the well-known correlation between morphology and
SFR. Most of the star-forming objects in this cluster have irregular
morphologies or irregular disky structures where star formation is
taking place. One of these, ID=6301, can be a possible merger with two
compact bright regions surrounded by a gas envelope. Another one,
ID=619, shows clear spiral-arm features and could also be a merger. In
contrast, most of the passive galaxies have morphologies typical of
elliptical and S0 galaxies, with no on-going star formation due to
their poor or zero gas content.

Passive early-type galaxies have red colors while star-forming
late-type galaxies are blue, (see Fig. \ref{cmd}), a manifestation of
the well-known correlation between stellar populations and
morphology. However, in Fig. \ref{cmd} we observe that a few
star-forming galaxies have colors as red as the passive cluster
members in the red sequence. One of these sources is ID=174, the
confirmed cluster member X-ray source (see section
\S\ref{spectroscopy}), with a $J-K_s$ color about 0.35 magnitudes
redder than the average color of red sequence members and whose
star-formation can be due to the central, dust-obscured AGN. The other
red, star-forming sources in the red sequence of Fig. \ref{cmd} are
IDs 248, 309, 339, 345 and 445. While IDs 248 and 339 show an
irregular disky appearance, IDs 309, 345 and 445 are characterized by
a prominent central bulge surrounded by a diffuse disk structure (see
Fig. \ref{mosaic}). We observe that IDs 309, 345 and 445 also have red
$i_{775}-z_{850}$ colors ($i_{775}-z_{850} \sim 0.95$) that locate
them in the locus occupied by red, passive galaxies in
Fig. \ref{colcol}. While the star-formation in these sources must be
occurring in their gas-rich regions, their red colors should arise
from the combined effect of old stellar populations and dust
reddening. One interesting object is ID=726. This source is very
compact and also shows a clear [OII] emission line in its spectrum, as
shown in Fig. \ref{mosaic}. In addition, its blue color (see table
\ref{phot_tab}) is consistent with the flat continuum of its
spectrum. With a magnitude $M_B=-20.74$, this object can be considered
in the class of luminous compact blue galaxies reported in the
literature \citep*[e.g.,][]{wjs04,hfe05,nkp06}. This object has a very
compact morphology (it is similar to the PSF), hence Postman et
al. did not provide a morphological T-type for it (see Table
\ref{phot_tab}).

\section{Discussion and Conclusions}\label{discussion}

We have combined an extensive multi-wavelength dataset to provide a
comprehensive picture of the X-ray luminous, massive cluster of
galaxies \CLhz\ at $z=1.237$. ACS data provides detailed morphological
information of galaxies and accurate photometry. By using FORS on the
VLT, we were able to spectroscopically confirm 38 cluster members. The
overall projected distribution of spectroscopic cluster members
presents a clear elongation in the East-West direction, consistent
with the elongation observed in the gas \citep{rte04} and dark matter
\citep{lrb05} distributions. As pointed out by \citet{rte04}, the
X-ray surface brightness distribution of the core of \CLhz\ has a
comet-like shape, resembling a cold front in low redshift clusters
\citep*[e.g.,][]{mmv01} produced by a merging process of a subclump in
the East-West direction.

The velocity dispersion $\sigma_v$ of the cluster members is
$747^{+74}_{-84}$ km s$^{-1}$. Assuming virial equilibrium we find
that the virial mass is $\sim 5 \times 10^{14} M_{\odot}$. Our mass
estimates are also consistent with the mass estimates from the X-ray
data \citep{rte04,etb04} and the weak-lensing mass map
\citep{lrb05}. Our value of $\sigma_v$ is also consistent with the
$L_x - \sigma_v$ relation for clusters, but turns out to be in
disagreement with the $\sigma_v - T_x$ relation \citep{xw00}. While
the observed $L_x$ and $T_x$ values follow the $L_x - T_x$ relation of
\citet{xw00}, the deviation from the $\sigma_v - T_x$ relation may be
the signature of a galaxy distribution still not in a fully dynamical
equilibrium.

Indeed, \CLhz\ has not yet attained an equilibrium state as suggested
by a more detailed analysis of the velocity distribution of cluster
members. \CLhz\ is the most distant cluster for which a substructure
analysis can be attempted. The kurtosis \citep{bgf91} indicates that
the velocity distribution of cluster members is not Gaussian at the
$>95$\% c.l.. The weighted gap procedure finds a significant gap in
the space of ordered velocities and a 3D KMM study detects
substructure in velocity at the $95.6$\% c.l.. The substructure is
composed by two groups of galaxies which are mostly distributed in a
East-West direction on the sky, as shown in Fig.
\ref{figkmm}. Twenty-six galaxies are confirmed in the
``low-velocity'' group (centered at $-354\pm159$ km s$^{-1}$ with
respect to the cluster mean velocity) with a velocity dispersion of
$486^{+47}_{-85}$ km s$^{-1}$ and located mostly to the West of the
cluster center. Twelve galaxies are confirmed in the ``high-velocity''
group (centered at $845\pm182$ km $s^{-1}$ with respect to the cluster
mean velocity) with a velocity dispersion of $426^{+57}_{-105}$ km
s$^{-1}$ and to the East of the cluster center. 

The existence of the above correlations between positions and
velocities of cluster galaxies is a footprint of real substructure.
In particular, it suggests that \CLhz\ is forming via a merger along
the E-W direction.  Going further, we estimate the mass contained
within $R_{200}$ for each group (see section \S\ref{vel_struct}). The
sum of these two masses is about half of the cluster mass, and it
should be considered as a lower limit of the latter. These groups may
be the virialized cores of two larger systems, with possibly $R
\gtrsim 2\ R_{200}$ and larger velocity dispersions. The fact that the
X-ray emission presents a single peak, although elongated in
morphology, indicates that these two groups have already started
virialization. This is supported by the degree of merging of the group
members as shown in Fig. \ref{figkmm}. Our substructure analysis has
thus found the remnant traces of two sub-clusters merging parallel to
the sky as suggested by the lack of a clear bimodal distribution of
the overall cluster velocity field.

By using the $\sigma_v-T_x$ relation for groups in \citet{xw00}, we
find that both groups have velocity dispersions consistent with the
same temperature of $\sim$2 keV. This temperature is inconsistent with
the overall cluster temperature of $\sim$6 keV, yet it corresponds to
about half the total cluster X-ray luminosity according to the group
$L_x-T_x$ relation \citep{xw00}. Thus, both groups seem to be
independently virialized and the sum of their X-ray luminosities would
be approximately equal to the total cluster $L_x$. The temperature
value of the overall ICM would have been the consequence of gas
particle interactions at an earlier epoch during the merger of the
groups. By the epoch of observation, the overall ICM is consistent
with being in an almost isothermal hydrostatic equilibrium in a single
potential well, while the groups are fossilized in the velocity
dispersion substructure. The agreement of the observed cluster $L_x$
and $T_x$ values with the $L_x-T_x$ relation and the deviation of the
cluster values from the $\sigma_v-T_x$ relation show that the gas in
the cluster is more relaxed than the galaxies. Since galaxies behave
more like dark matter particles, they still need more time to reach
the $\sigma_v-T_x$ relation expected for the cluster. In summary, the
elongated shape of the cluster structure (in all its components:
baryons and dark matter) and the substructure in velocity leads us to
conclude that \CLhz\ has not yet reached a final virial state: we may
be seeing gravitational collapse of two or more sub-clusters along a
high density filament.

In this work we present an improved analysis of the near-IR
color-magnitude diagram of galaxies in \CLhz, based on ISAAC data
which have been reprocessed in order to optimise image quality and
obtain more accurate photometry over the entire region covered by
near-IR observations. We use only spectroscopic cluster members and
the analysis extends over more than $1$ Mpc from the cluster center,
in contrast to previous studies. The fit to the red cluster sequence,
including all passive spectroscopic members, yields slope and scatter
values consistent with the result of \citet{bfp03} and
\citet{lrd04}. This indicates that the early-type galaxy cluster
population has formed most of its stellar content at $z\sim3$ and has
passively evolved since down to the epoch of observation, as
previously shown for $z\sim1$ clusters
\citep{vds03,bfp03,lrd04,bhf06,mbs06,mhb06}. If we restrict the fit of
the cluster red sequence to spectroscopic members brighter than $\sim
K^*_s$, the slope becomes significantly shallower, suggesting that the
CM relation of passively evolving systems seems to be non-linear,
although this could also be due to a selection effect. The [OII]
emission line observed in 17 cluster galaxies allows us to estimate a
lower limit value of $0.7 M_{\odot}$ yr$^{-1}$ for the median SFR in
the cluster. Only 5 out of 25 spectroscopic members that are brighter
than $K^*_s+1.5$ show ongoing star formation. Three star-forming
galaxies have colors as red as galaxies in the red cluster sequence of
both $i_{775}-z_{850}$ vs. $z_{850}$ (optical) and $J_s-K_s$ vs. $K_s$
(near-IR) color-magnitude diagrams. One additional star-forming galaxy
is observed in the red sequence of the near-IR color-magnitude diagram
but not in the optical one, suggesting that we are observing a
dust-rich galaxy. Finally, a highly obscured AGN is observed to be a
spectroscopic member of \CLhz, and we do not find any evidence of
``red mergers'' like those observed in the galaxy cluster MS1054
\citep{vdff99}.

The projected distribution of passive and star-forming galaxies in
this cluster shows a spectrum-density relation that qualitatively
resembles the observed morphology-density relation at $z\sim1$
\citep{ste05,pfc05}: passive, early-type galaxies dominate the cluster
core while star-forming, late-type galaxies are found in the outskirts
of the cluster. From a dynamical point of view, we are witnessing
hierarchical structure formation: we are observing a merger of two
large groups of galaxies into a more massive structure.

\acknowledgments

This work would not have been possible without the dedicated efforts
of ESO staff, in both Chile and Europe. ACS was developed under NASA
contract NAS5-32865. We are grateful to K. Anderson, D. Magee,
J. McCann, S. Busching, A. Framarini, and T. Allen for their
invaluable contributions to the ACS project. R.D. acknowledge the
hospitality and support of ESO in Garching. SAS's work was performed
under the auspices of the U.S. Department of Energy, National Nuclear
Security Administration by the University of California, Lawrence
Livermore National Laboratory under contract No. W-7405-Eng-48.


\clearpage

\begin{deluxetable}{lcccccc}

\tablecaption{A summary of the VLT spectroscopic data obtained on
\CLhz. Grism/Filter corresponds to the grism/order sorting filter
combination used and the exposure times are given in units of
hours. The (*) indicates observations carried out with the mosaic of
two 2k $\times$ 4k MIT CCDs.\label{spec_tab}}

\tablewidth{0pt}
\tablehead{
\colhead{Mask name} & \colhead{Date} & \colhead{Telescope} & \colhead{Instrument} & \colhead{Grism/Filter} & \colhead{Exp. time} & \colhead{Mask}
}
\startdata
m1  &  Mar. 2001 &   UT2 &         FORS2        &     300I/OG590  &      4.9    &       MXU \\
m2  &  Mar. 2001 &   UT2 &         FORS2        &     300I/OG590  &      4.0    &       MXU \\
m3  &  Mar. 2001 &   UT2 &         FORS2        &     300I/OG590  &      0.9    &       MXU \\
m3  &  Apr. 2001 &   UT2 &         FORS2        &     300I/none   &      4.5    &       MXU \\
m4  &  Mar. 2001 &   UT2 &         FORS2        &     300I/OG590  &      1.7    &       MOS \\
m4  &  Mar. 2001 &   UT2 &         FORS2        &     150I/GG435  &      1.4    &       MOS \\
m5  &  Mar. 2001 &   UT2 &         FORS2        &     150I/GG435  &      1.3    &       MOS \\
m6  &  Apr. 2001 &   UT2 &         FORS2        &     300I/none   &      4.5    &       MXU \\
m7  &  Apr. 2001 &   UT2 &         FORS2        &     150I/none   &      2.0    &       MOS \\
m8  &  Apr. 2001 &   UT2 &         FORS2        &     150I/none   &      2.0    &       MOS \\
m9  &  Apr. 2001 &   UT2 &         FORS2        &     150I/none   &      1.0    &       MOS \\
m10 &  Apr. 2001 &   UT2 &         FORS2        &     300I/none   &      2.5    &       MXU \\
m11 &  Oct. 2002 &   UT4 &         FORS2$^{*}$  &     300I/none   &      6.5    &       MXU \\
m12 &  Feb. 2003 &   UT4 &         FORS2$^{*}$  &     300I/none   &      3.7    &       MXU \\
m13 &  Feb. 2003 &   UT4 &         FORS2$^{*}$  &     300I/none   &      3.0    &       MXU \\
m14 &  Apr. 2004 &   UT4 &         FORS2$^{*}$  &     300I/none   &      4.7    &       MXU \\
m15 &  Apr. 2004 &   UT4 &         FORS2$^{*}$  &     300I/none   &      4.7    &       MXU \\

\enddata
\end{deluxetable}

\begin{deluxetable}{llllllllll}
\tabletypesize{\scriptsize}

\tablecaption{\small Spectroscopically-confirmed cluster members.
R.A. and DEC. (J2000) coordinates are in (h:m:s) and
(${}^o$:${}^{'}$:${}^{''}$), respectively. Column 4 shows the redshift
and its formal error as obtained from the cross-correlation.  Column 5
shows the correlation coefficient, R, as defined in
\citet{td79}. Column 6 indicates the most prominent spectral features
identified in the spectrum$^{a}$. Column 7 shows the emission line
flag assigned to the object. A value of 0 corresponds to galaxies
without emission lines and a value of 1 corresponds to galaxies with
narrow emission lines. The last three columns are the integrated [OII]
line flux (in $\times 10^{-17} \ erg \ s^{-1} \ \ cm{-2}$), the [OII]
line equivalent width (EW; in \AA) and the star-formation rate (SFR;
in $M_{\odot} \ yr^{-1}$) respectively.
\label{spec_phot_tab}}

\tablewidth{0pt}
\tablehead{\colhead{ID} & \colhead{R.A.} & \colhead{DEC.} & \colhead{$z$} & \colhead{R} & \colhead{Spec. Feat.$^{a}$} & \colhead{E. L.} & \colhead{Flux [OII]} & \colhead{EW [OII]} & \colhead{SFR} \\ }

\startdata
SEE & EDITION & FROM & JOURNAL &  &  &  &  &  &  \\
\enddata

\tablenotetext{a}{Most prominent spectral features:
FeI($\lambda$3581), HeI($\lambda$3614), [OII]($\lambda$3727),
H10($\lambda$3750), H9($\lambda$3770), H8($\lambda$3799),
MgI($\lambda$3834), H6($\lambda$3889), the CaII lines
(H($\lambda$3934) and K ($\lambda$3969)), the decrement at 4000 \AA\
(D4000), H$_{\delta}$($\lambda$4102) and CaI($\lambda$4227).}

\end{deluxetable}

\begin{deluxetable}{lcccccccc}
\tabletypesize{\normalsize}

\tablecaption{Photometric catalog of spectroscopically confirmed
cluster members. The galaxy ID in the catalog is given in the first
column. Magnitudes in the AB system, corrected for galactic reddening,
and the corresponding errors are given in the following columns: B-,
V- and R-band magnitudes were obtained with VLT/FORS, the $i_{775}$-
and $z_{850}$-band magnitudes are from HST/ACS and the near-IR
magnitudes are from the reprocessed VLT/ISAAC data. A value of 99
has been asigned to unavailable photometry. The last column gives the
morphological T-type \citep{dvdvc76} of the galaxy as determined by
\citet{pfc05}.\label{phot_tab}}

\tablewidth{0pt}

\tablehead{\colhead{ID} & \colhead{$K^{Tot}_s$} & \colhead{$B-V$} & \colhead{$V-R$} & \colhead{$V-i_{775}$} & \colhead{$R-K_s$} & \colhead{$i_{775}-z_{850}$} & \colhead{$J_s-K_s$} & \colhead{T-type}}

\startdata
SEE & EDITION  & FROM  & JOURNAL  &  &  &  &  &  \\
\enddata

\end{deluxetable}

\begin{deluxetable}{llllcccc}

\tablecaption{X-ray point sources, in the cluster field of view,
detected by Chandra. Right Ascension (R.A.) and Declination (DEC.)
J2000 coordinates are in (h:m:s) and (${}^o$:${}^{'}$:${}^{''}$),
respectively. The spectroscopic redshift of the source is given in
column 4. The flux, in units of $erg \ s^{-1} \ \ cm{-2}$, in the
[0.5-2] and [2-10] keV bands of Chandra are given in columns 5 and 6,
respectively. The rest frame X-ray luminosity, in units of $erg \
s^{-1}$, in the [0.5-2] and [2-10] keV bands are given in columns 7
and 8, respectively.\label{xray_tab}}

\tablewidth{0pt}
\tablehead{\colhead{ID} & \colhead{R.A.} & \colhead{DEC.} & \colhead{$z$} & \colhead{$f_x$[0.5-2]} & \colhead{$f_x$[2-10]} & \colhead{$L_x$[0.5-2]} & \colhead{$L_x$[2-10]}}
\startdata

425  & 12:52:58.14 &  -29:26:49.3 &  0.4694  & 5.1E-16   & 2.3E-15  & 3.3E+41  & 1.5E+42  \\
69   & 12:53:03.60 &  -29:28:28.4 &  0.4762  & 1.1E-15   & 6.8E-15  & 7.3E+41  & 4.6E+42  \\
7006 & 12:52:56.96 &  -29:30:10.9 &  0.7313  & 1.4E-15   & 3.3E-15  & 2.4E+42  & 5.7E+42  \\
661  & 12:53:01.50 &  -29:25:38.8 &  0.7439  & 1.4E-14   & 2.6E-14  & 2.6E+43  & 4.8E+43  \\
217  & 12:53:03.64 &  -29:27:42.1 &  0.7533  & 3.3E-16   & 9.0E-16  & 6.1E+41  & 1.7E+42  \\
3043 & 12:53:00.39 &  -29:29:17.2 &  0.8220  & 2.9E-15   & 1.1E-14  & 6.5E+42  & 2.4E+43  \\
355  & 12:52:42.48 &  -29:27:03.0 &  0.8440  & 4.2E-16   & 9.5E-16  & 9.9E+41  & 2.3E+42  \\
612  & 12:52:59.93 &  -29:25:18.4 &  1.1764  & 1.1E-15   & 4.5E-15  & 5.5E+42  & 2.2E+43  \\
174  & 12:52:49.82 &  -29:27:54.9 &  1.2382  & 1.9E-16   & 1.6E-15  & 1.1E+42  & 8.6E+42  \\
321  & 12:52:40.36 &  -29:27:14.1 &  1.3479  & 1.6E-16   & 5.6E-15  & 1.1E+42  & 3.6E+43  \\
374  & 12:52:50.08 &  -29:27:00.7 &  1.5198  & 1.6E-15   & 3.5E-15  & 1.4E+43  & 3.0E+43  \\

\enddata
\end{deluxetable}

\begin{deluxetable}{lllcc}

\tablecaption{Spectroscopic data of field galaxies. The object ID in
the catalog is given in the first column. Right Ascension (R.A.)  and
Declination (DEC.) J2000 coordinates are given in the second and third
column respectively. RAs are given in (h:m:s) and DECs in
(${}^o$:${}^{'}$:${}^{''}$). Columns 4 and 5 show the redshift and the
formal error in redshift as obtained from the cross-correlation. The
table has been ordered from low to high redshift. A prominent
``structure'' in redshift can be seen at redshift $z \sim 0.74$ (see
section \S\ref{redshifts}). Objects belonging to this ``structure''
are marked with a star. \label{spec_field_tab}}

\tablewidth{0pt}
\tablehead{\colhead{ID} & \colhead{R.A.} & \colhead{DEC.} & \colhead{$z$} & \colhead{$Err. \ z$} \\ }

\startdata
SEE & EDITION  & FROM  & JOURNAL  &  \\
\enddata

\end{deluxetable}

\begin{deluxetable}{ccccc}

\tablecaption{Color-magnitude fitting. Only spectroscopic members without
[OII] emission are used to fit the CM relation (solid red line in
Fig. \ref{cmd}). \label{cmfit}}

\tablewidth{0pt}
\tablehead{
\colhead{Fit \#} & \colhead{Slope} & \colhead{Color at $K_s=18.5$} & \colhead{Intrinsic Scatter} & \colhead{Comments}
}
\startdata
1      &  $-0.048\pm0.015$   &   1.85                &   0.070            & \citep{lrd04} \\
2      &  $-0.042\pm0.022$   &   1.87                &   0.061            & This work     \\
\enddata

\end{deluxetable}


\clearpage

\begin{center}
\begin{figure}

\caption{(SEE EDITION FROM JOURNAL) Projected distribution of cluster
members on the sky (North is up and East is to the left). Circles are
non star-forming galaxies and triangles are galaxies showing
[OII]($\lambda$3727) in their spectrum. The background image
corresponds to the FORS2 coverage in BVR, the darker image corresponds
to the $z_{850}$ ACS data and the dashed lines show the field covered
by ISAAC. The overall shape of the distribution of confirmed galaxy
members is clearly elongated in the East-West direction.}

\label{galdist}
\end{figure}
\end{center}

\begin{center}
\begin{figure}
\plotone{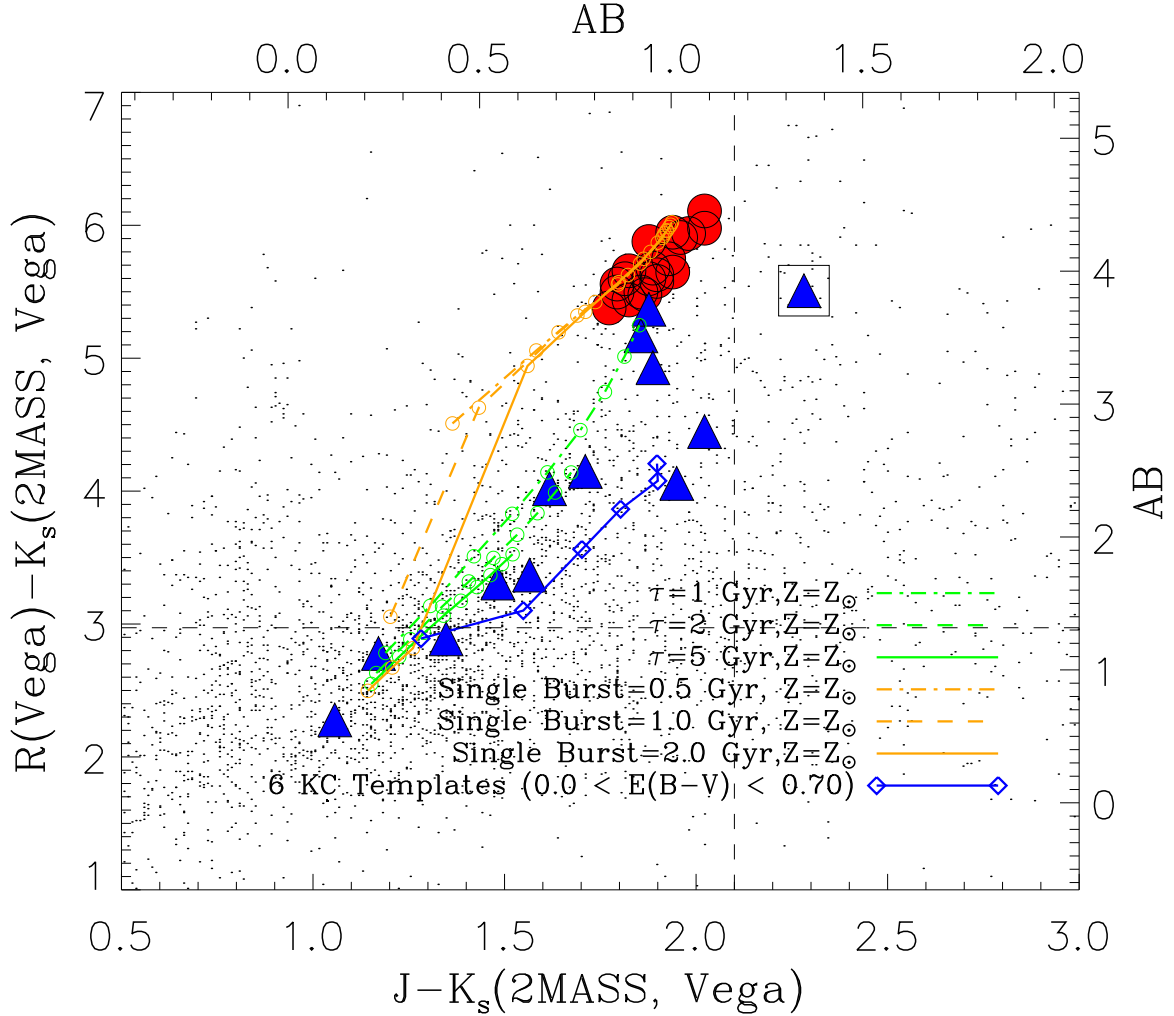}

\caption{\small Color-color selection of the spectroscopic
sample. Galaxies with $K_s< 21$, $J-K_s<2.1$ and $R-K_s>3$ were
targeted. The different color tracks correspond to different
evolutionary models, computed with Bruzual \& Charlot's code, and to 6
KC templates with different $E(B-V)$ color excess (see text for
details). The square indicates the spectroscopically confirmed cluster
AGN (ID=174). Filled circles indicate passive members while filled
triangles indicate emission line [OII]($\lambda 3727$) members. Small
dots are objects in our full photometric catalog.}

\label{color_selection}
\end{figure}
\end{center}

\begin{center}
\begin{figure}
\plotone{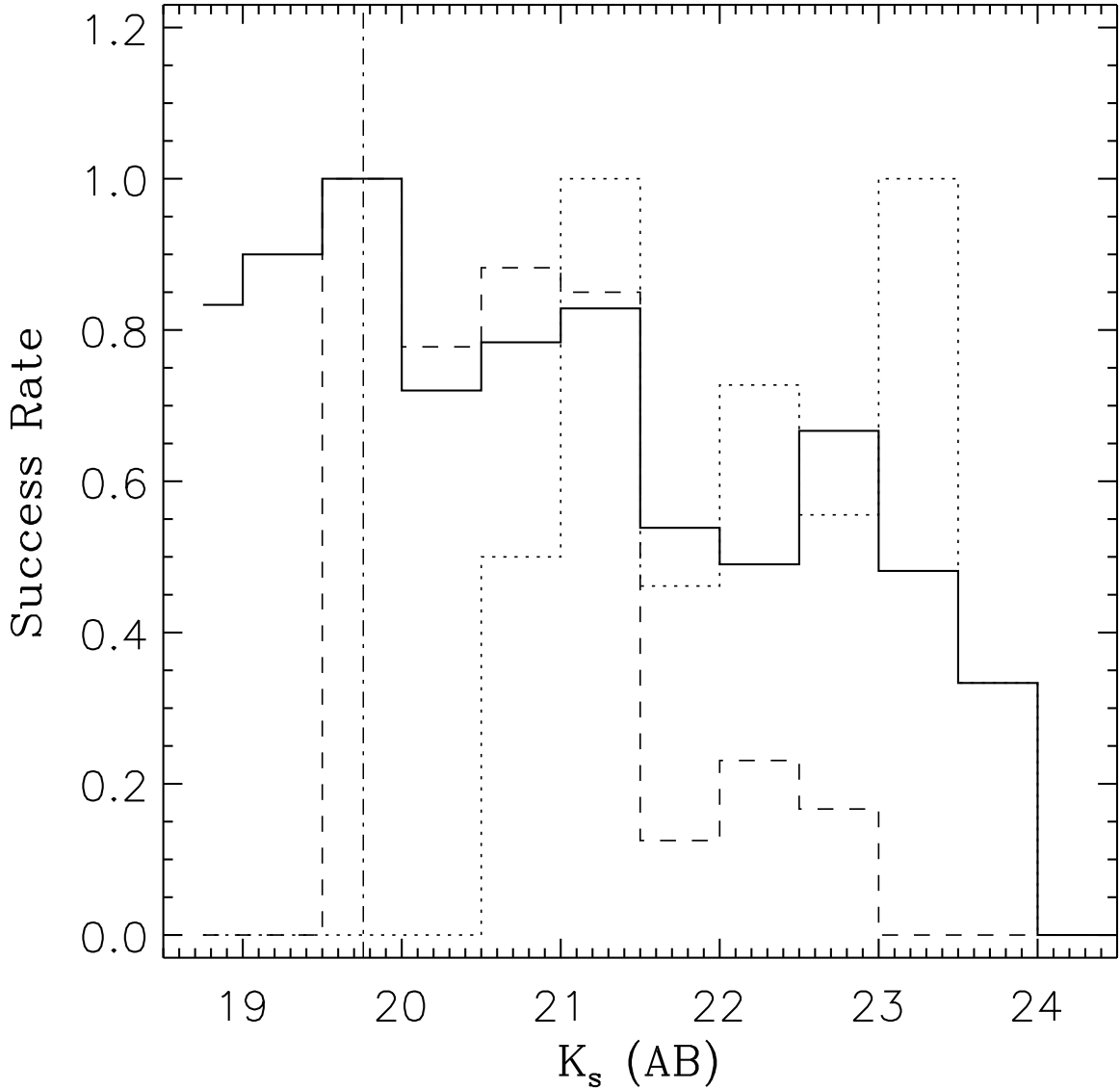}

\caption{The success rate of the spectroscopic survey, defined as the
ratio of the number of objects with spectroscopic redshifts to the
number of objects that were targeted for spectroscopy as a function of
$K_s$ magnitude, is shown by the solid histogram. The data were binned
in $\Delta K_s=0.5$ mag intervals. This distribution includes only
objects for which an estimate of their $K_s$ magnitude is
available. The dashed histogram shows the success rate only
considering our main color-color selection: $R-K_s > 3$ and $J-K_s <
2.1$, while the dotted histogram shows the same ratio only considering
the secondary color-color selection: $I-z_{850} > 0.4$, $I-z_{850} <
0.85$, $V-I > 0.2$, $V-I < 1.2$, $V-I > [2.4\ (I-z_{850})-1.12]$ and
$V-I < [7.0\ (I-z_{850})-2.3]$. This secondary selection was intended
to target Balmer-Break and star-forming galaxies. The dot-dashed line
corresponds to the $K_s$ magnitude of the brightest cluster member.}

\label{ssr}
\end{figure}
\end{center}

\begin{center}
\begin{figure}
\plotone{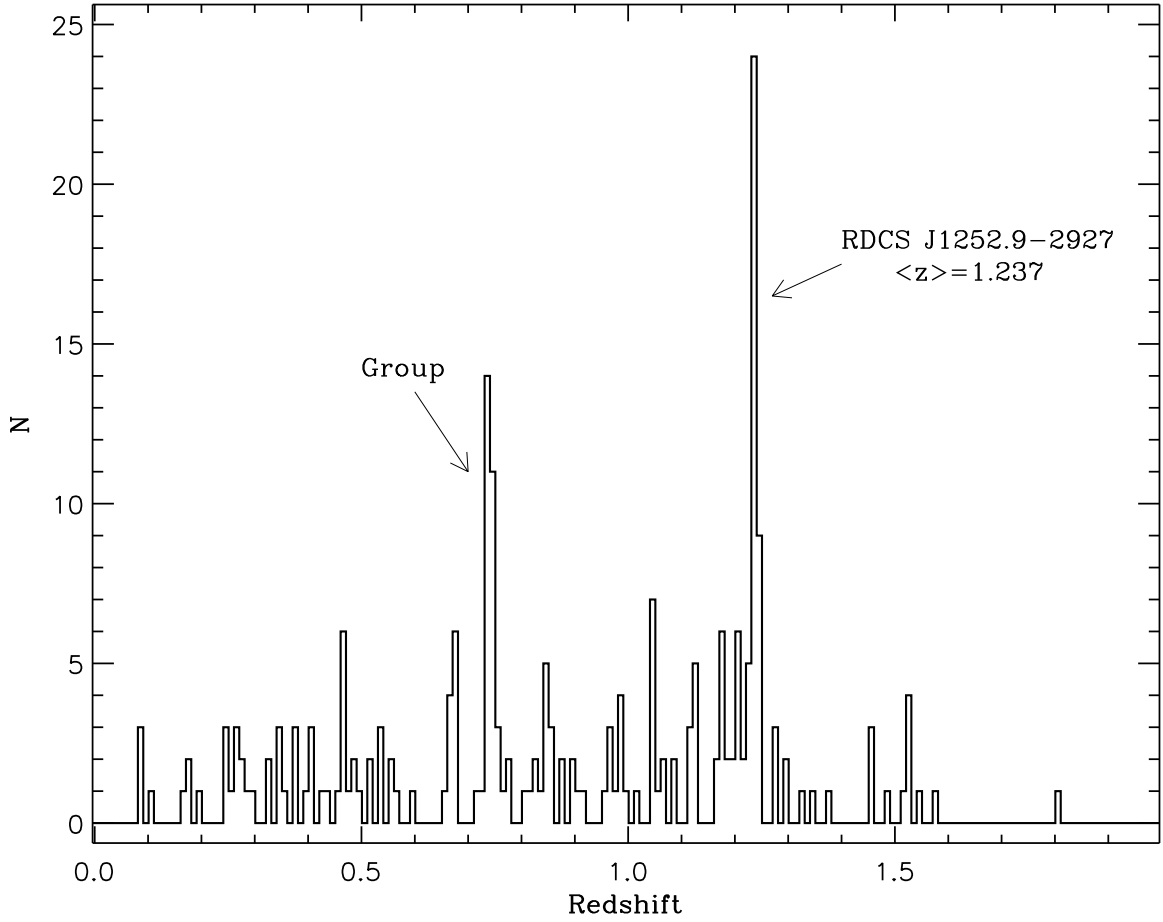}

\caption{Redshift distribution of the 227 galaxies with a secure
estimate of their redshifts. The bin size of the histogram is $\Delta
z=0.01$.~\CLhz\ appears as the strongest peak of the distribution with
a median redshift of $z=1.2373$ (3$\sigma$-clipped mean of
$\bar{z}=1.2371\pm0.0009$), with 38 galaxies within the range $1.22 <
z < 1.25$. A number of smaller peaks in the survey distribution are
observed along the line of sight to the cluster, where a significant
peak at $z\simeq0.74$ is detected. This ``group'', composed of 31
galaxies, has a 3$\sigma$-clipped mean redshift of
$\bar{z}=0.7429\pm0.0024$.}

\label{z_survey}
\end{figure}
\end{center}

\begin{center}
\begin{figure}[!htb]

\includegraphics[scale=0.37]{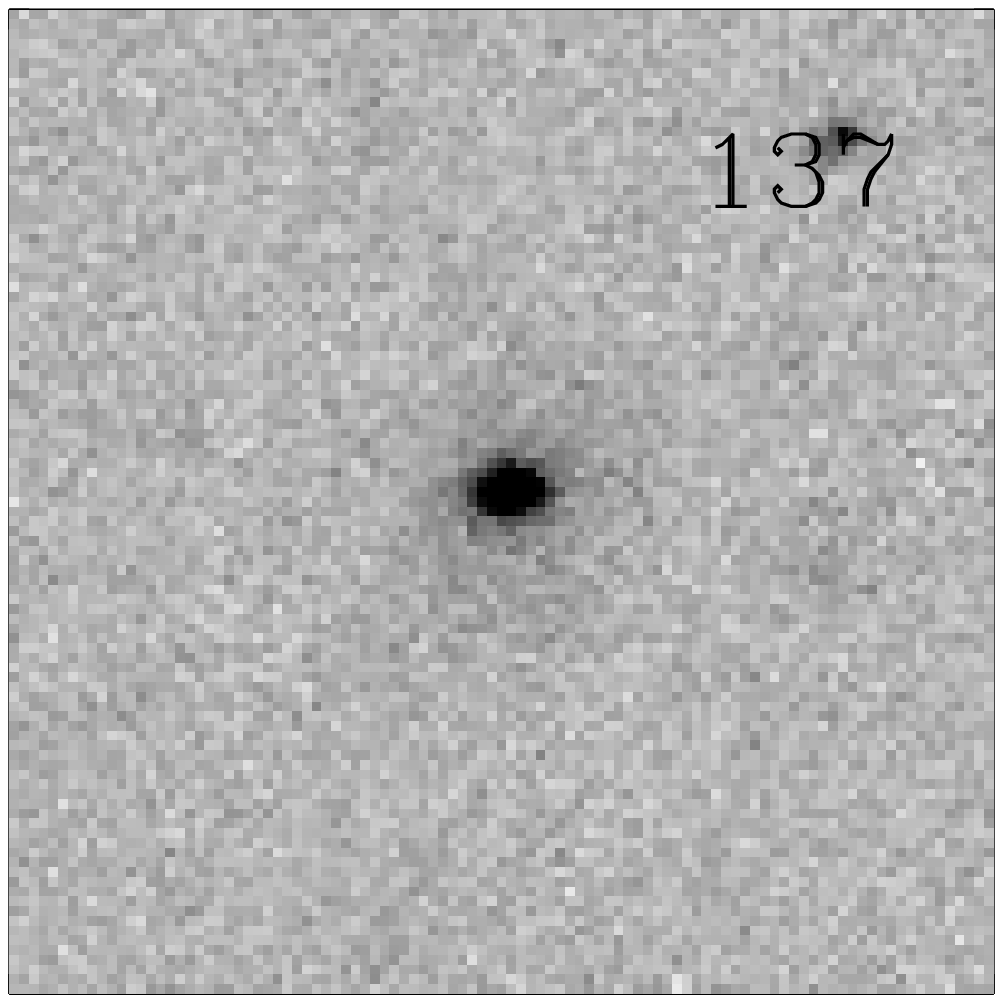}
\includegraphics[scale=0.37]{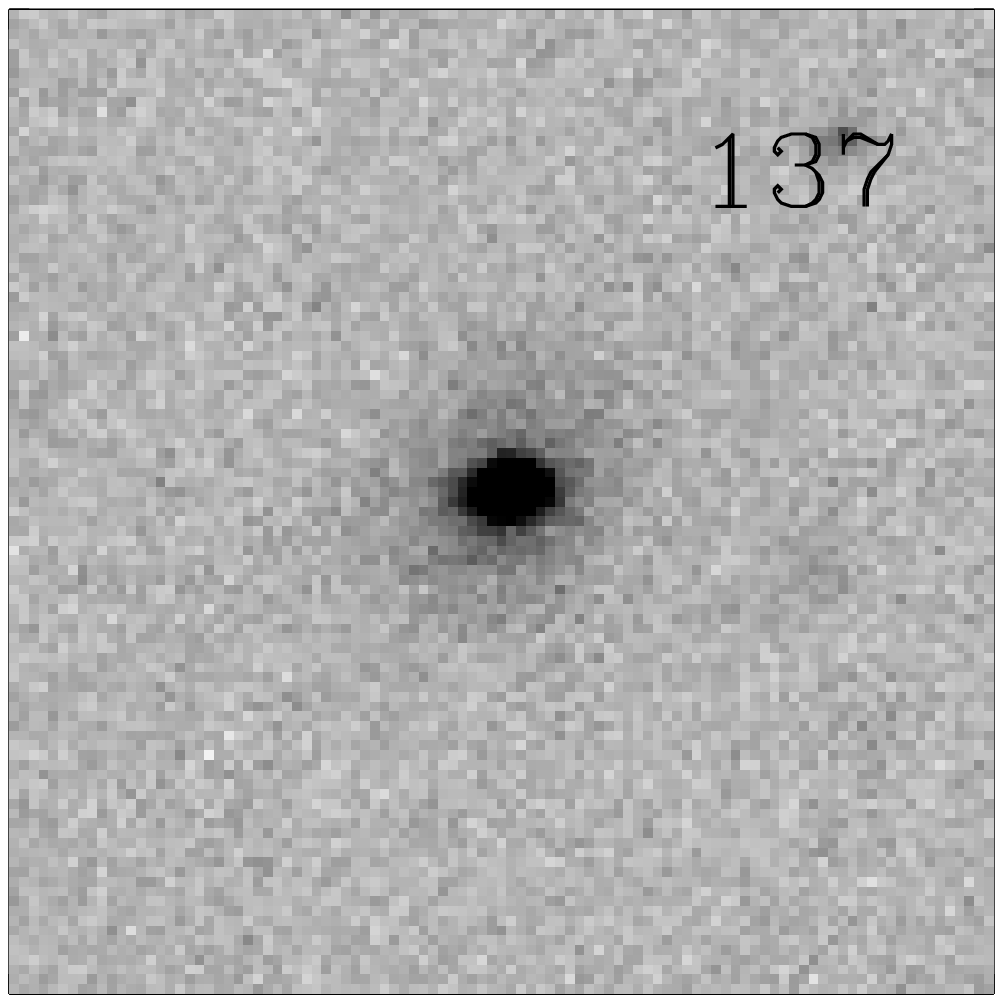}
\includegraphics[scale=0.32]{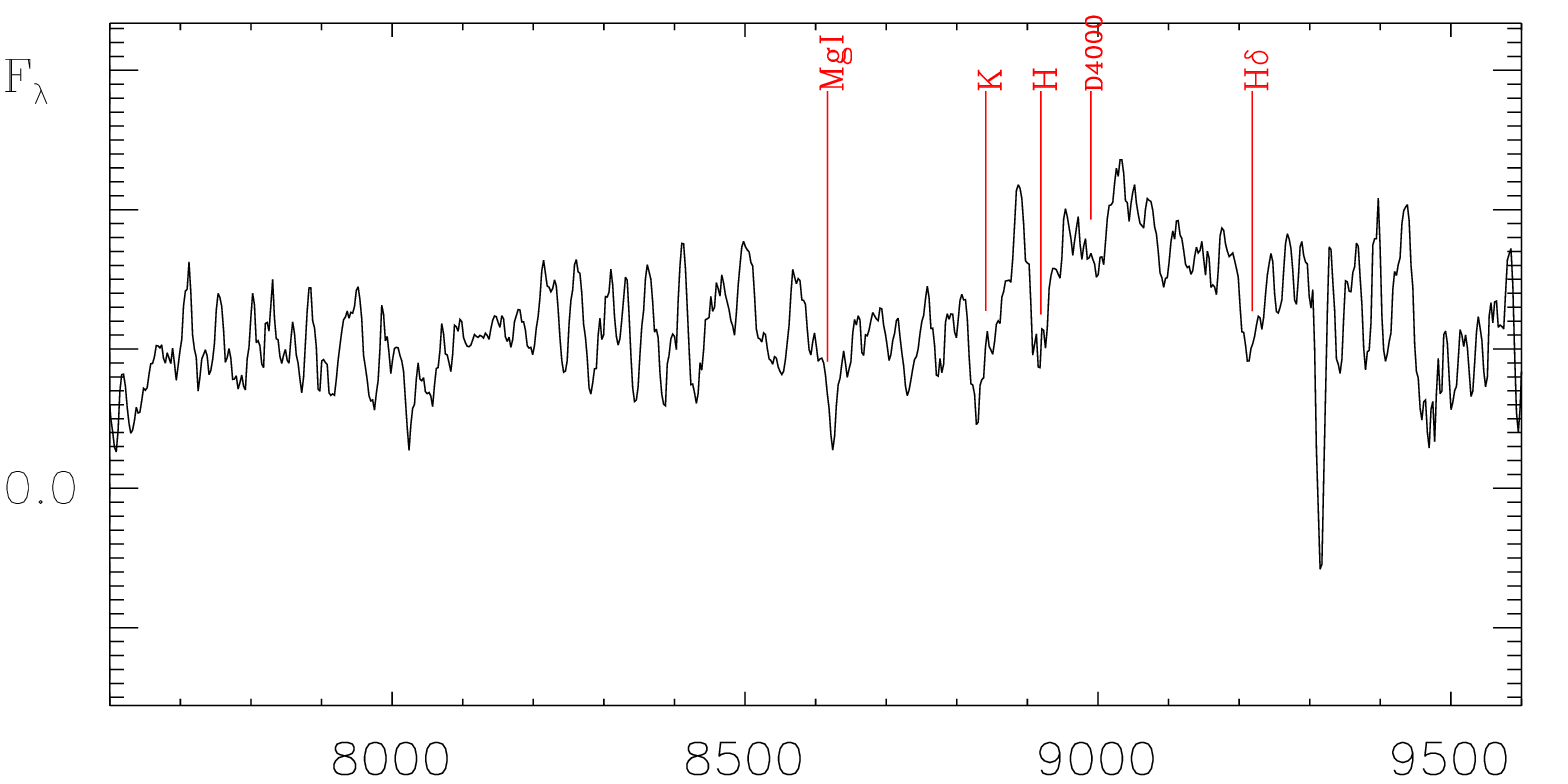}
\vspace{0.1cm}

\includegraphics[scale=0.37]{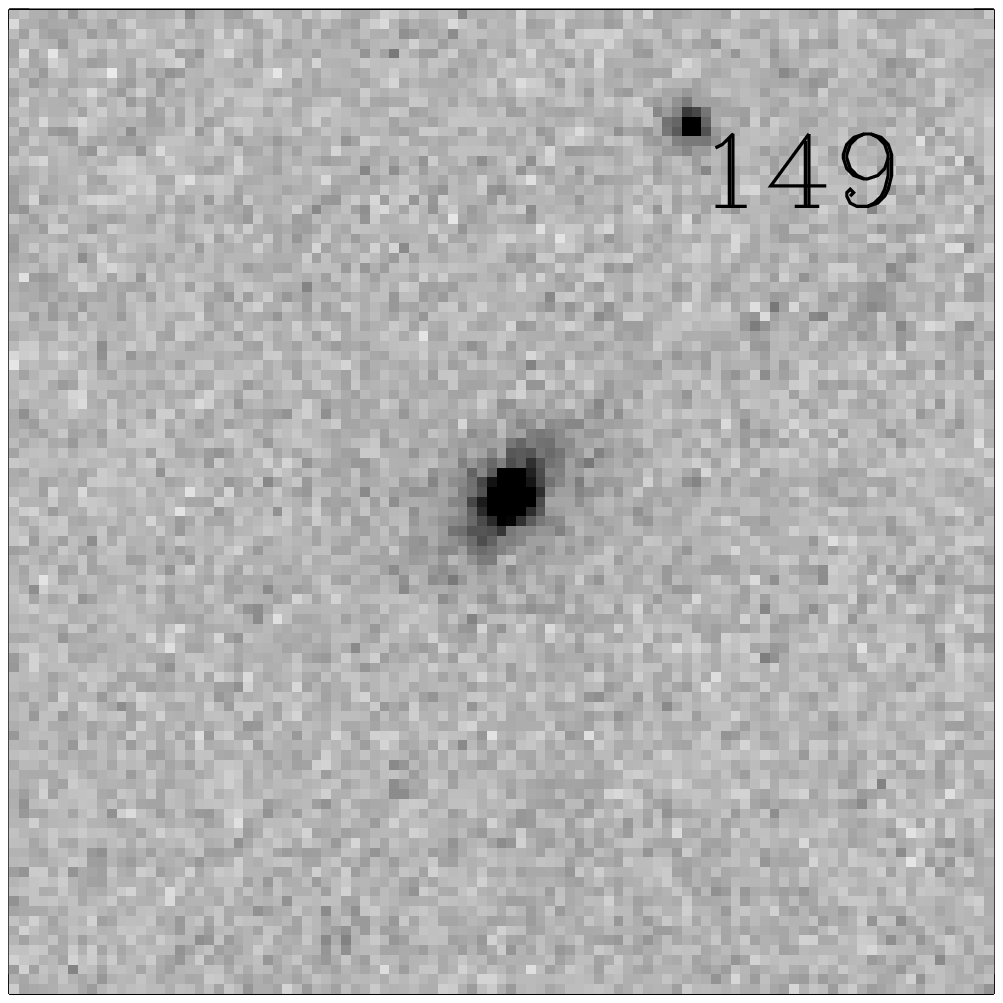}
\includegraphics[scale=0.37]{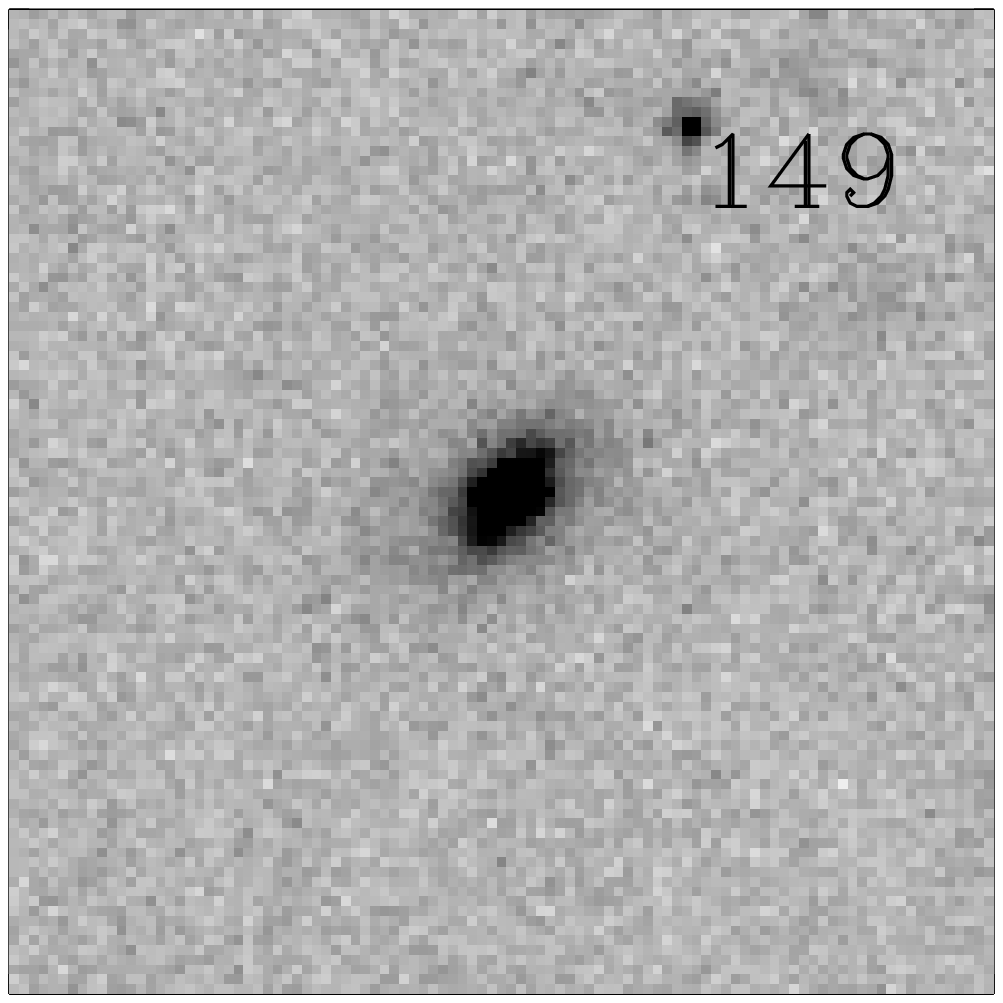}
\includegraphics[scale=0.32]{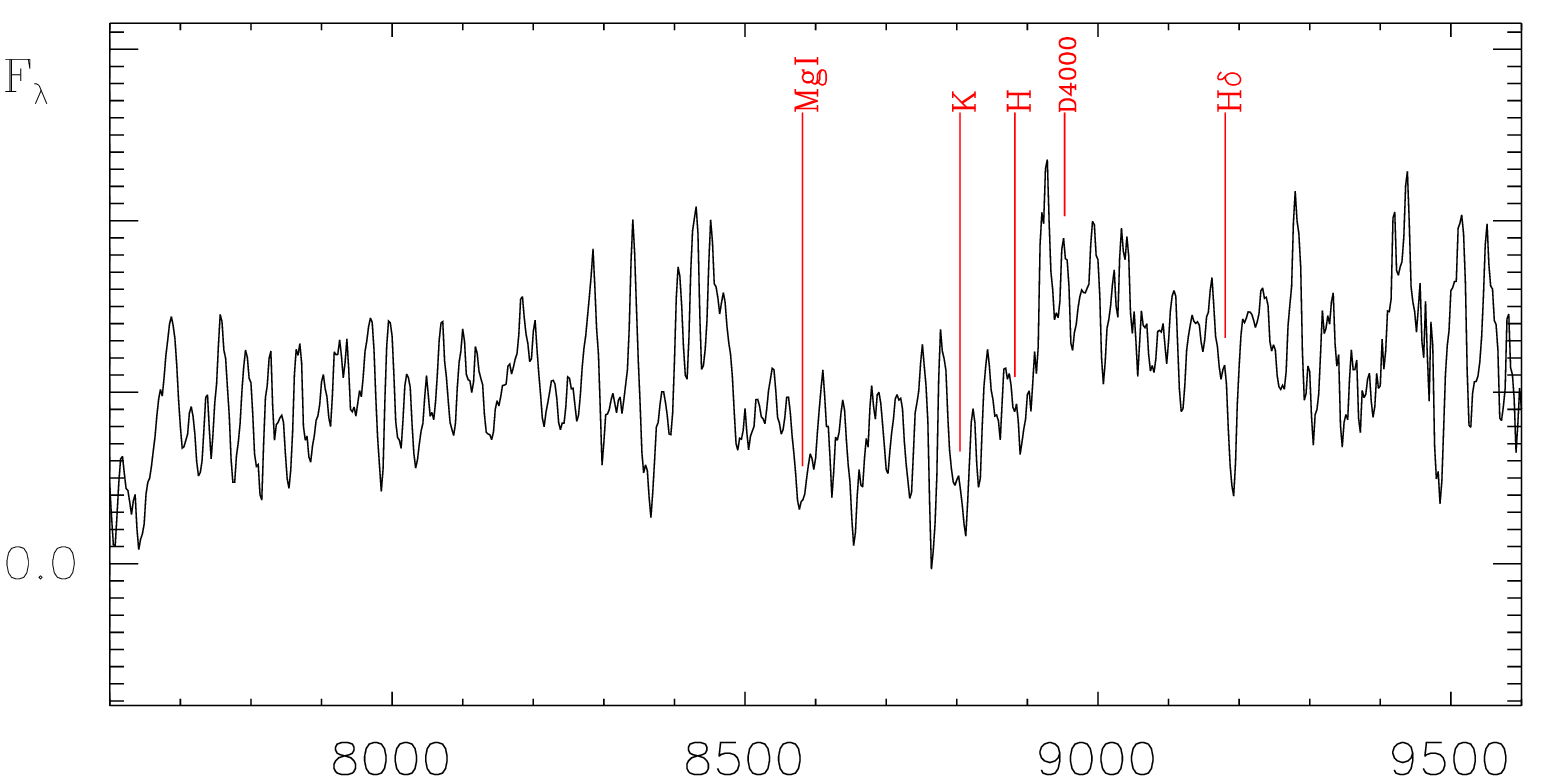}
\vspace{0.1cm}

\includegraphics[scale=0.37]{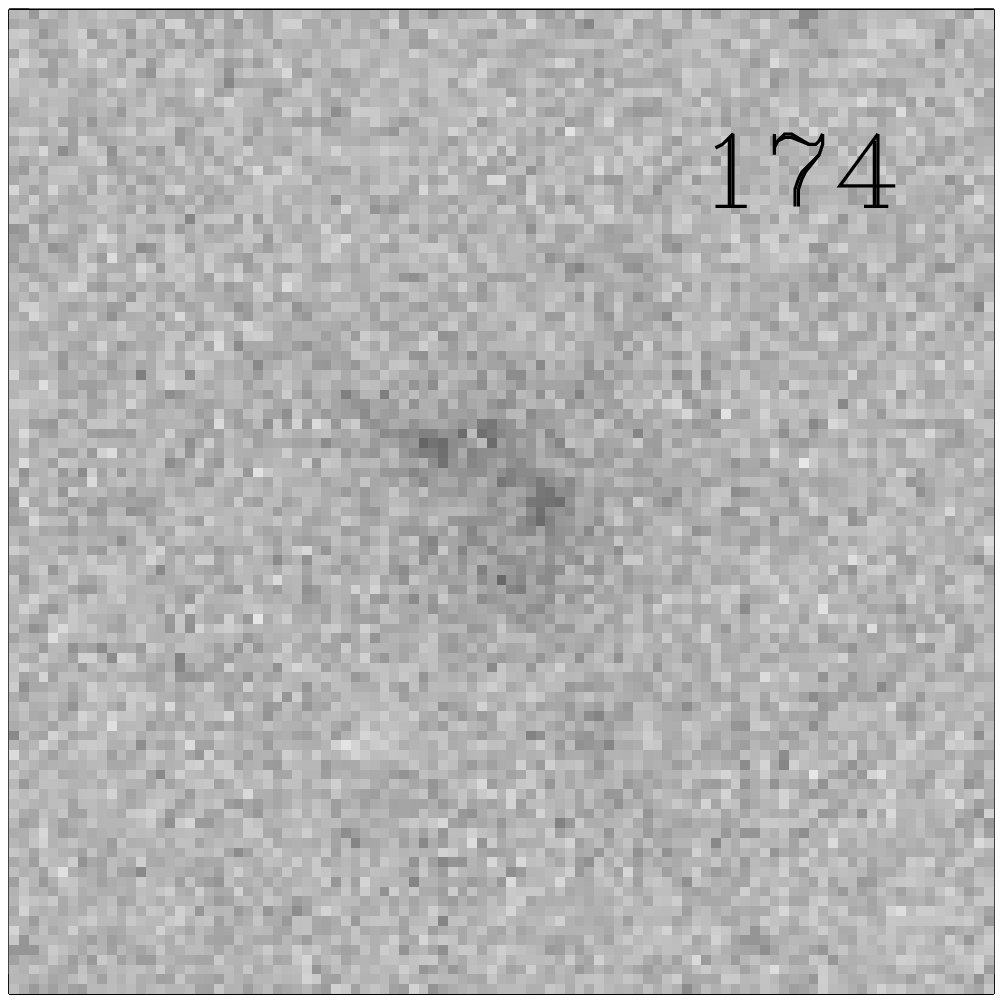}
\includegraphics[scale=0.37]{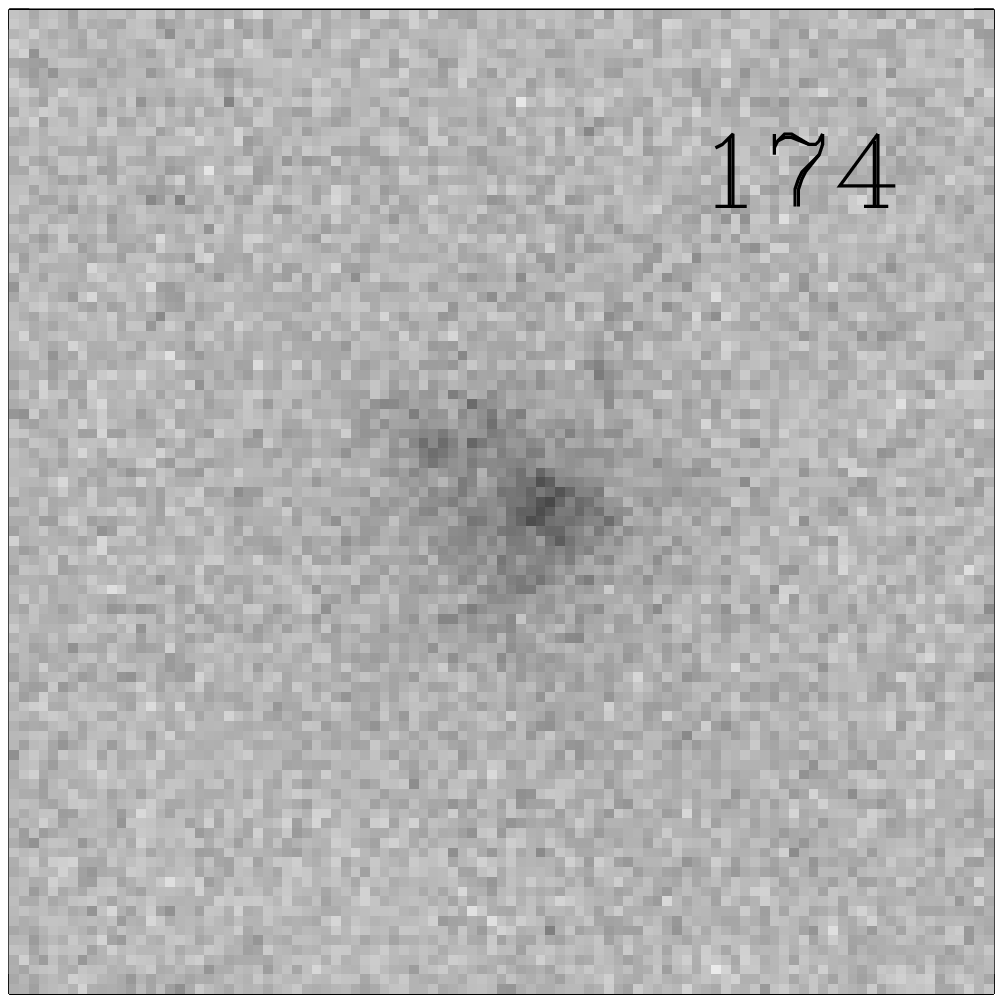}
\includegraphics[scale=0.32]{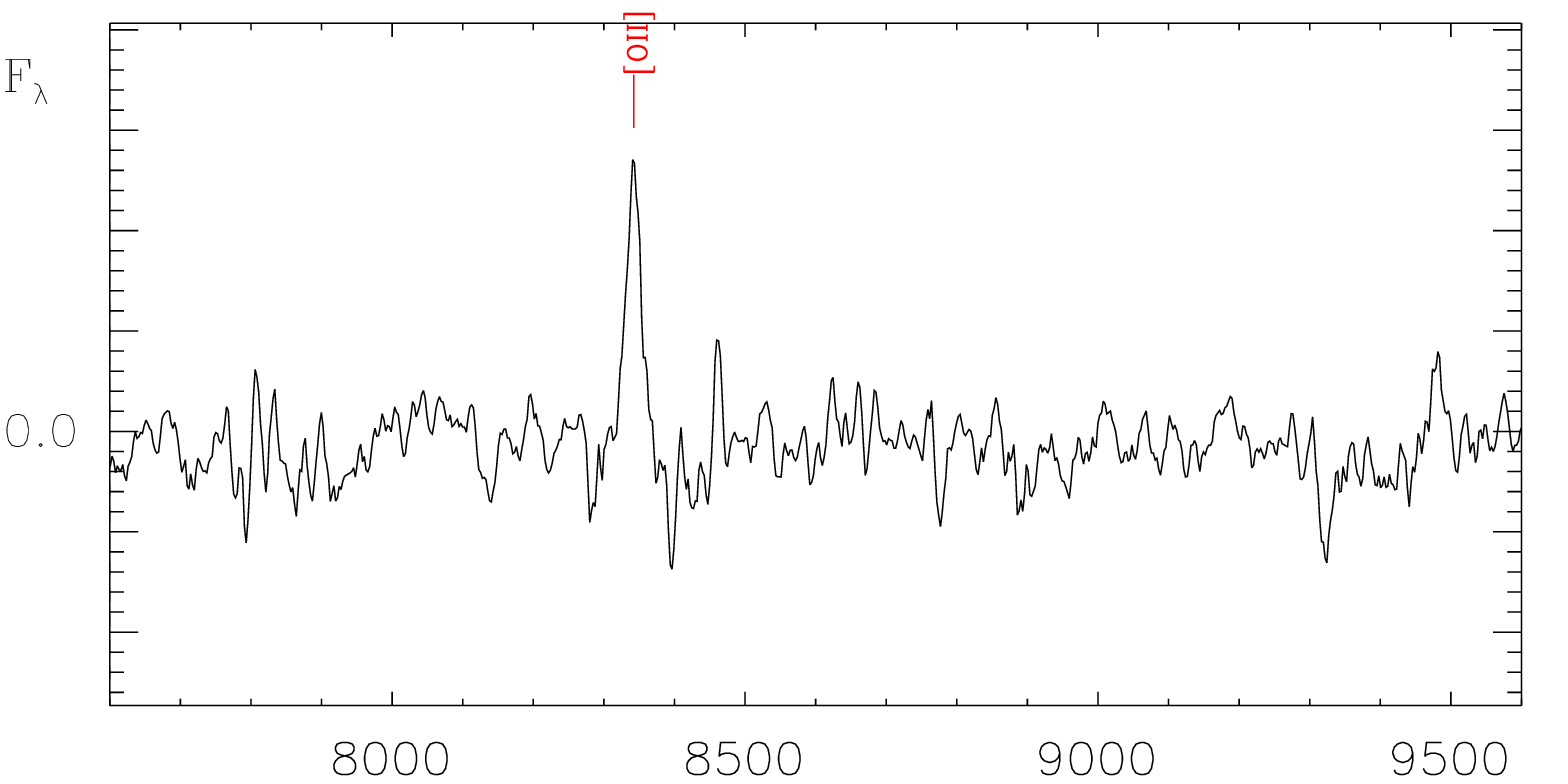}
\vspace{0.1cm}

\includegraphics[scale=0.37]{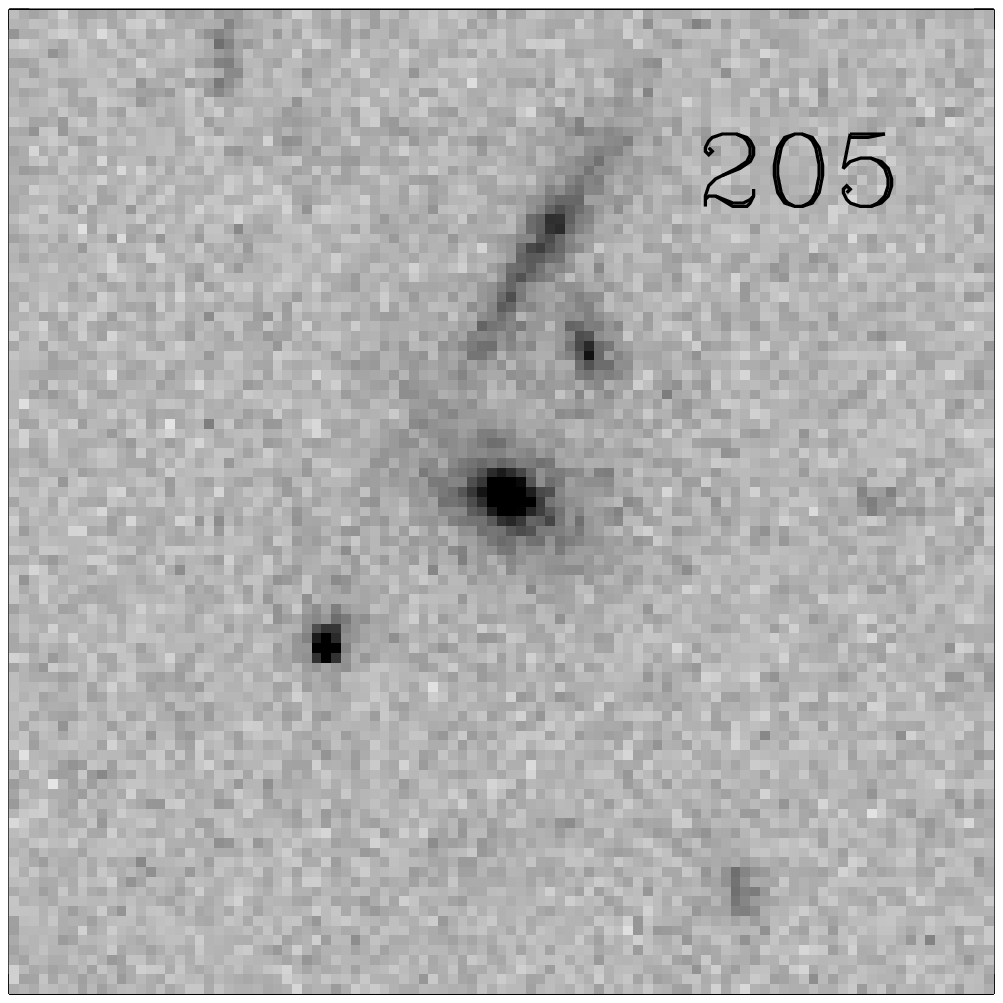}
\includegraphics[scale=0.37]{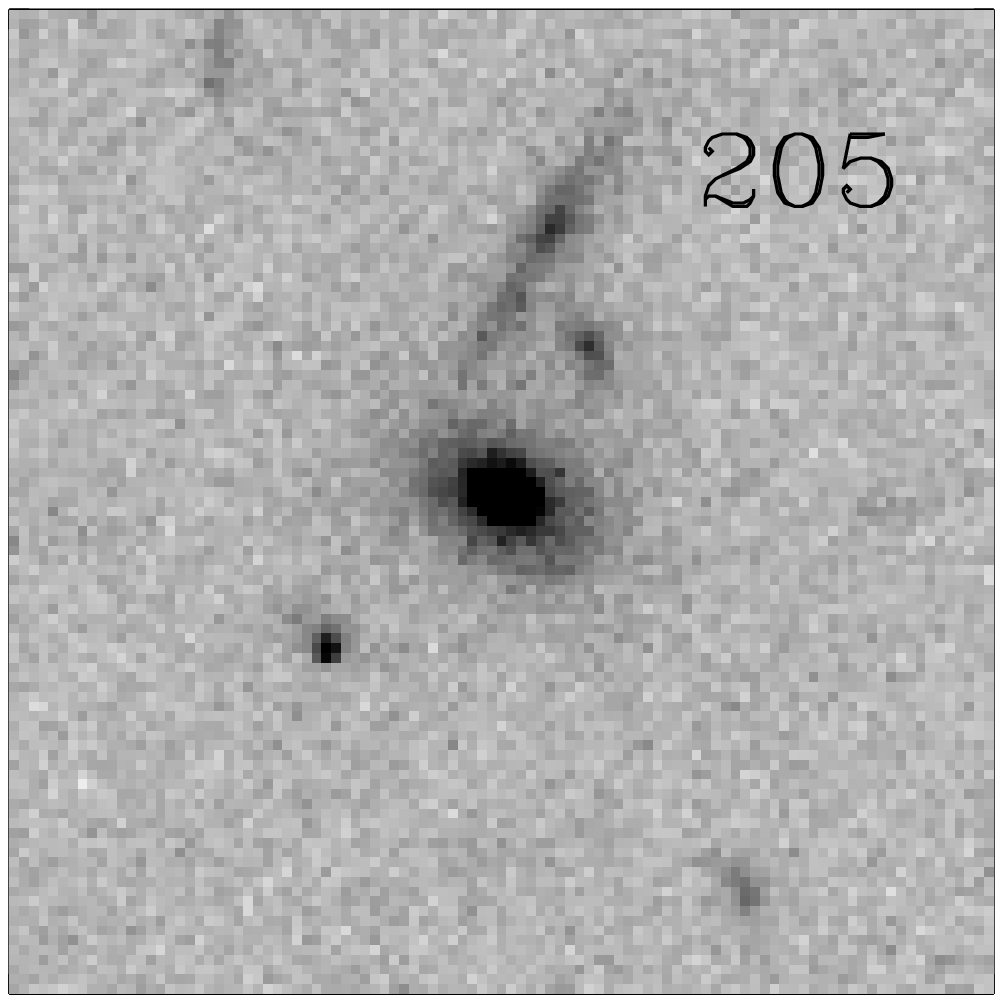}
\includegraphics[scale=0.32]{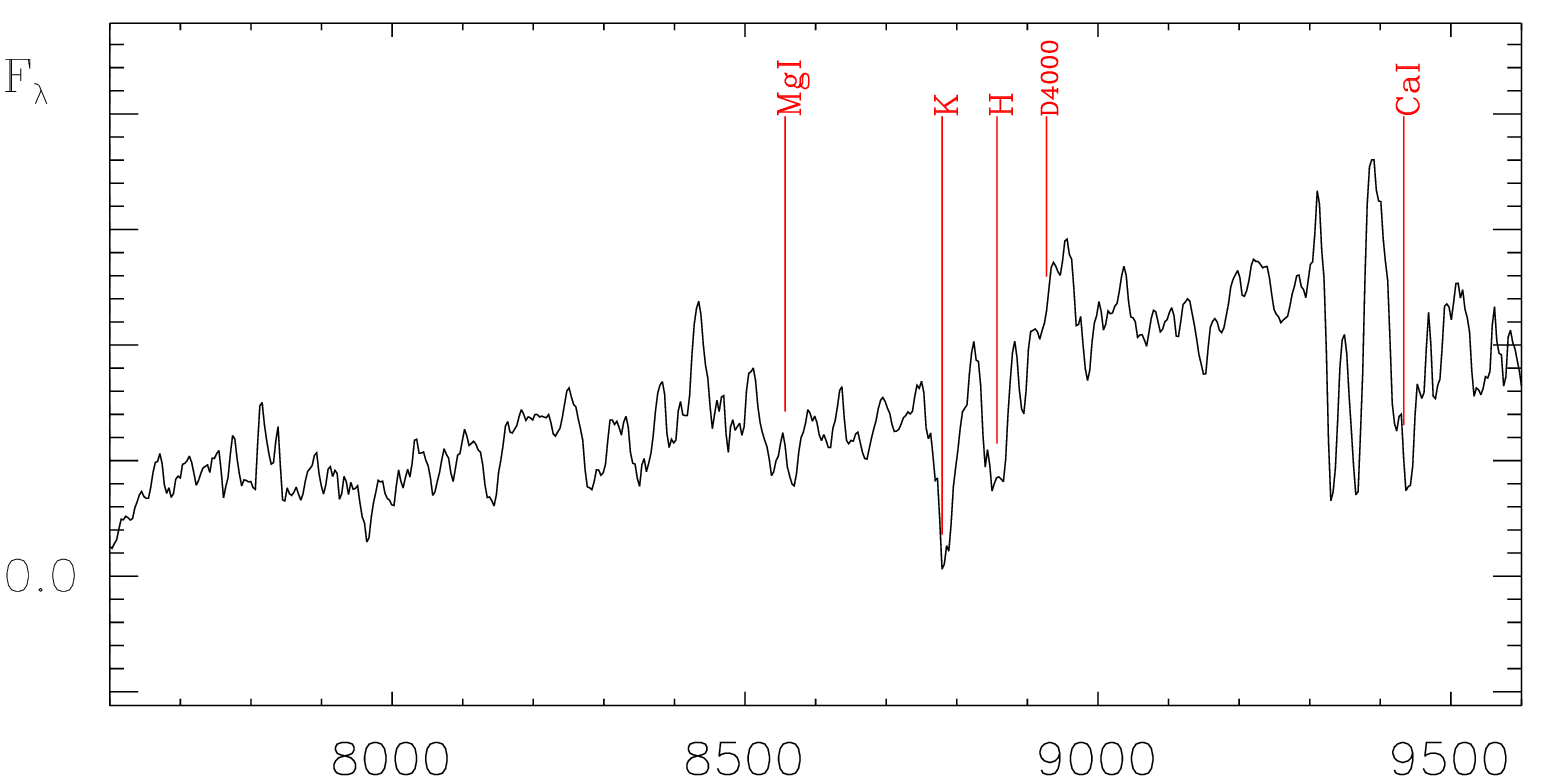}
\vspace{0.1cm}

\includegraphics[scale=0.37]{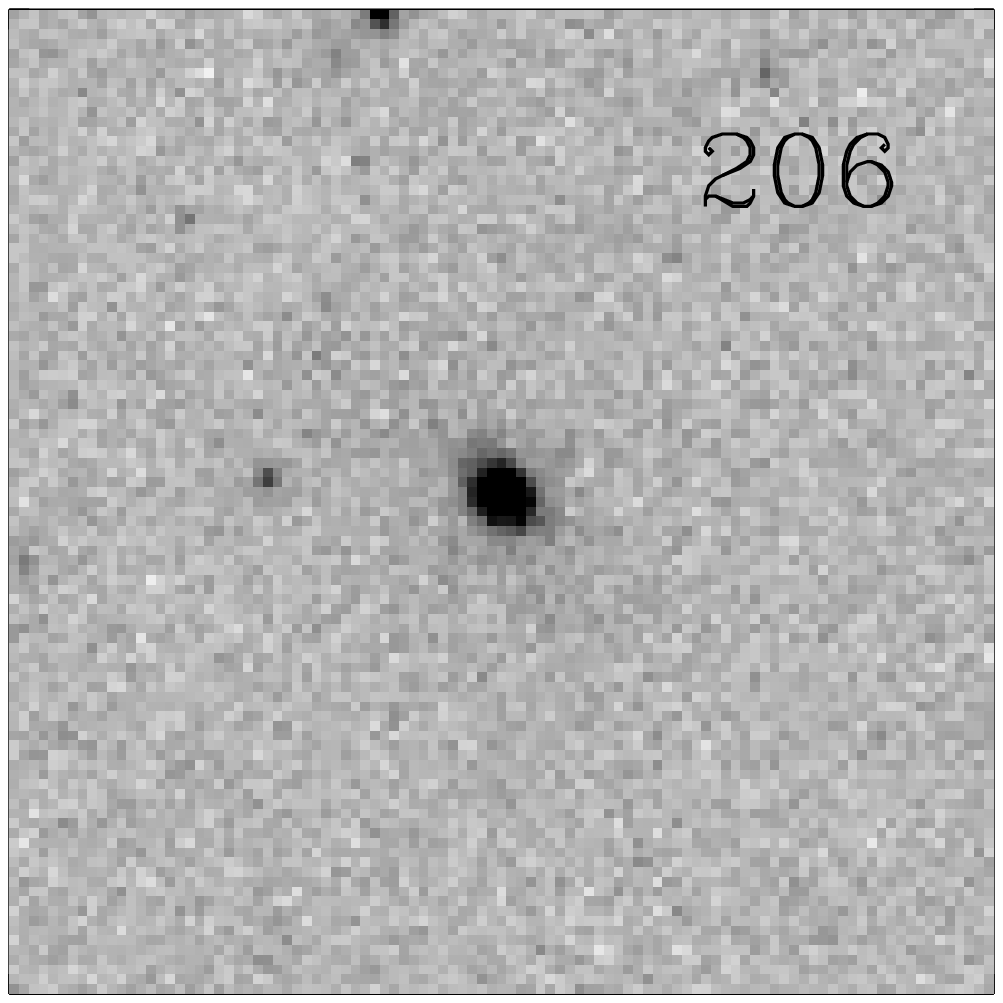}
\includegraphics[scale=0.37]{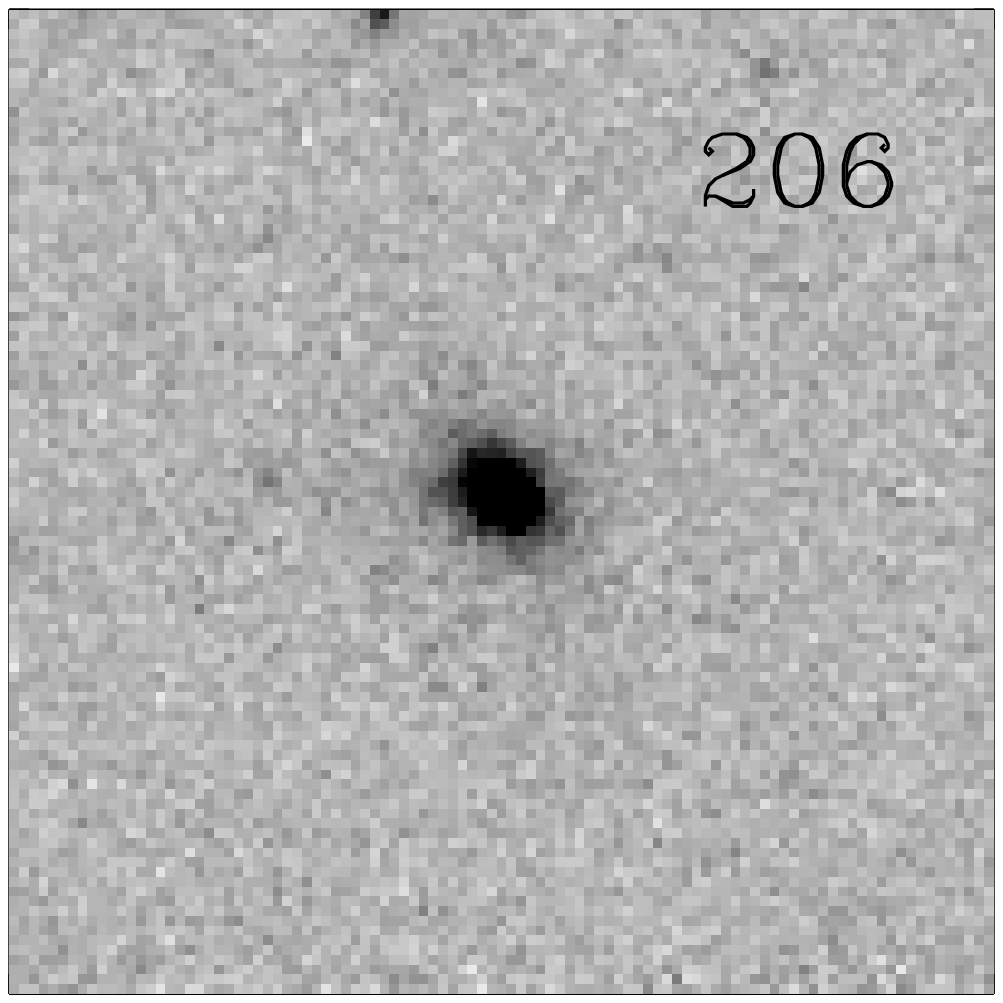}
\includegraphics[scale=0.32]{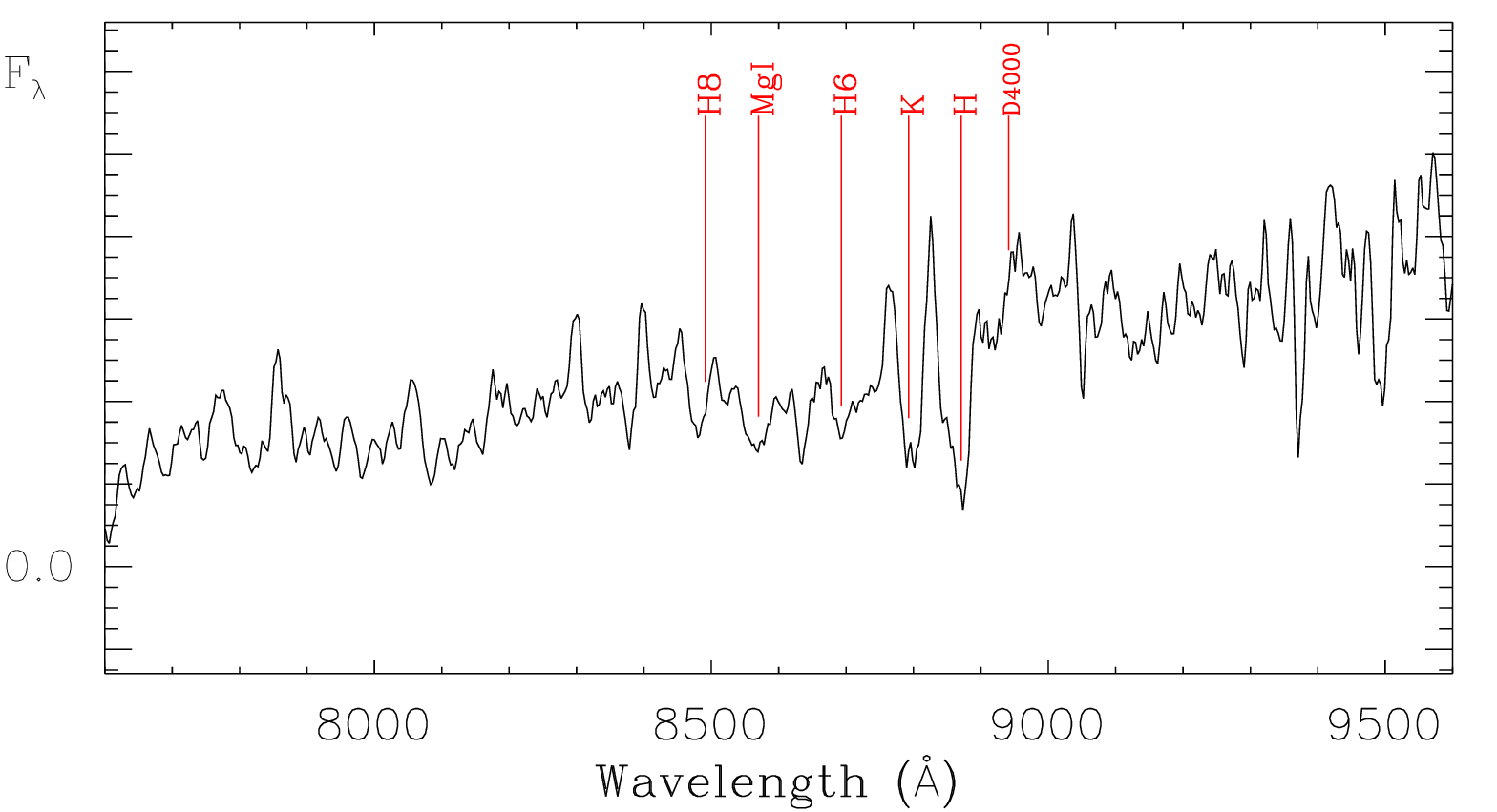}
\vspace{0.3cm}

\caption{(SEE EDITION FROM JOURNAL FOR THE FULL FIGURE) The sample of
37 spectroscopic cluster members in the ACS field of view. The left
and middle panels show ACS cutouts in the $i_{775}$ and
$z_{850}$ bands, respectively. Each cutout is 5\arcsec\ a side and the
number in the panels corresponds to the galaxy ID. The right panel shows the
corresponding VLT/FORS spectrum. Most prominent spectral features
identified in each spectrum in the wavelength range shown here are
marked (see footnote in table \ref{spec_phot_tab} for the list of
spectral features). The spectra have been smoothed by 5 pixels ($1$
pix $\sim 2.5$\AA).}

\label{mosaic}
\end{figure}
\end{center}

\begin{center}
\begin{figure}[!htb]
\plotone{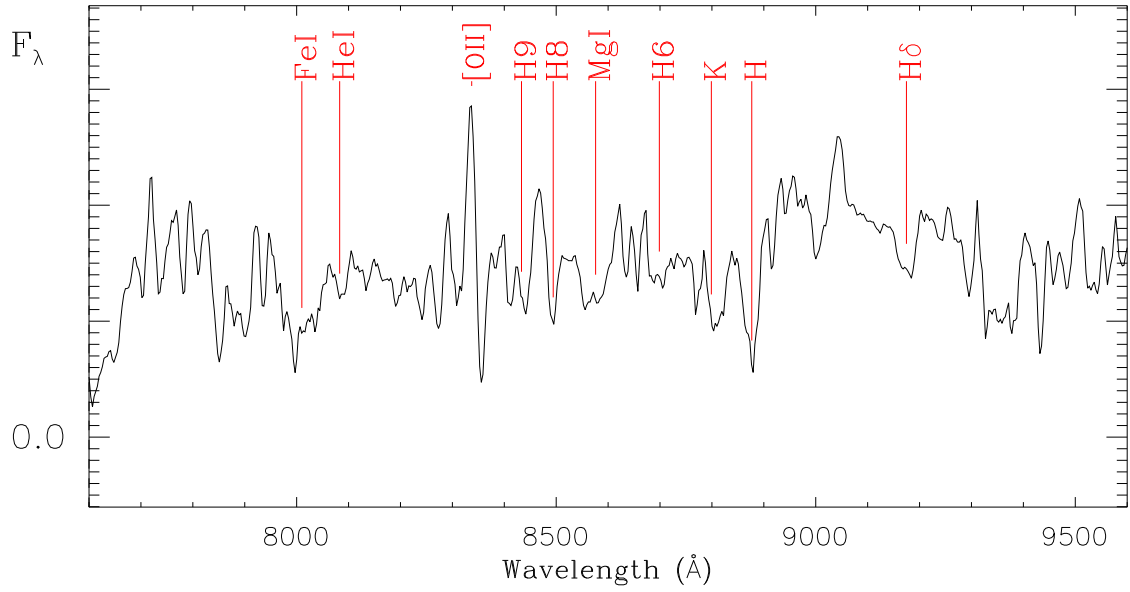}
\vspace{0.3cm}

\caption{VLT/FORS spectrum of cluster member ID=7001. This object was
discovered outside the ACS field of view. Most prominent spectral
features in the wavelength range shown here are marked (see footnote
in table \ref{spec_phot_tab} for the list of spectral features). The
spectrum has been smoothed by 5 pixels ($1$ pix $\sim 3.1$\AA).}

\label{g7001}
\end{figure}
\end{center}

\begin{center}
\begin{figure}[!htb]

\includegraphics[scale=0.35]{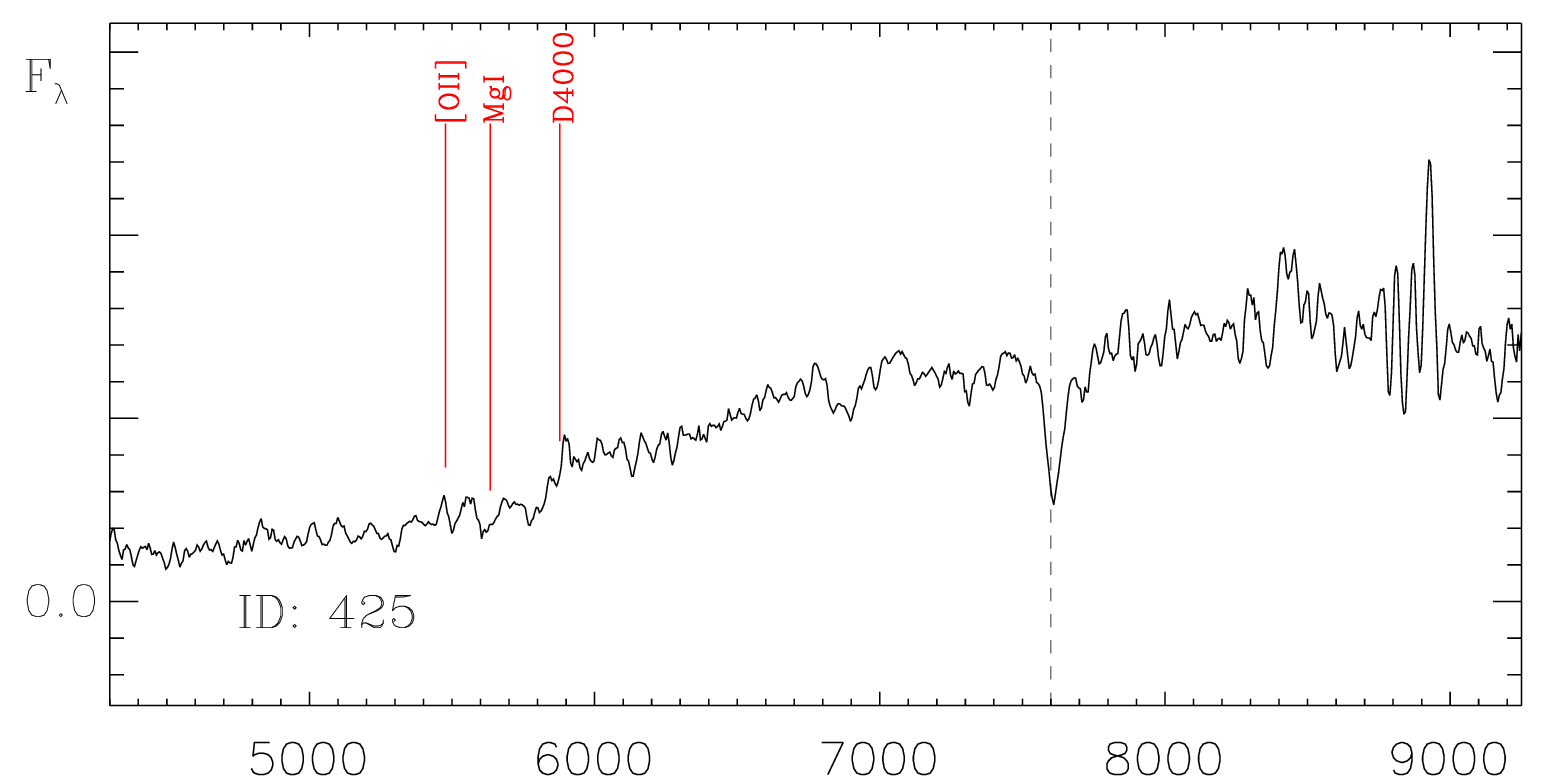}
\includegraphics[scale=0.35]{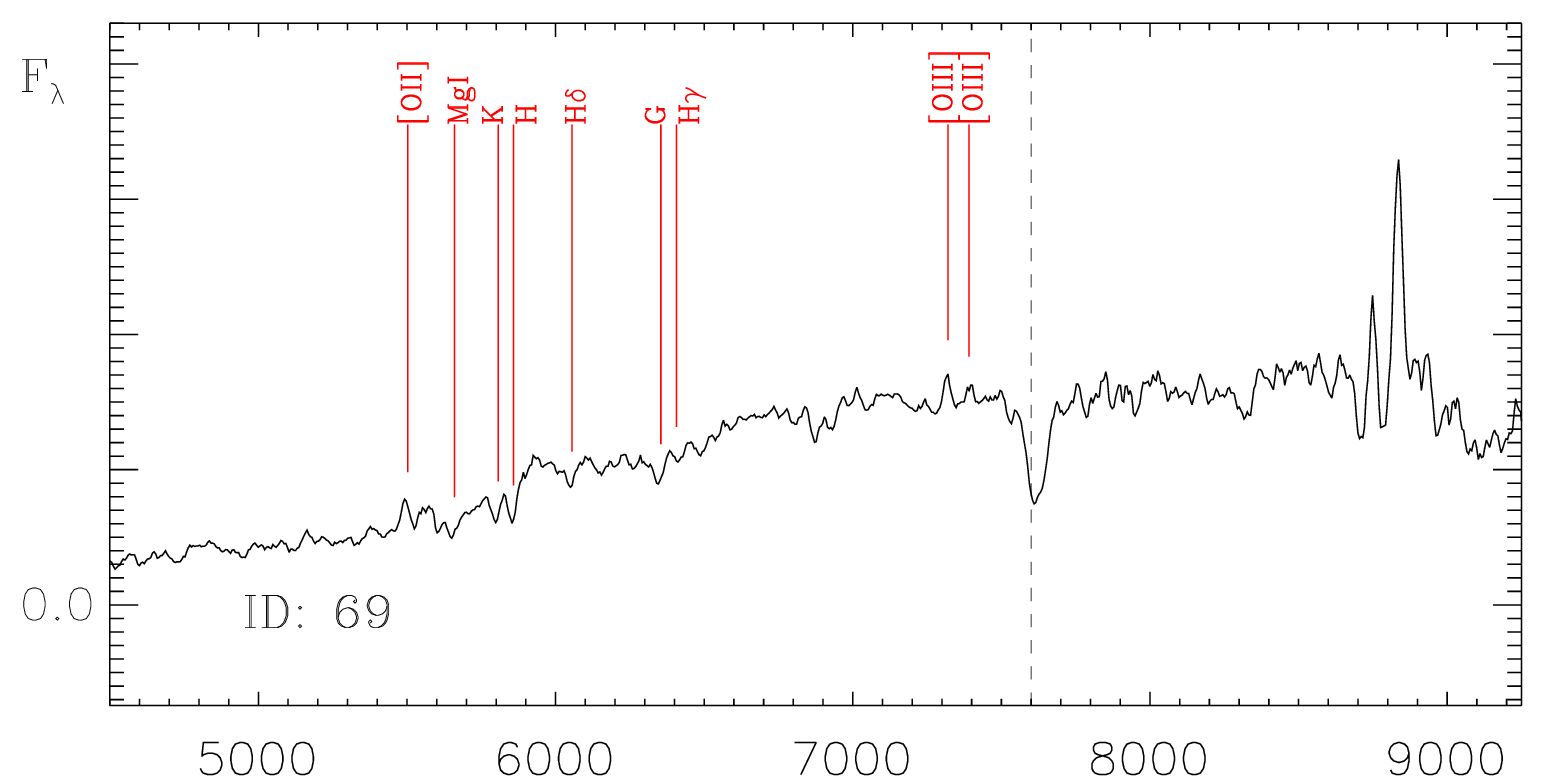}
\includegraphics[scale=0.35]{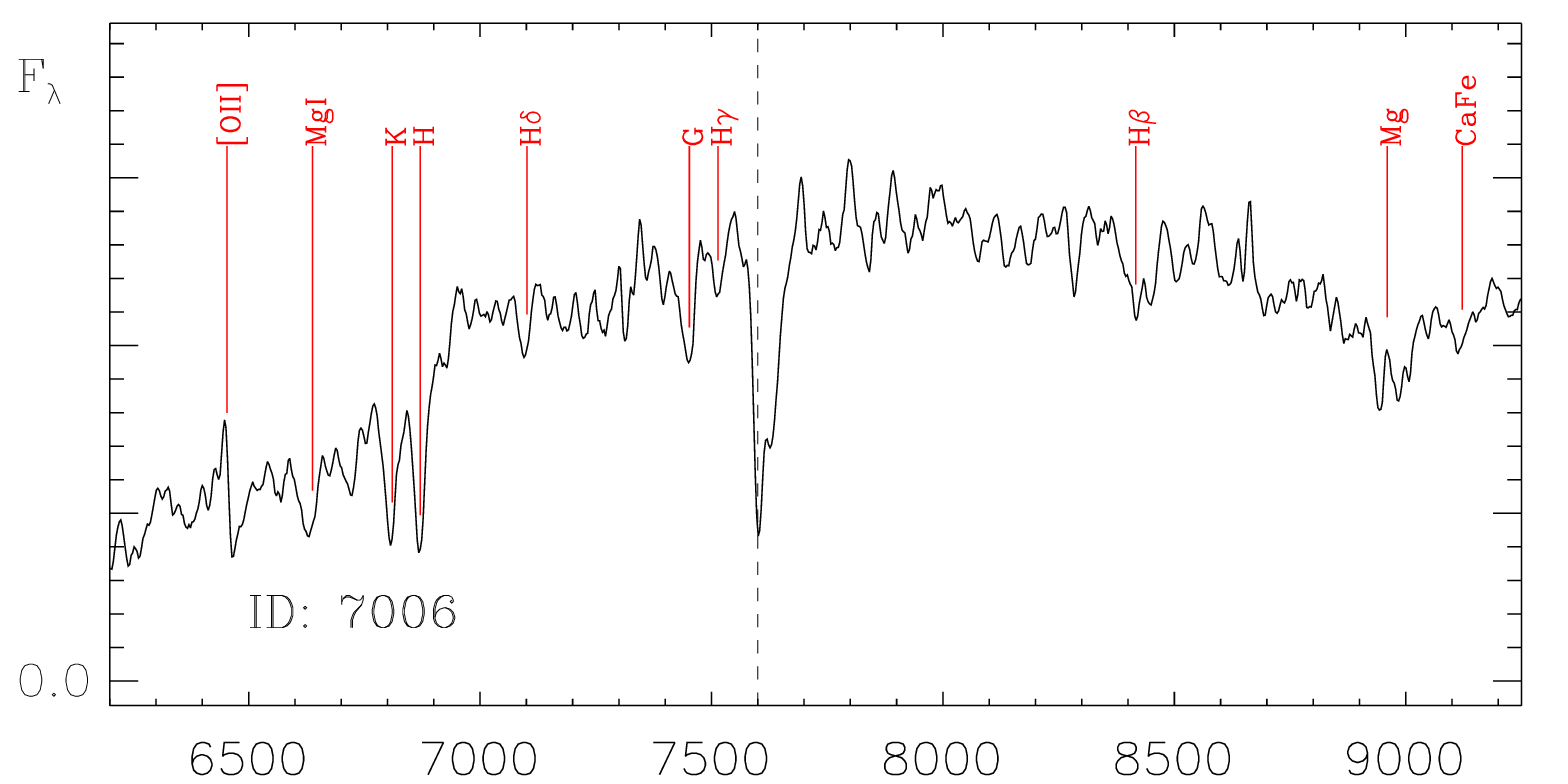}
\includegraphics[scale=0.35]{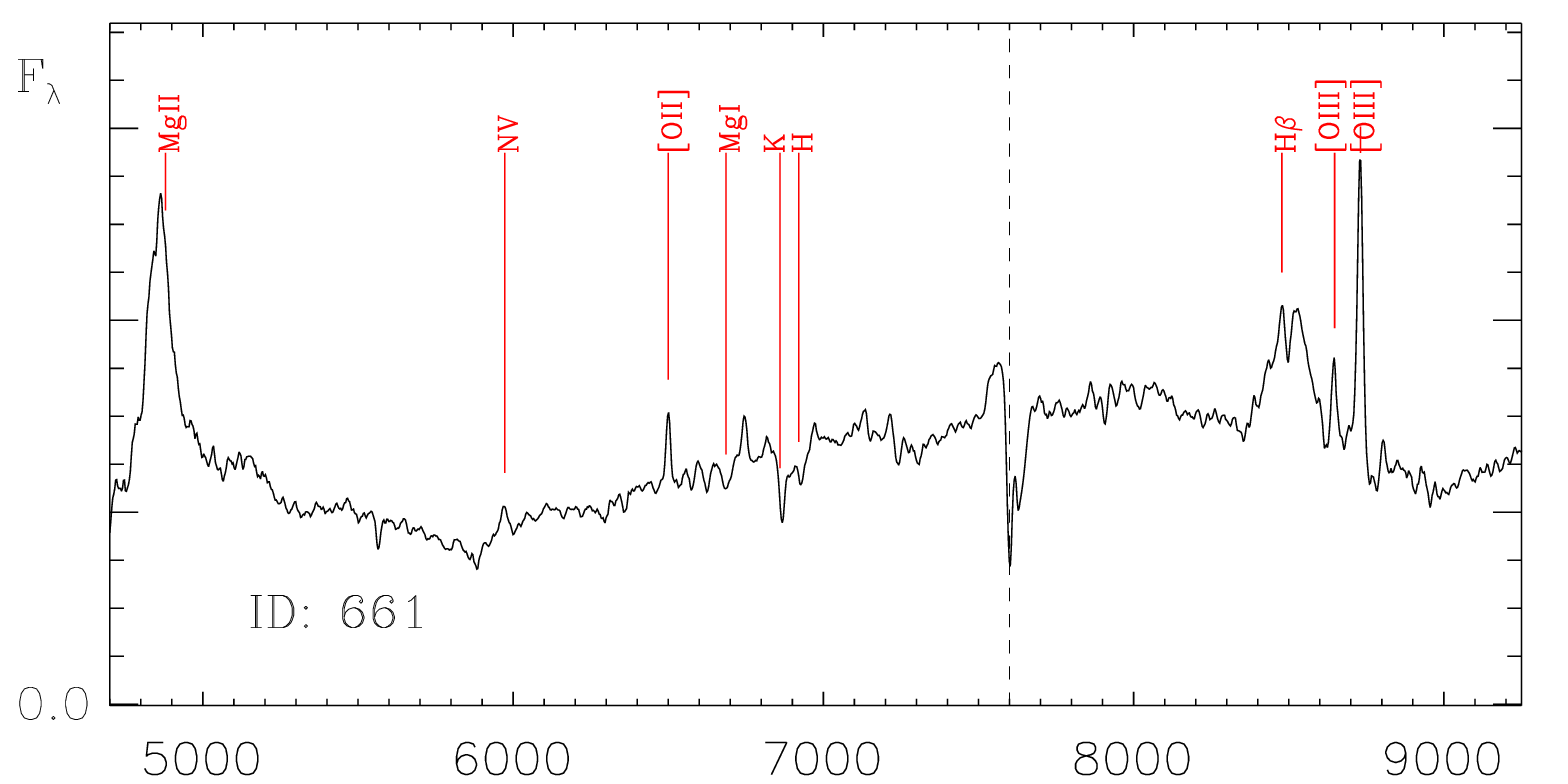}
\includegraphics[scale=0.35]{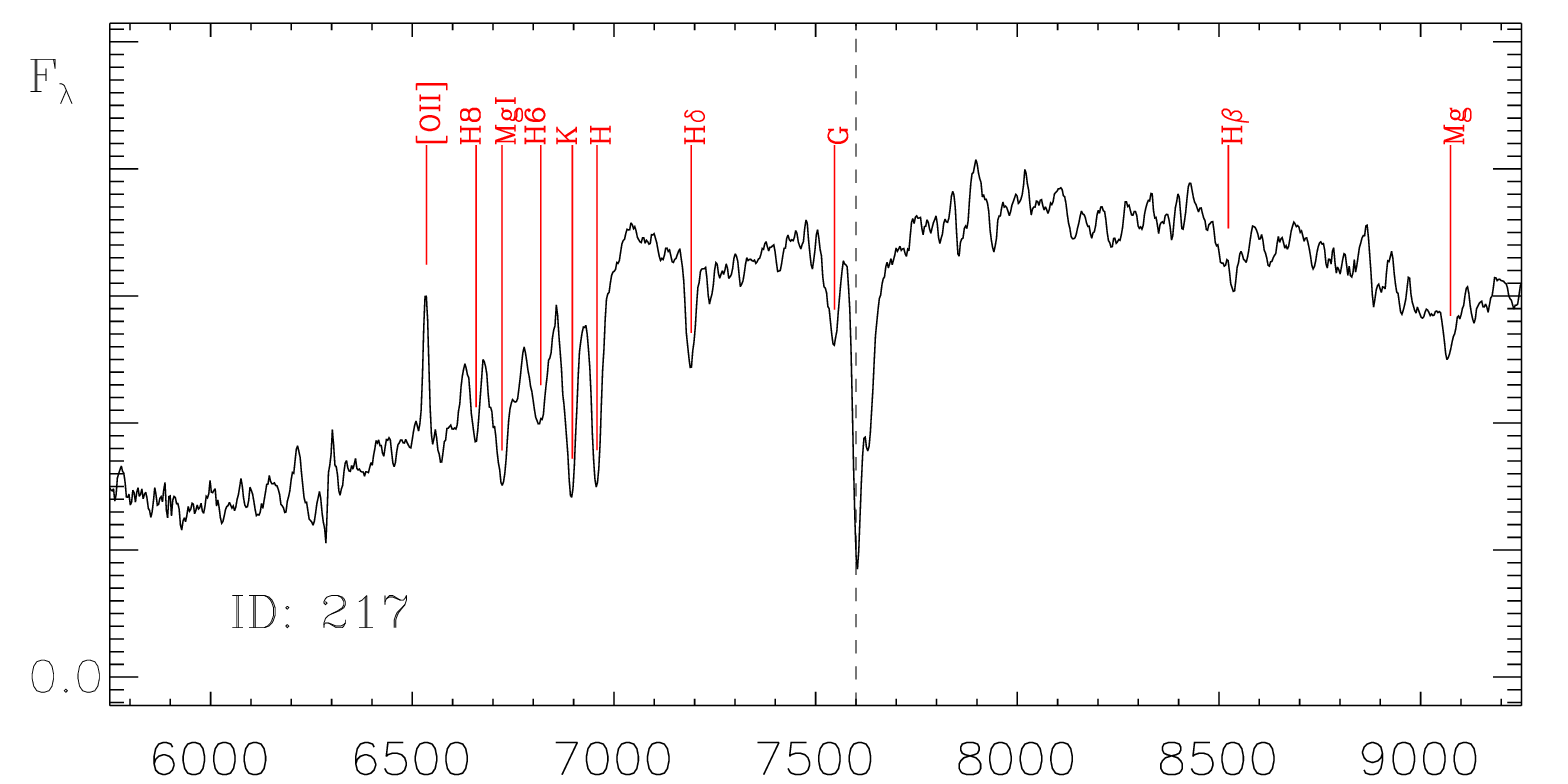}
\includegraphics[scale=0.35]{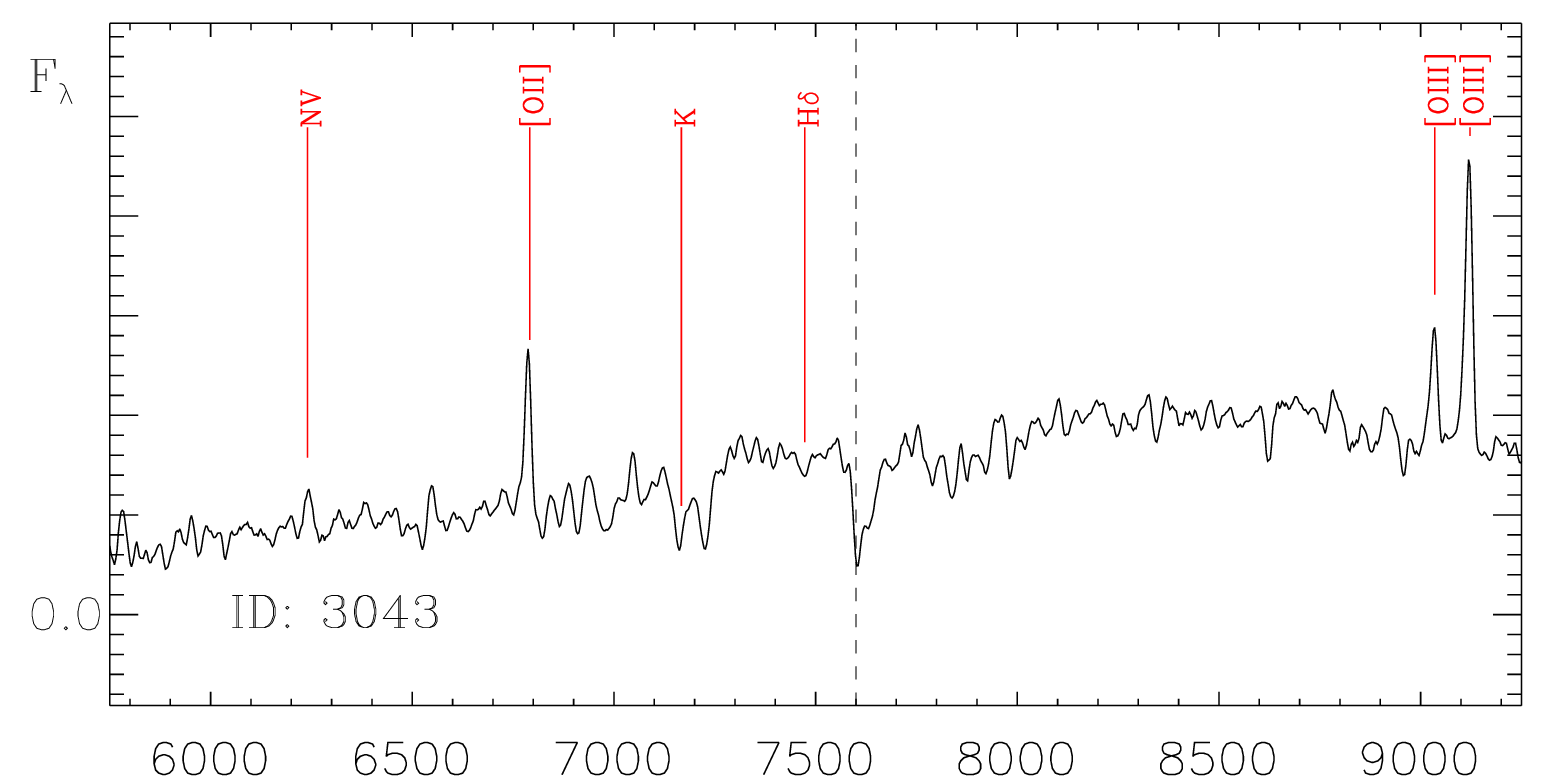}
\includegraphics[scale=0.35]{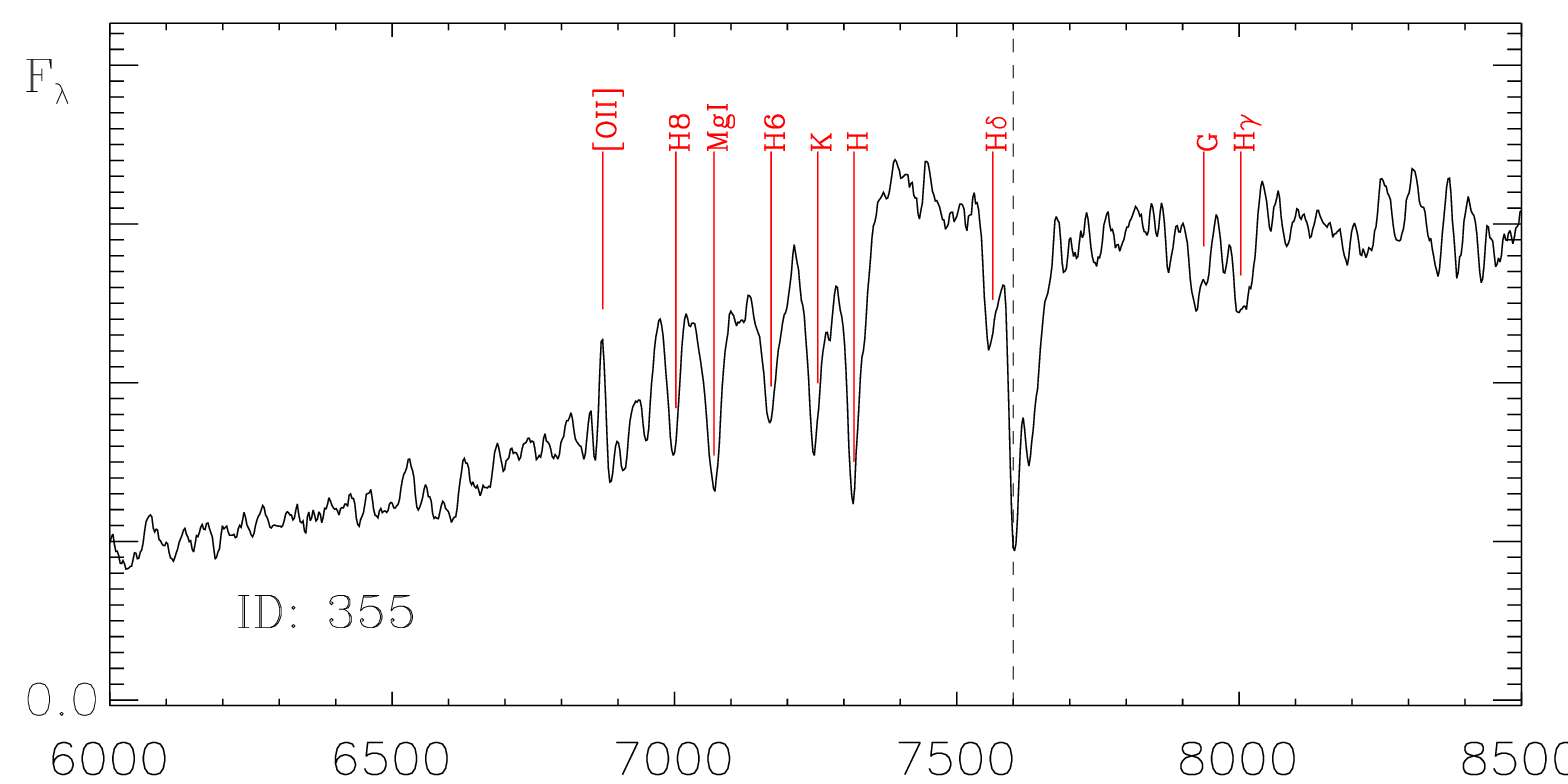}
\includegraphics[scale=0.35]{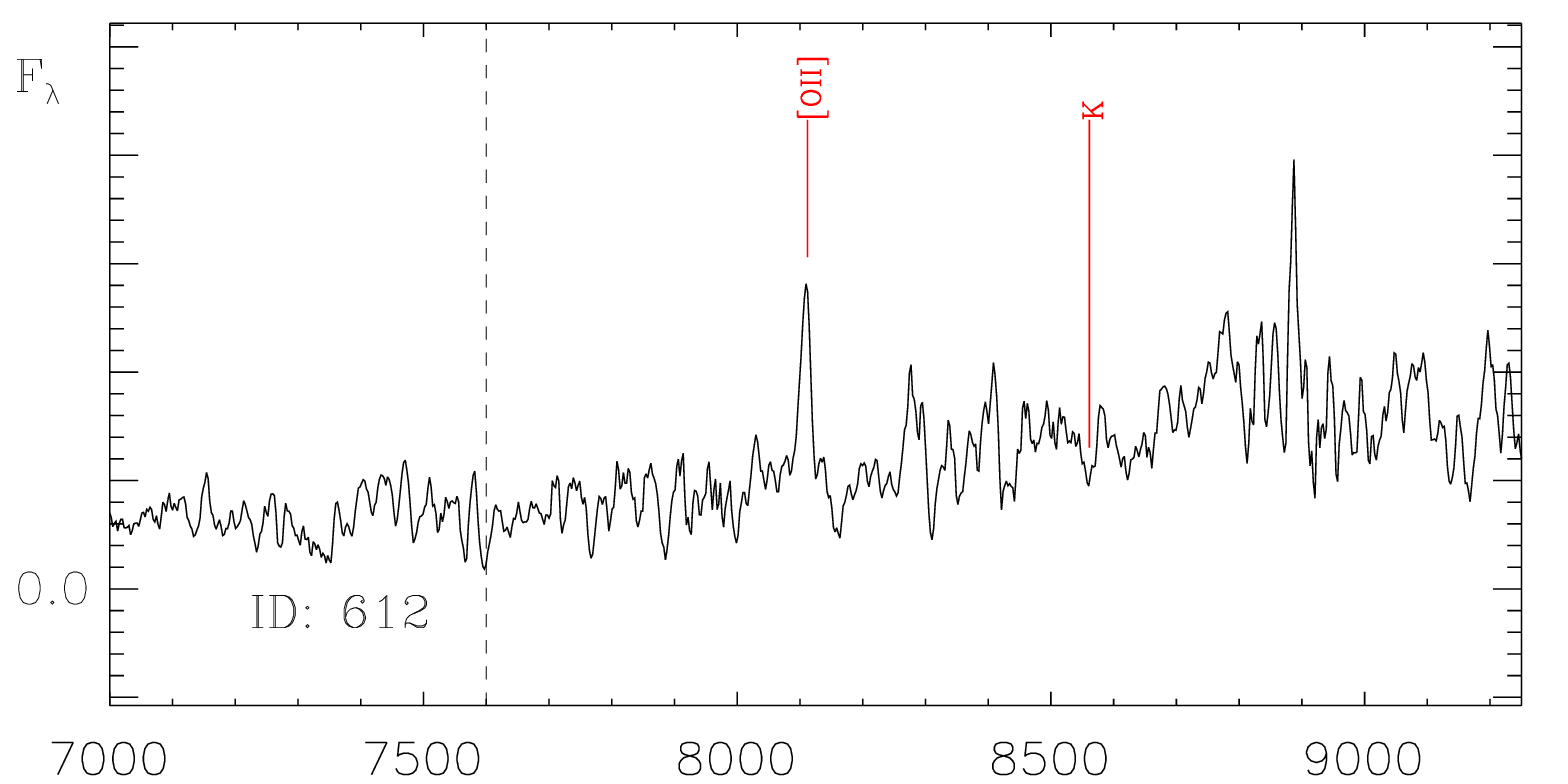}
\includegraphics[scale=0.35]{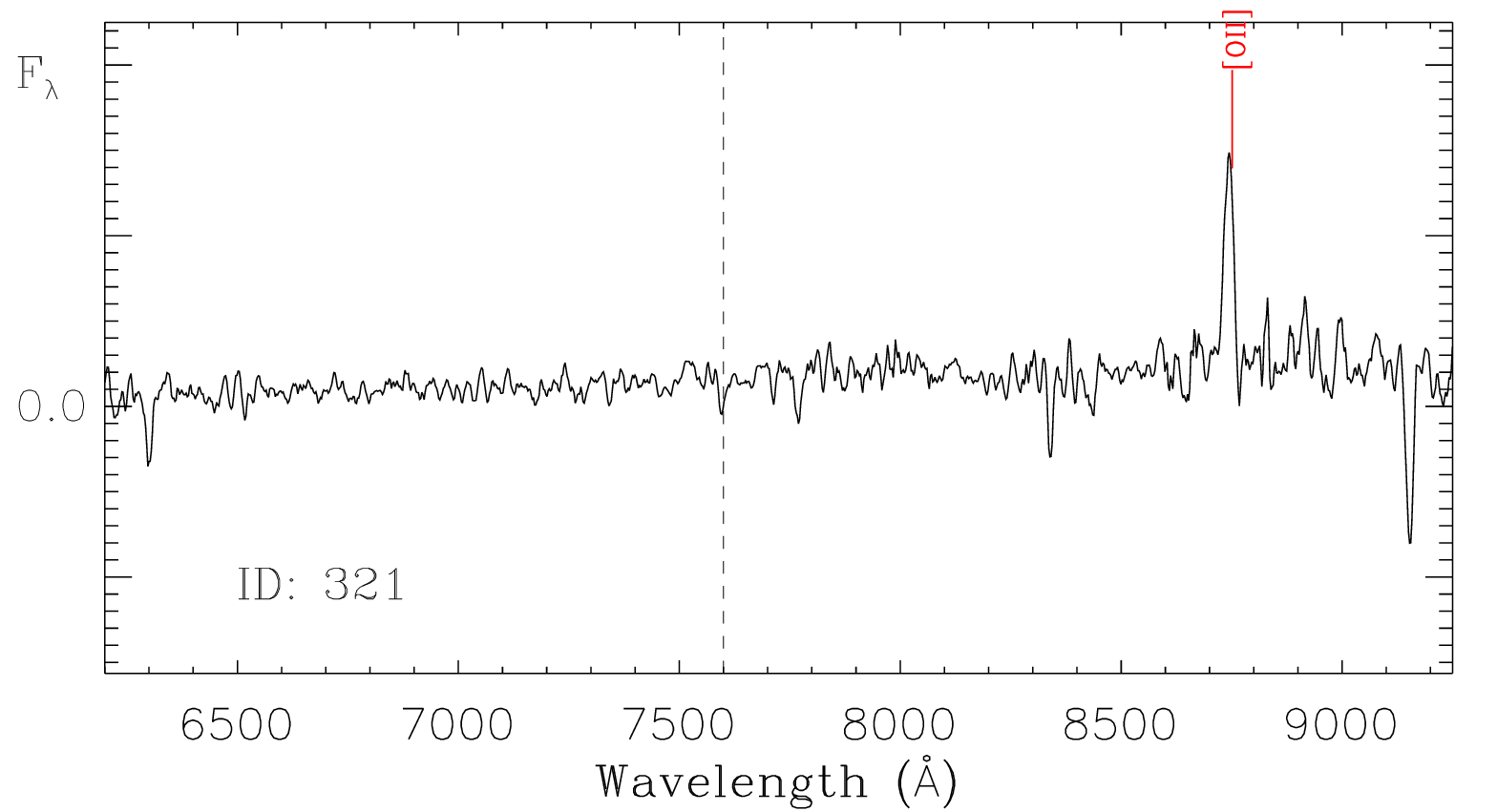}\hspace{0.5cm}
\includegraphics[scale=0.35]{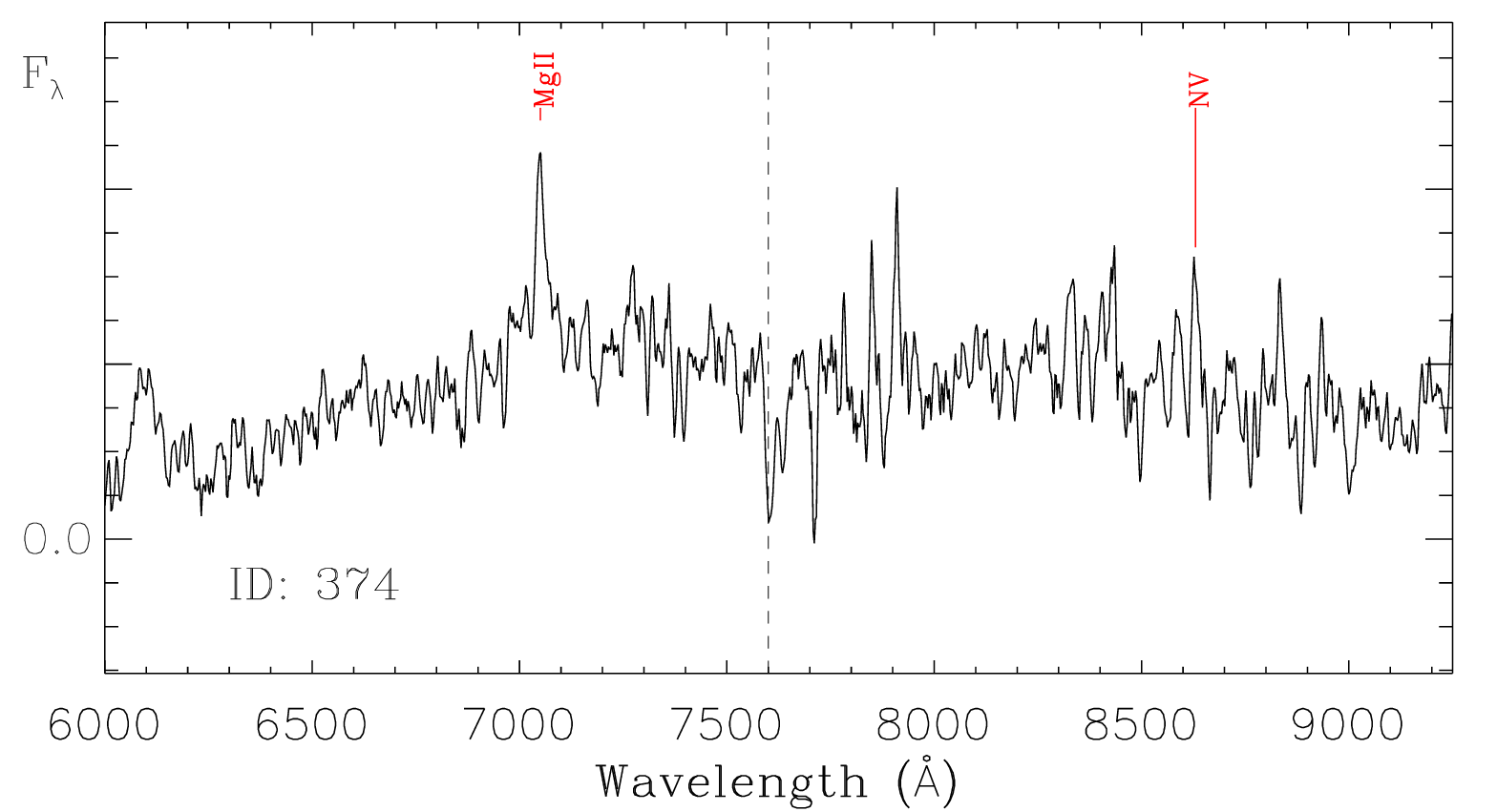}
\vspace{0.5cm}

\caption{\small Optical spectrum of spectroscopically identified non
cluster member AGN in the field of view of \CLhz. Most prominent
spectral features are indicated: MgII($\lambda$2798), NV,
[OII]($\lambda$3727), H8, MgI($\lambda$3834), H6, CaII(H),CaII(K),
D4000, H$\delta$, G, H$\gamma$, H$\beta$, [OIII], Mg and CaFe. The
vertical dashed line indicates the telluric A-band feature at
7600\AA. The spectra have been smoothed by 5 pixels (1 pix $\sim$
3.0\AA).}

\label{mosaic_agn}
\end{figure}
\end{center}

\begin{center}
\begin{figure}
\plotone{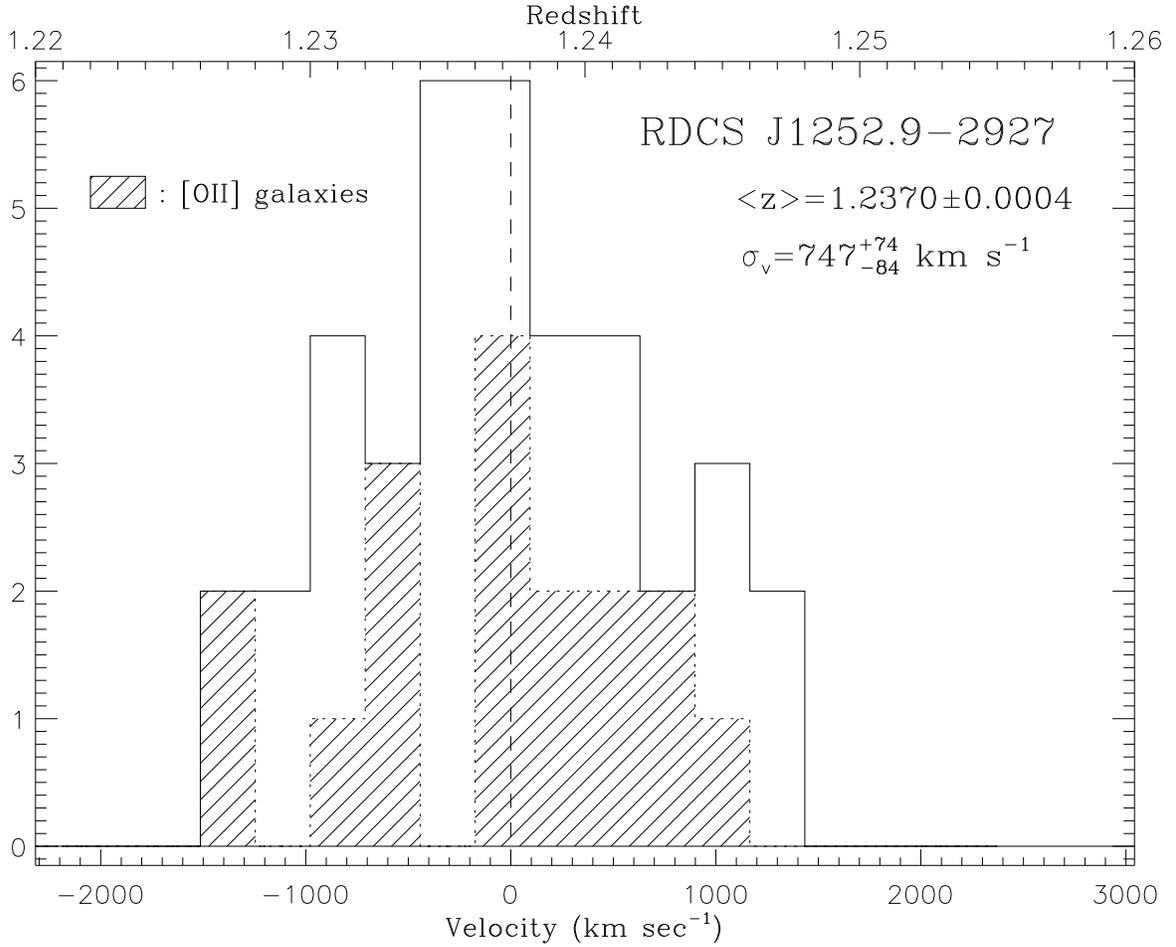}

\caption{Redshift and velocity distribution of cluster members. The
y-axis indicates the number of objects, and the histogram bin size has
been set to $\Delta z=0.002$. Assuming a Gaussian distribution in
velocity of the spectroscopic members, the median redshift of the
distribution is $z=1.2373$ (vertical dashed line) and the global
velocity dispersion is $747^{+74}_{-84}$ km s$^{-1}$, based on the
biweight estimator \citep{bfg90}. The hatched area indicates the
distribution of star-forming ([OII]($\lambda$3727)) members. The mean
redshift and velocity dispersion of passive and [OII] cluster members
are consistent with each other and with the overall cluster values
within the uncertainties.}

\label{z_cluster}
\end{figure}
\end{center}

\begin{center}
\begin{figure}
\plotone{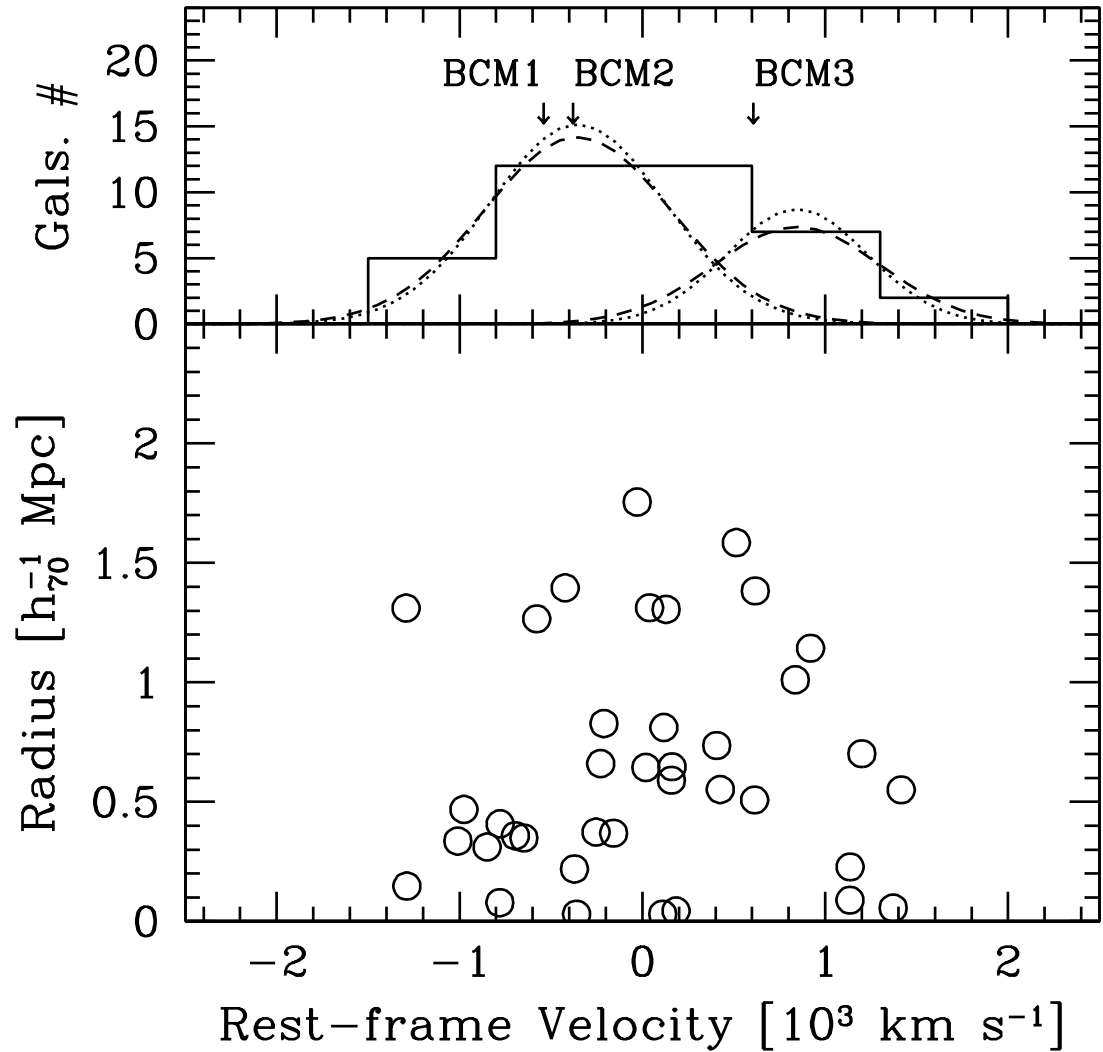}

\caption{ {\it Lower panel}: rest-frame velocity vs. projected
distance from the cluster center \citep*[taken as the X-ray
center;][]{rte04} of the 38 galaxies assigned to the cluster. {\it
Upper panel}: velocity histogram of the 38 galaxies assigned to the
cluster (solid).  Velocities of the three brightest galaxies are
pointed out.  The two dashed Gaussians correspond to the 3D KMM
partition (KMM1 and KMM2 from left to right) while the dotted
Gaussians indicate the two groups detected by the WGAP procedure
(WGAP1 and WGAP2 from left to right). The 3D diagnostics is, in
general, the most sensitive indicator of the presence of
substructure. Therefore, we consider the KMM groups as a robust
characterization of the cluster substructure.}

\label{radial_profile}
\end{figure}
\end{center}

\begin{center}
\begin{figure}
\plotone{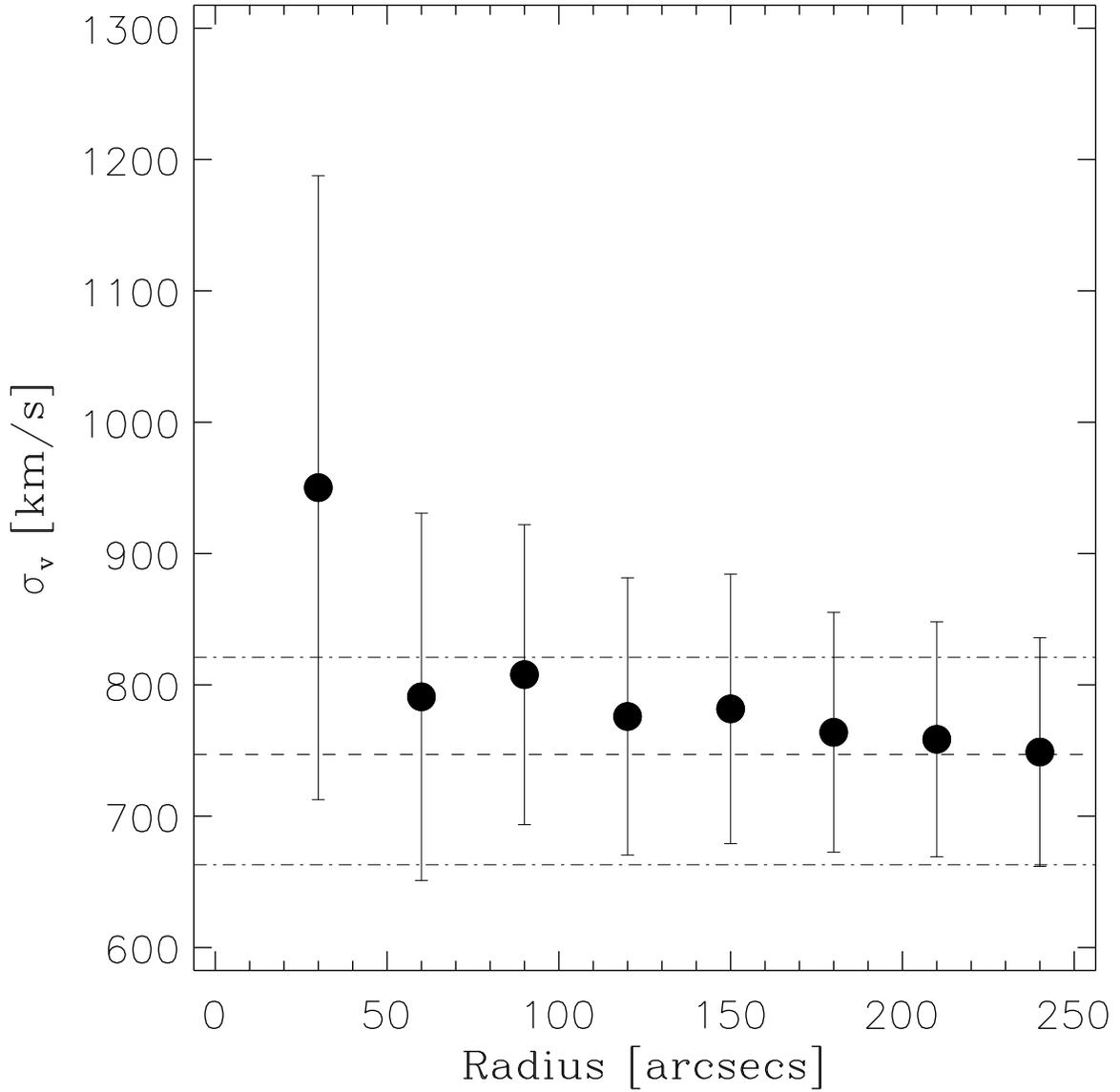}

\caption{Rest-frame integrated velocity dispersion profile of
\CLhz. At the cluster redshift ($z=1.237$), $0.5$ $\mathrm{arcmin}$
corresponds to $250$ kpc. The dashed line indicates the overall
rest-frame cluster velocity dispersion ($\sigma_v = 747^{+74}_{-84} \
km \ s^{-1}$) obtained from the biweight estimator \citep{bfg90} with
the corresponding error bars as dot-dashed lines. The solid dots are
the integrated rest-frame velocity dispersion calculated at the given
radius by using all the galaxies within that radius \citep{gfg96}.
These values were obtained by using Tukey's biweight estimator
\citep{ptv92}, correcting for velocity errors \citep{ddd80}. Error
bars\ are obtained from the fractional uncertainty in estimate for
$\sigma_v$ \citep{t97}. The cluster velocity dispersion values are
observed to be robust at radii larger than $170$ $\mathrm{arcsec}$
(1.4 Mpc), i.e., at these radii, any velocity anisotropy of cluster
galaxies does not affect the value of $\sigma_v (< R)$.  }

\label{sigma_profile}
\end{figure}
\end{center}


\begin{center}
\begin{figure}
\plotone{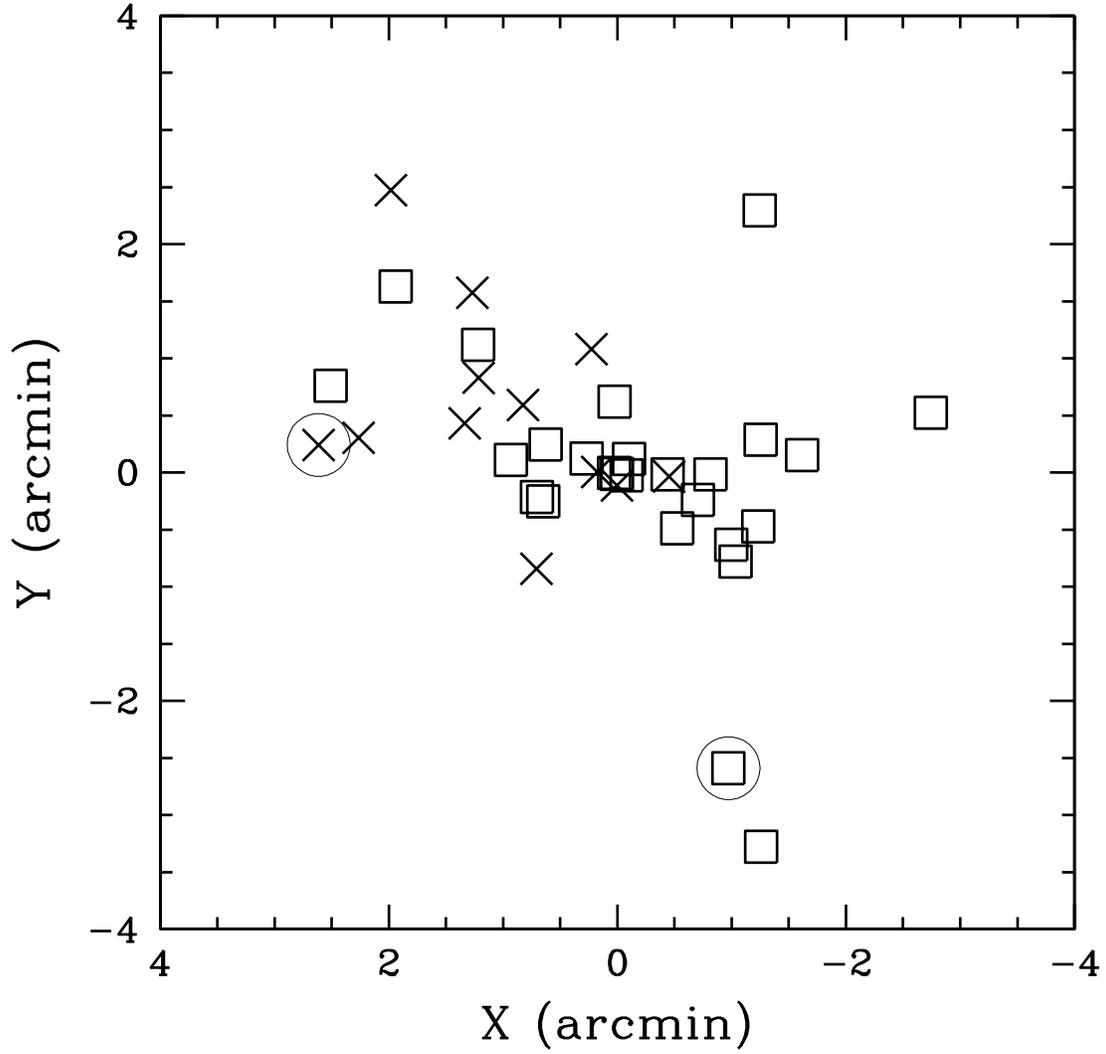}

\caption{Projected distribution on the sky of the 38 member galaxies:
open squares and crosses indicate galaxies assigned to KMM1 and KMM2
groups, respectively. The two large open circles indicate the two
galaxies which gives the difference in membership with WGAP1 and WGAP2
groups. The plot is centered on the X--ray cluster center.  KMM1 is
mostly located to the West of the cluster center while KMM2 is mostly
to the East, and the degree of merging observed from galaxy positions
shows that these two groups have already started virialization in the
main cluster potential.}

\label{figkmm}
\end{figure}
\end{center}

\begin{center}
\begin{figure}
\plotone{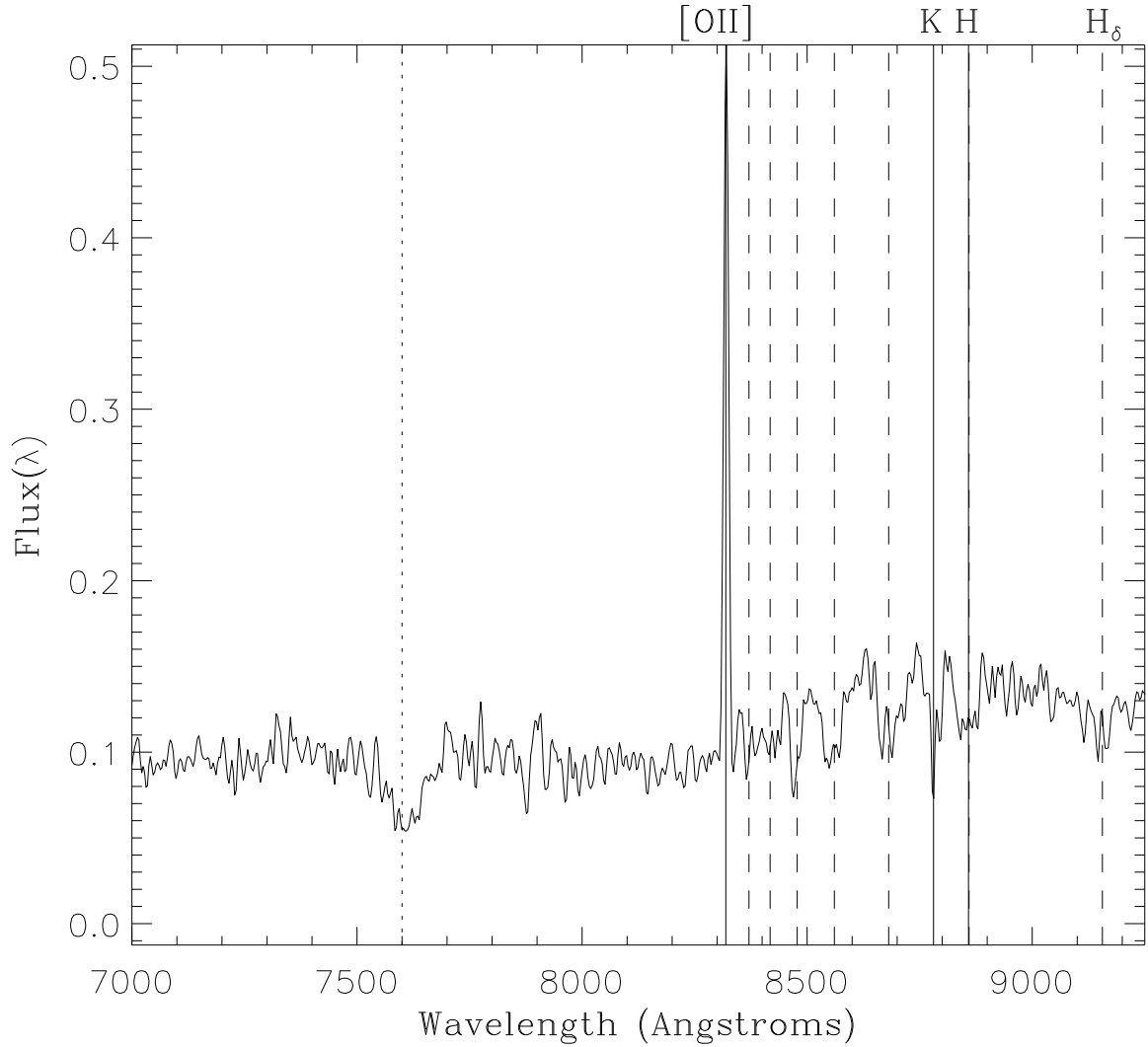}

\caption{Co-added spectrum of the 17 spectroscpically confirmed
star-forming members in \CLhz. The most prominent spectral features in
the displayed wavelength range are indicated. From right to left, the
dashed lines are features of the Balmer series: $H_{\delta}$,
$H_{\epsilon}$ (next to the $Ca II$ H line), and the higher order
features H6, H7, H8, H9 and H10. The solid lines are the [OII]
($\lambda$3727) emission feature and the $Ca II$ K and $Ca II$ H
absorption features. The dotted line indicates the A band telluric
feature at 7600 \AA. The spectrum has been smoothed by 5 pixels (1 pix
$\sim$ 3.1\AA).}

\label{sf_coadded}
\end{figure}
\end{center}

\begin{center}
\begin{figure}[!htb]

\includegraphics[scale=0.65]{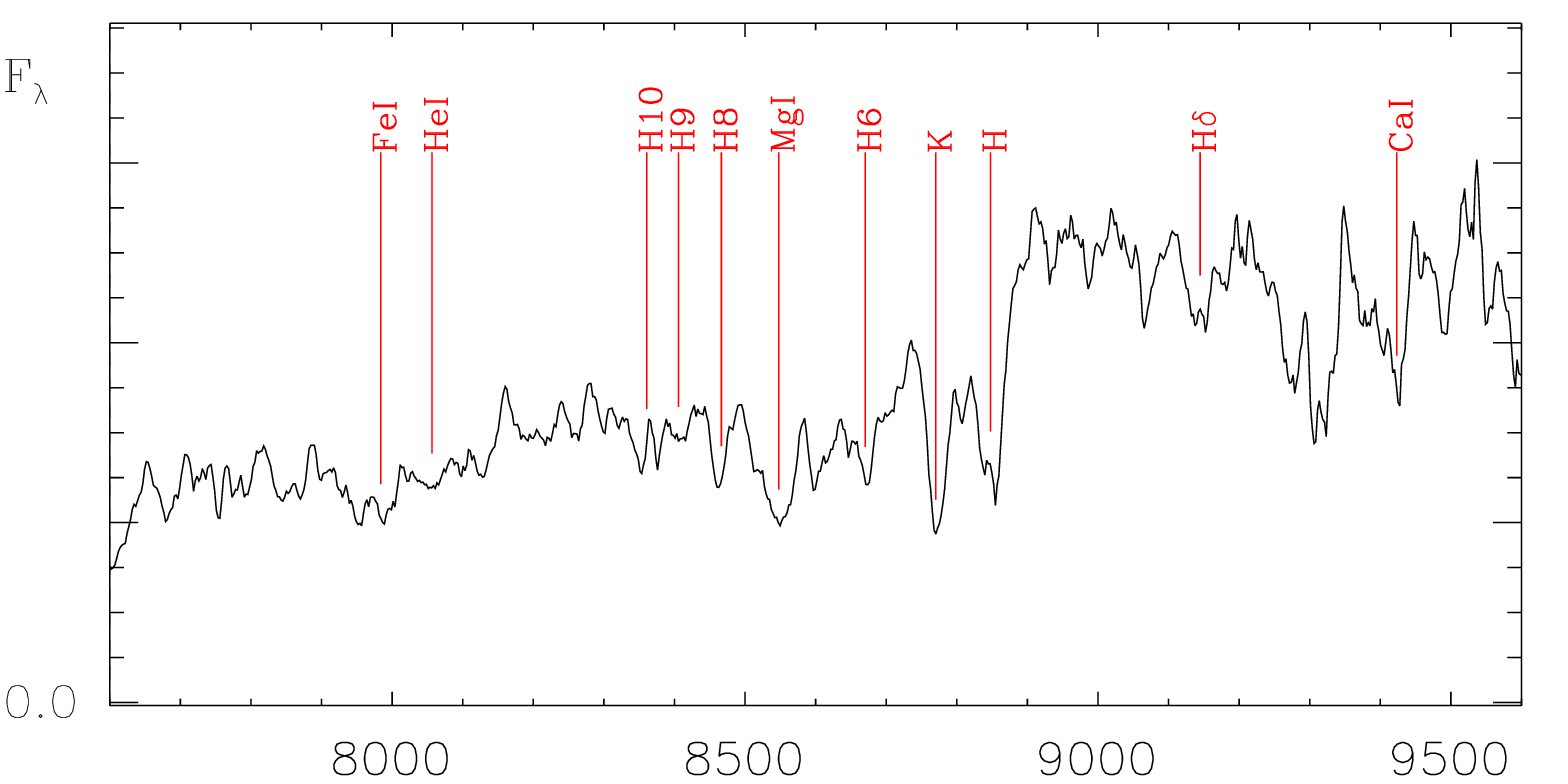}
\includegraphics[scale=0.65]{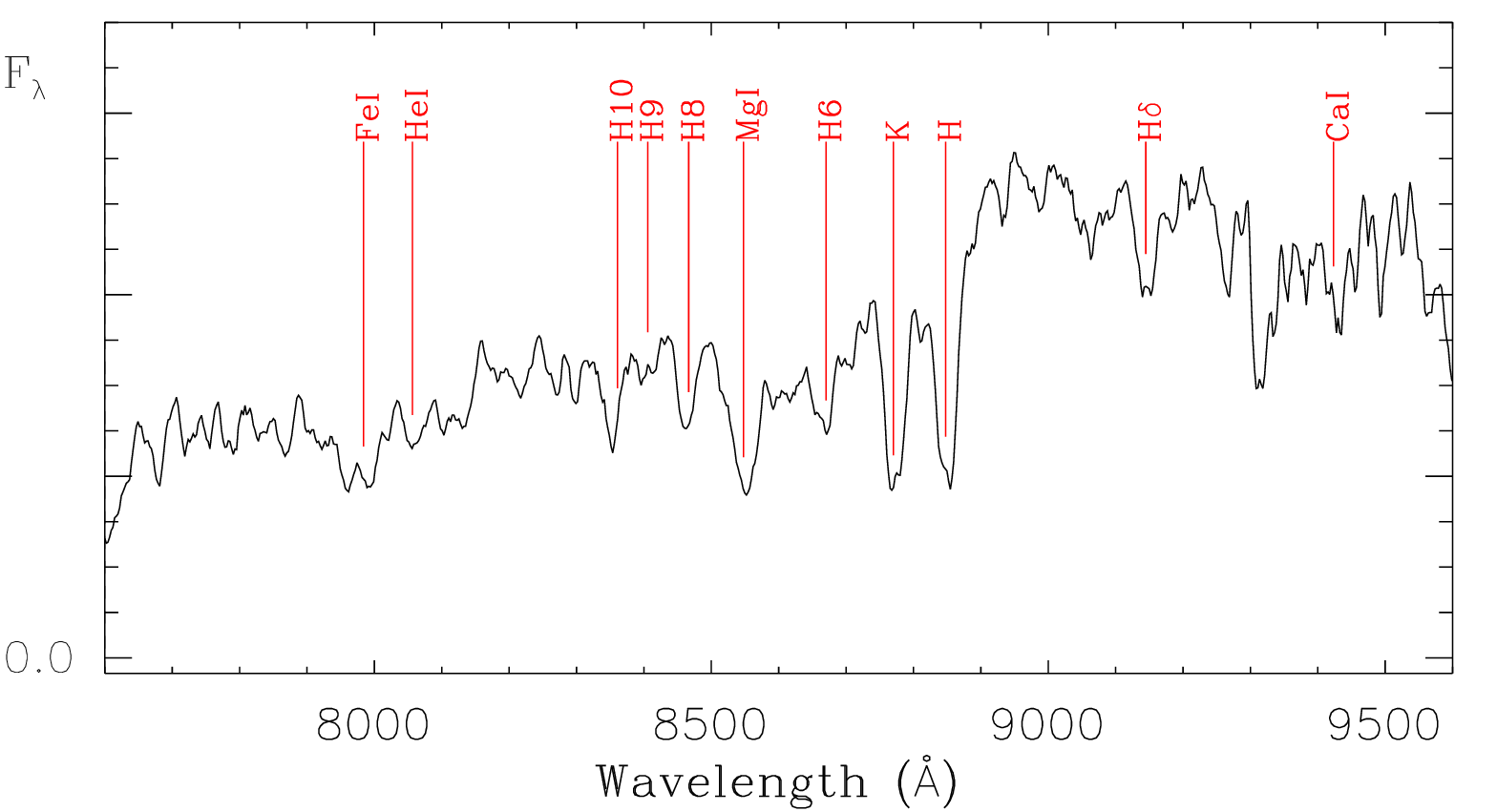}
\vspace{0.5cm}

\caption{Co-added spectrum of the 10 (top) and 20 (bottom) brightest
(in $K_s$) passive cluster members.  Prominent Balmer absorption
features become visible by co-adding spectra, indicating the existence
of post-starburst stellar populations in passive early-type galaxies
at this redshift (see footnote in table \ref{spec_phot_tab} for a list
of spectral features). The spectra have been smoothed by 5 pixels (1
pix $\sim$ 2.5\AA).}

\label{stack_membs}
\end{figure}
\end{center}

\begin{center}
\begin{figure}
\plotone{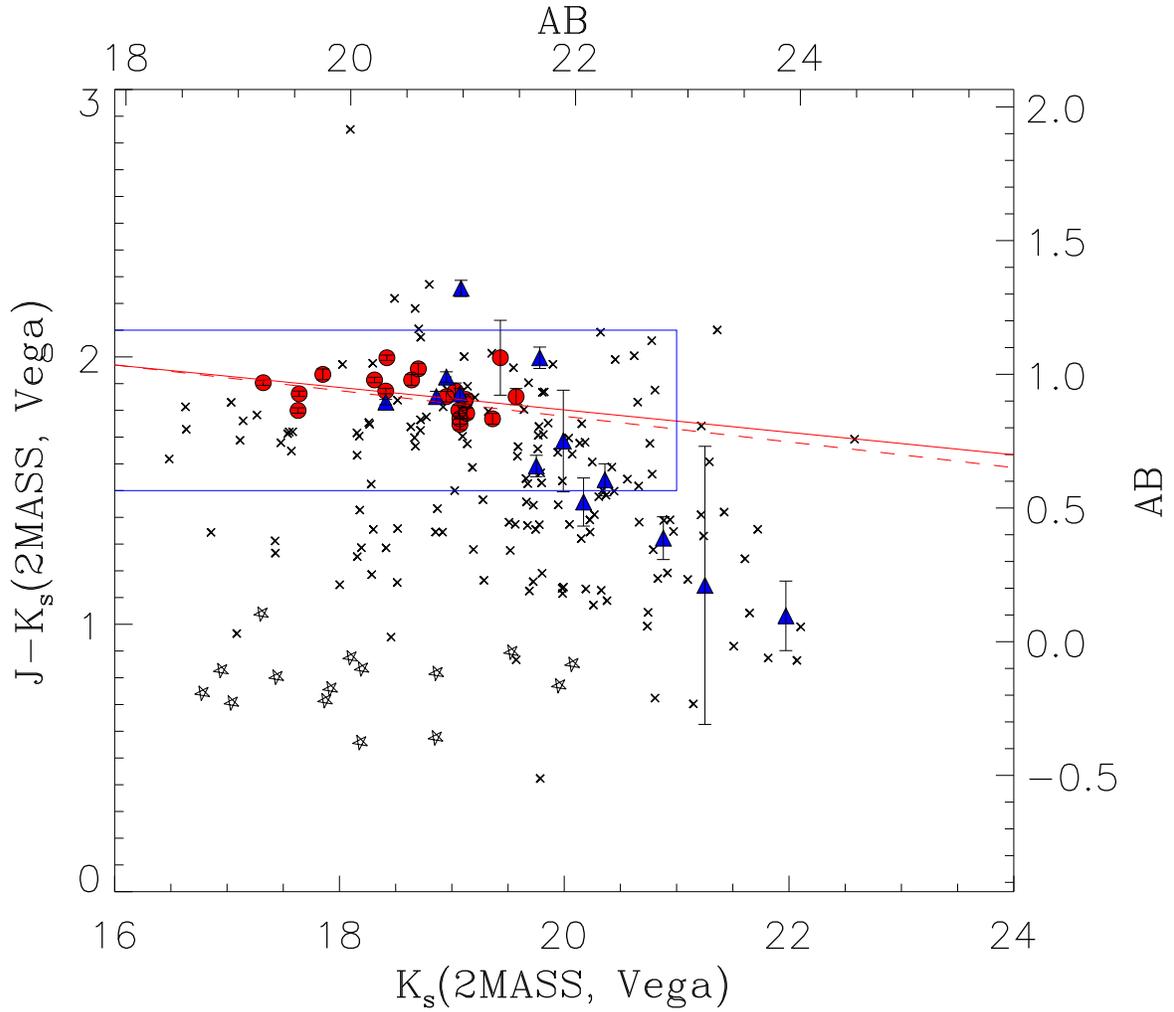}

\caption{Near-IR (reprocessed ISAAC data) color-magnitude diagram
(CMD) of spectroscopic cluster members. Colors and magnitudes are
shown in the 2MASS \citep{c01,cwm03} and AB systems. The star symbols
represent spectroscopically confirmed stars. Crosses are non-cluster
members, i.e., objects with redshift $z \leq 1.22$ or $z \geq
1.25$. Filled red circles are spectroscopic cluster members (objects
with redshift in the range $1.22 < z < 1.25$) without detectable
[OII]($\lambda$3727) emission. The filled blue triangles are cluster
members with [OII] emission. The dotted red line is the fit published
in \citet{lrd04} including only galaxies within 20\arcsec\ ($0.17$
Mpc) of the cluster center and within the blue rectangle. The solid
red line (fit \#2 in Table \ref{cmfit}) is the fit presented in this
work using only cluster members without [OII] (filled red
circles). The slope and scatter about the two fits are listed in Table
\ref{cmfit}. The reddest star-forming member in the diagram
corresponds to the confirmed AGN (ID=174). In this study we extend the
analysis of the color-magnitude diagram of \CLhz\ to more than $1$
Mpc in radius from the cluster center, and show that the red sequence
extends to more than $0.5$ Mpc in clustercentric radius.}

\label{cmd}
\end{figure}
\end{center}

\begin{center}
\begin{figure}
\plotone{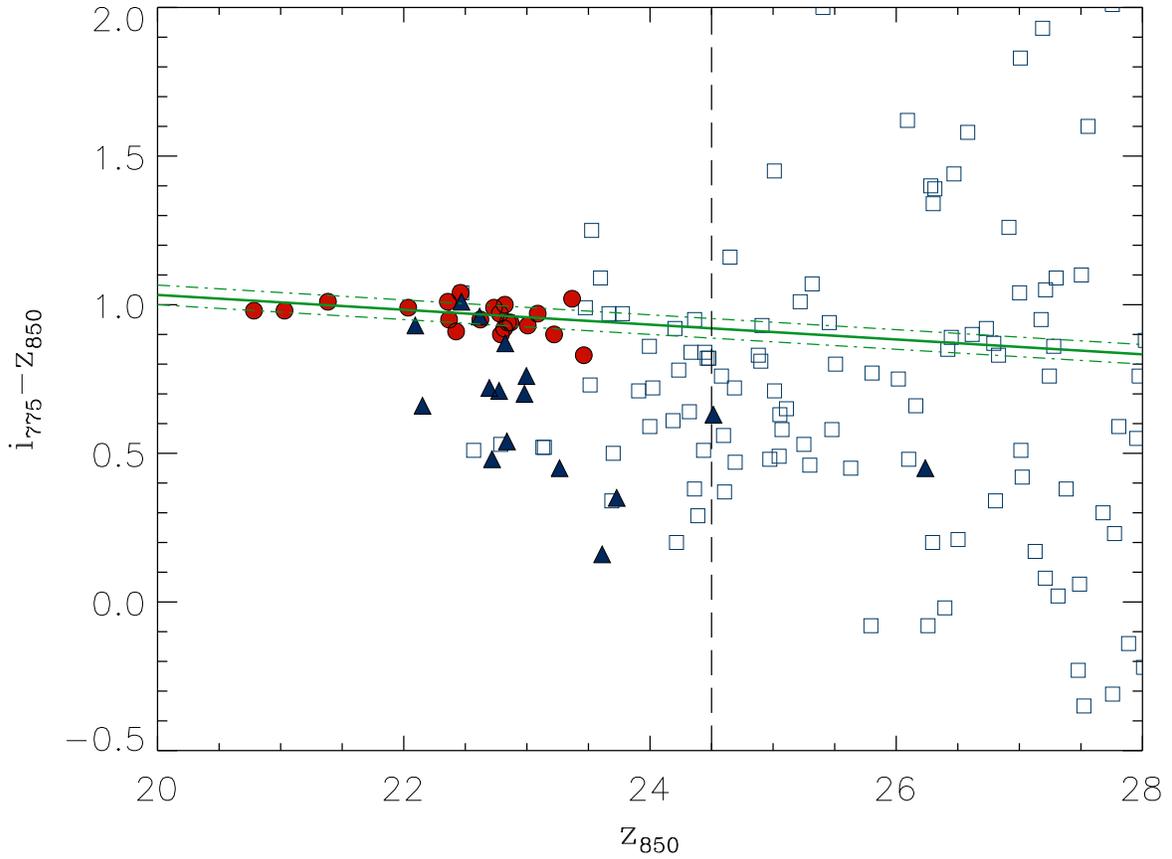}

\caption{ACS color-magnitude diagram (CMD) of spectroscopic cluster
members (circles and triangles) and photometric members
(squares). Photometric redshifts are computed by using the Bayesian
method \citep{b00,bfb04}. Colors and magnitudes are computed in the AB
system.  Red circles correspond to passive members while blue
triangles correspond to [OII] emission line galaxies. The best-fit
color-magnitude relation and scatter from \citet{bfp03} are shown in
green. Only objects within the central 1 Mpc region around the cluster
center and with $1.14 < z_{phot} < 1.34$ are shown. A color-magnitude
relation for early-type galaxies is seen down to $z_{850}=24.5$
(dashed vertical line). At fainter magnitudes this relation seems to
be truncated in the bandpasses shown here, although this cutoff
magnitude can be affected by uncertainties in the photometric
redshifts.}

\label{cmd_acs}
\end{figure}
\end{center}

\begin{center}
\begin{figure}
\plotone{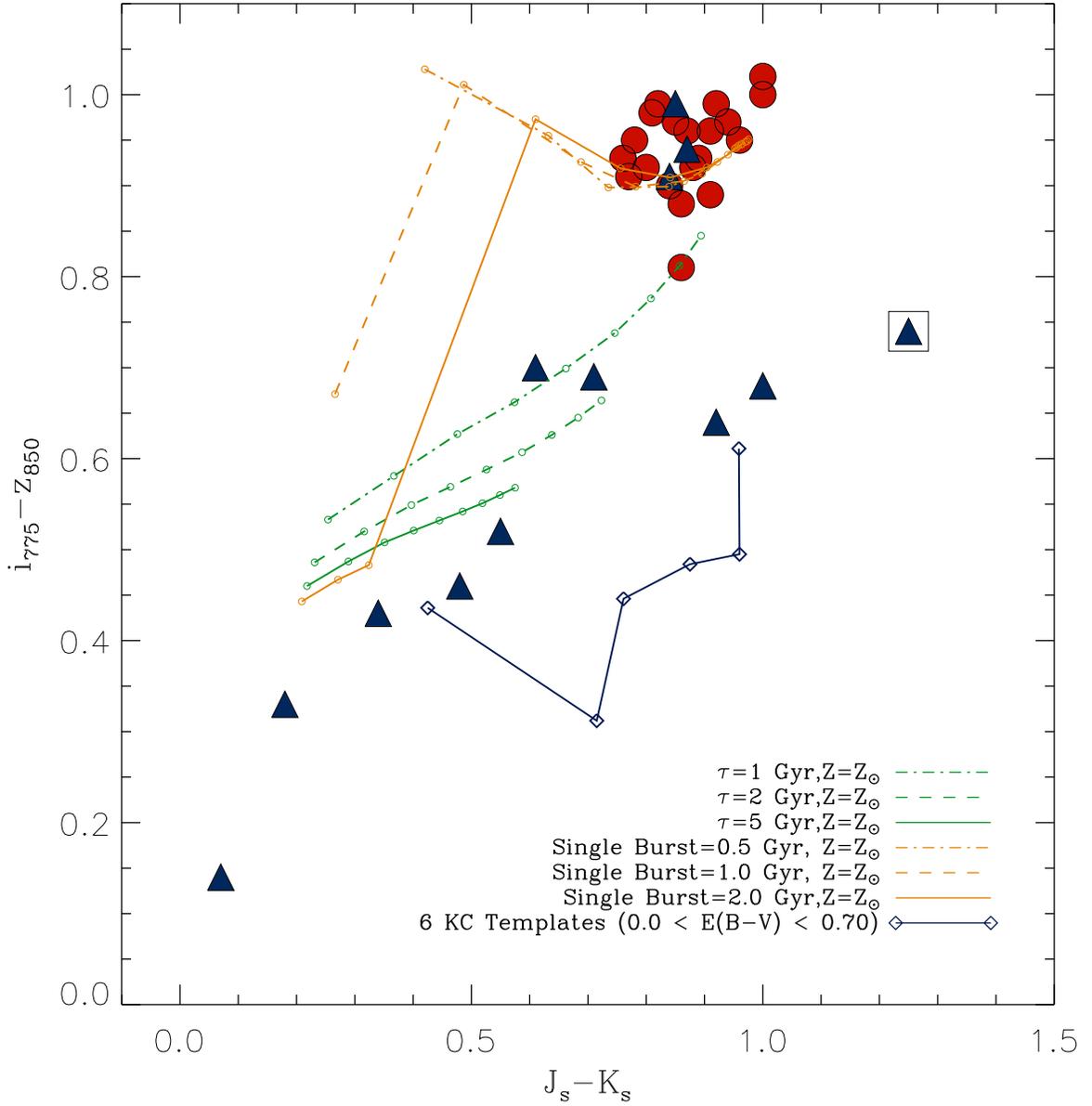}

\caption{Distribution of spectroscopic cluster members in color-color
space. All magnitudes are AB. Red filled circles correspond to passive
galaxies and blue filled triangles to [OII] emission galaxies. The
open square indicates the only known AGN at the cluster redshift
(ID=174). Model tracks are the same as those used in
Fig. \ref{color_selection}, spanning from 1 Gyr to 5 Gyr with
intervals of 0.5 Gyr (see text). In this plot we show at the same time
the color of passive and star-forming galaxies in both ACS and ISAAC
filters. We note that this color-color diagram improves the selection
of passive, red objects with respect to a $J-K$ color cut only.}

\label{colcol}
\end{figure}
\end{center}

\end{document}